\documentclass[12pt]{article}

\usepackage{graphicx,psfrag,epsf,color}
\usepackage{amsmath,amssymb,amsfonts}
\usepackage{array}
\usepackage{cite}
\usepackage{multirow}
\usepackage{rotating}
\usepackage{slashed}
\usepackage{bm}

\usepackage{enumerate}

\usepackage{tensor}
\usepackage{mathtools}
\usepackage[colorlinks]{hyperref}
\usepackage{slashed}
\usepackage[dvipsnames]{xcolor}
\usepackage{cleveref}
\usepackage{dsfont}
\hypersetup{pageanchor=false,linkcolor=NavyBlue,citecolor=NavyBlue,urlcolor=RoyalPurple}

\usepackage[colorinlistoftodos]{todonotes} 
\presetkeys{todonotes}{fancyline, color=blue!15}{}

\newcounter{MBQ}

\newcounter{RSQ}

\newcounter{PHQ}

\newcounter{PHT}

\newcounter{PHA}

\newcommand{\be}{\begin{equation}}
\newcommand{\ee}{\end{equation}}
\newcommand{\bea}{\begin{eqnarray}}
\newcommand{\eea}{\end{eqnarray}}
\newcommand{\bi}{\begin{itemize}}
\newcommand{\ei}{\end{itemize}}
\newcommand{\ben}{\begin{enumerate}}
\newcommand{\een}{\end{enumerate}}
\newcommand{\bt}{\begin{tabular}}
\newcommand{\et}{\end{tabular}}
\newcommand{\lc}{\left[}
\newcommand{\rc}{\right]}
\newcommand{\lp}{\left(}
\newcommand{\rp}{\right)}
\newcommand{\ket}[1]{\left\lvert#1\right\rangle}
\newcommand{\bra}[1]{\left\langle#1\right\rvert}

\newcommand{\np}{n_+}
\newcommand{\nm}{n_-}

\def\de{d}
\newcommand{\chris}[2]{\tensor{\Gamma}{^{#1}_{#2}}}
\newcommand{\jac}[2]{\frac{\partial y^{#1}}{\partial x^{#2}}}
\newcommand{\jacin}[2]{\frac{\partial x^{#1}}{\partial y^{#2}}}
\newcommand{\brac}[1]{\left[#1\right]}
\newcommand{\Riem}[2]{\tensor{R}{^{#1}_{#2}}}

\newcommand{\hinv}{\mathfrak{h}}

\newcommand{\delp}{n_+\partial}
\newcommand{\nip}{n_{i+}}
\newcommand{\nim}{n_{i-}}
\newcommand{\ch}{\mathfrak{h}}
\definecolor{navy}{rgb}{0.0,0.0,0.5}

\newcommand{\nn}{\nonumber}

\numberwithin{equation}{section}

\begin{document}
\allowdisplaybreaks

\begin{titlepage}

\begin{flushright}
{\small
TUM-HEP-1378/21\\
March 10, 2022 \\
arXiv:2112.04983 [hep-ph]
}
\end{flushright}

\vskip1cm
\begin{center}
{\Large \bf Soft-collinear gravity beyond the leading power}
\end{center}

  \vspace{0.5cm}
\begin{center}
{\sc Martin~Beneke,$^{a}$ \sc Patrick~Hager,$^{a}$ and Robert~Szafron$^{b}$} 
\\[6mm]
{\it ${}^a$Physik Department T31,\\
James-Franck-Stra\ss e~1, 
Technische Universit\"at M\"unchen,\\
D--85748 Garching, Germany}\\[6mm]
{\it ${}^b$Department of Physics, Brookhaven National Laboratory,\\ 
Upton, N.Y., 11973, U.S.A.}
\end{center}
\vskip1cm

\begin{abstract}
\noindent
We construct ``soft-collinear gravity'', the effective field 
theory which describes the interaction of collinear and soft gravitons with matter (and themselves), to all orders in the soft-collinear power expansion. Despite the absence of collinear divergences in gravity at leading power, the construction exhibits remarkable similarities with soft-collinear effective theory of 
QCD (gauge fields). It reveals an emergent soft background gauge symmetry, which allows for a manifestly gauge-invariant representation of the interactions in terms of a soft covariant derivative, the soft Riemann tensor, and a covariant generalisation of the collinear light-cone gauge metric field. 
The gauge symmetries control both the unsuppressed collinear field components and the inherent inhomogeneity in $\lambda$ of the invariant objects to all orders, resulting in a consistent expansion.
\end{abstract}

\end{titlepage}


\section{Introduction}

The quantum field theory of gravity is currently  
fully understood only as a low-energy effective field theory 
(EFT) below the Planck scale. Compared to 
non-abelian gauge theory, perturbative gravity is still 
a complicated subject. This motivates the consideration 
of its structure in certain limits, in particular, 
as gravitons are massless, the collinear and 
soft ones. The corresponding soft-collinear EFT (SCET) 
for gravity describes how energetic matter particles 
(or gravitons themselves) interact with soft 
gravitons or split off collinear ones. 

It has been noted long ago that the soft or eikonal 
limit of gravity is similar to that of gauge theory 
\cite{Weinberg:1965nx}, while collinear emission is 
rather different. In fact, collinear divergences are 
completely absent  \cite{Akhoury:2011kq}. An effective 
field theory (``soft-collinear gravity''), which makes 
this difference evident, has first been constructed 
in \cite{Beneke:2012xa}, following the development of 
SCET for QCD \cite{Bauer:2000yr,Bauer:2001yt,Bauer:2002nz,Beneke:2002ph,Beneke:2002ni}. The absence of collinear 
divergences makes the leading-power soft-collinear gravity Lagrangian for matter rather peculiar, as the purely collinear interactions are lacking. Yet, similar to gauge theories, certain components of the collinear graviton field are not suppressed (and, unlike in gauge theories, even enhanced) 
by power-counting. These fields must be controlled to all 
orders to ensure a consistent EFT. In QCD (gauge theories), this is done with the help of collinear Wilson lines. A direct analogue of collinear Wilson lines does not exist for SCET gravity, but a conceptually similar construction ensures that the unsuppressed components of the collinear field do not couple to the sources and can be gauged away or controlled in the Lagrangian. On the other hand, soft Wilson lines naturally also appear in 
gravity \cite{Naculich:2011ry,White:2011yy,Beneke:2012xa}.

These differences and similarities suggest that one should 
understand them from a more general perspective and to higher orders in the soft-collinear expansion. In the case of SCET for QCD, the subleading Lagrangian interactions were derived early on, and there exists a formalism which allows one to construct the soft-collinear Lagrangian to any order in the power expansion \cite{Beneke:2002ni}. Important ingredients in this construction are the multipole expansion around the light-cone and the implementation of separate collinear and soft gauge symmetries, which are power-counting homogeneous. A similar construction and understanding is not available for gravity, and there is very little work \cite{Beneke:2012xa, Okui:2017all,Chakraborty:2019tem} concerned with subleading-power topics.  

In this paper, we continue the investigation of soft-collinear graviton interactions with matter and self-interactions by going beyond the leading terms in the soft-collinear expansion. Our primary motivation is to 1) provide a conceptually (though not necessarily technically) transparent formulation of the collinear and soft limit in the sense of a systematic power expansion, and 2) to exhibit the differences and similarities of gravity and gauge theories. Our main finding is that despite the apparently very different leading-power collinear physics, SCET for gauge theory and gravity is much more similar than might have been expected. In both cases, the collinear sector can be expressed in a manifestly gauge-invariant fashion through fields representing the physical degrees of freedom, ``dressed'' by collinear ``Wilson lines''. The soft interactions take a transparent form after the light-front multipole expansion. A soft background gauge symmetry living on the light-cone emerges in QCD and in gravity, and the interactions can be expressed entirely in terms of the corresponding background-field covariant derivative or directly in terms of the gauge-invariant field-strength (Riemann) tensor. In the present work we explain and derive this remarkable structure of soft-collinear gravity. As an immediate consequence, we provided a new derivation of the gravitational soft theorem up to sub-subleading order in \cite{Beneke:2021umj}, which also shows that the universality of the next-to-soft and next-to-next-to-soft terms in gravity is a simple consequence of the above mentioned emergent soft gauge invariance. 

These results add to the growing body of evidence of 
structural similarities of the theories of massless spin-1  
(gauge theory) and spin-2 particles (gravity).
A direct relation between both theories has been understood for some time in the form of the gauge-gravity double copy \cite{Bern:2008qj,Bern:2010yg}, for a comprehensive review see \cite{Bern:2019prr}. 
Using the double copy, it is possible to compute gravitational scattering amplitudes from gauge-theory ones.
There are also hints of this double copy already at the Lagrangian level
\cite{Bern:1999ji,Beneke:2021ilf}, so a soft-collinear effective Lagrangian for gravity could prove useful to further identify connections between QCD SCET and gravity. Since one only employs physical degrees of freedom in the effective theory, this could provide a more direct link to the properties of on-shell scattering amplitudes.

The outline of this paper is as follows. In section 2, we review the construction of SCET for gauge theories. We use the example of scalar QCD, which has not been discussed in the literature before. Section 3 briefly introduces the QFT of gravity as a theory of a massless  spin-two tensor field minimally coupled to the scalar field. In section 4, we derive the purely collinear Lagrangian for gravity. The paper's main result is given in section 5, where we present the multipole expansion and provide the background field construction for the soft-collinear interactions. This leads to the complete scalar SCET gravity Lagrangian, which includes both collinear and soft graviton fields, 
which we provide explicitly to order $\lambda^2$ in the SCET expansion parameter. A number of appendices collect additional details on useful gauges and the derivation of the soft-collinear Lagrangian.


\section{SCET for scalar QCD}\label{sec::SQCD}

It turns out that SCET gravity can be constructed along 
very similar lines as the gauge theory (QCD) case. 
Since this is not obvious from the beginning, we describe 
first the construction of the soft-collinear effective theory for scalar QCD (sQCD), i.e.~a  scalar field $\phi$ coupled to a non-abelian gauge field $A_\mu$. We consider the scalar 
theory for two reasons. We will later consider gravity coupled to scalar matter. Furthermore, the SCET sQCD has actually never been constructed beyond the leading power before, so it is interesting to see the similarities and differences to the fermionic case \cite{Beneke:2002ni}.

Most of the results presented in this section are applications of \cite{Beneke:2002ph,Beneke:2002ni} to the scalar field, but we provide a comprehensive introduction to most concepts, as well as some intuition that has not been appreciated before, as these are useful for the more complicated gravitational setting.
The SCET expert reader may skip this section.

We begin with the basic kinematical definitions, define the field content and discuss the effective-theory gauge symmetry in detail. After this, we proceed with the construction of the effective Lagrangian and give some remarks on the source basis using the $N$-jet operator as a prototypical example.
More details concerning the SCET construction can be found in \cite{Bauer:2000yr,Bauer:2001yt,Bauer:2002nz,Beneke:2002ph,Beneke:2002ni}, for the operator basis see \cite{Kolodrubetz:2016uim,Moult:2017rpl,Feige:2017zci,Beneke:2017ztn,Beneke:2018rbh,Beneke:2019kgv}.

\subsection{Power-counting}
\label{sec::SQCD::PowerCounting}

To begin with the construction of the effective theory, we first clarify the power-counting rules. We introduce two light-like vectors $n_{\pm}^\mu$,\footnote{In case of multiple relevant collinear directions, we introduce pairs of light-like vectors $n_{i\pm}^\mu$, one for each direction.} which satisfy
\begin{equation}
    \np \nm = 2\,,\quad n_{\pm}^2 = 0\,.
\end{equation}
The collinear momentum $p$ is decomposed in this basis as
\begin{equation}
    p^\mu = \np p \,\frac{\nm^\mu}{2} + \nm p \,\frac{\np^\mu}{2} + p_{\perp}^\mu\,.
\end{equation}
We expand the theory in the small parameter $\lambda \sim 
p_\perp/(\np p)\ll 1$, where the components of the collinear momentum-vector scale as
\begin{equation}\label{eq::SQCD::MomentumPowerCounting}
(\np p, p_{\perp}, \nm p) \sim (1,\lambda,\lambda^2)Q\,,
\end{equation}
and $Q$ is the hard scale of the process, often set to 1 
in the following.
In addition, the theory features the soft matter and 
gluon modes, whose momenta scale as 
$k_s\sim\lambda^2 Q$. This setting, with soft virtualities 
much smaller than the collinear ones, is commonly referred 
to as a SCET$_{\rm I}$ problem. We shall construct 
SCET gravity for this situation, corresponding to \cite{Bauer:2000yr,Bauer:2001yt,Bauer:2002nz,Beneke:2002ph,Beneke:2002ni} for QCD.

\subsection{Field content}\label{sec::SQCD::FieldContent}

In scalar QCD, the full-theory fields are given by a scalar $\phi$ in the fundamental representation of $\mathrm{SU}(N)$, and by the gluon field $A_\mu$.
Unlike in traditional Wilsonian EFTs, we do not integrate out heavy particles, but rather certain regions of loop momentum or phase-space, or modes corresponding to fluctuations with definite scaling.
This means that the effective theory makes use of multiple fields describing the same particle species, but with different fluctuations adapted to different kinematic regions.
The hard region is given by momenta scaling as $p_\mu\sim Q$,
and we integrate out this region.
We are then left with (multiple) collinear regions, as well as the soft one.
Hence, we split both full-theory fields $\phi$, $A_\mu$ into their collinear and soft modes, integrating out the hard region. 
In the EFT, this is manifested by a collinear field $\phi_{c}$ ($\phi_{c_i}$ in case of multiple collinear directions) and a soft field $\phi_s$ for each full-theory field $\phi$.
We can view this split as a Lagrangian-level implementation of the method of regions \cite{Beneke:1997zp}.

Our approach is based on the position-space formulation of SCET developed in \cite{Beneke:2002ph,Beneke:2002ni}.
In this formalism, we assign a scaling to the position 
argument of the fields, deduced from $e^{i p\cdot x} 
\sim 1$. Collinear fields then depend on $\nm x\sim 1$, $x_\perp\sim \lambda^{-1}$, $\np x\sim\lambda^{-2}$, and the 
$\lambda$-scaling reflects the characteristic distance 
over which the fields exhibit substantial variations.
Consequently, this also gives the scaling of derivatives 
acting on the collinear fields, which scale like a 
collinear momentum, i.e.
\begin{equation}
    \np\partial\phi_c \sim \phi_c\,,\quad \partial_\perp\phi_c\sim\lambda\phi_c\,,\quad \nm\partial \phi_c\sim\lambda^2\phi_c\,.
\end{equation}
Soft fields vary only over the large distance $x_s\sim\lambda^{-2}$, and derivatives of soft fields scale homogeneously as
\begin{equation}
    \partial_\mu\phi_s\sim\lambda^2\phi_s\,.
\end{equation}

From their respective two-point functions, we can find the scaling of collinear and soft fields themselves \cite{Beneke:2002ph}.
For the scalar field, we have
\begin{equation}\label{eq::SQCD::ScalarTwoPoint}
    \bra{0} T (\phi(x)\phi(y))\ket{0} = \int \frac{d^4p}{(2\pi)^4}\,e^{-ip(x-y)}\,\frac{i}{p^2 + i\varepsilon}\,,
\end{equation}
and inserting the power-counting \eqref{eq::SQCD::MomentumPowerCounting}, we find the scaling
\begin{equation}\label{eq::SQCD::ScalarScaling}
    \phi_c \sim \lambda\,,\quad \phi_s\sim\lambda^2\,,
\end{equation}
for collinear fields $\phi_c$ and soft fields $\phi_s$, respectively.
For gluon fields, we find in general covariant gauge 
\begin{equation}
    \bra{0} T (A_\mu(x)A_\nu(y))\ket{0} = \int\frac{d^4p}{(2\pi)^4} \,e^{-ip(x-y)}\,\frac{i}{p^2+i\varepsilon}\lc -g_{\mu\nu} + (1-\alpha)\frac{p_\mu p_\nu}{p^2}\rc,
\end{equation}
and see that gluon fields scale like momenta, i.e.
\begin{equation}
    \np A_c \sim 1\,,\quad A_{c\perp}\sim\lambda\,,\quad \nm A_c \sim \lambda^2\,,\quad  A_{s\mu} \sim \lambda^2\,.
\label{eq:gaugefieldlambdascaling}
\end{equation}
Finally, we need to fix the power-counting of the $d^4x$ measure in the effective action. The scaling depends on the presence of collinear or soft fields in the integrated Lagrangian terms: it is $\lambda^{-4}$ if there are collinear fields present and $\lambda^{-8}$ for purely soft field products.

\subsection{Gauge symmetries}
\label{sec::SQCD::GaugeSymmetry}

The soft and collinear fields not only differ in their 
power-counting but also in their gauge transformation. It is the latter that allows us to give these modes a physical interpretation in terms of collinear fluctuations in a soft background.

To understand how this arises, consider first the full-theory gluon field $A_\mu$. This field transforms under a gauge transformation as 
\begin{equation}\label{eq::SQCD::Full_Gluon_Transformation}
    A_\mu \to U A_\mu U^\dagger + \frac{i}{g} U\lc \partial_\mu\,, U^\dagger\rc .
\end{equation}
We now introduce the split 
\begin{equation}\label{eq::SQCD::GluonSplit}
    A_\mu(x) = A_{c\mu}(x) + A_{s\mu}(x)\,,
\end{equation}
and realise the original gauge symmetry \eqref{eq::SQCD::Full_Gluon_Transformation} in the form of two separate gauge symmetries -- one that acts solely on the collinear fields, and another which transforms both, the soft and collinear fields. This is very similar to the treatment of $A_{s}$ as a background field, and $A_c$ as a fluctuation on top of this background.

To be precise, we require the transformations
\begin{equation}\label{eq::SQCD::FullGaugeTransformations}
\begin{aligned}
    &\text{collinear:} & A_{c}&\to U_c A_c U_c^\dagger + \frac{i}{g} U_c\lc D_s\,, U_c^\dagger\rc\,,\;  
    & \phi_c &\to U_c\phi_c\,,\\
    &   & A_{s}&\to A_s\,,
    & \phi_s &\to \phi_s\,,\\
    &\text{soft:}    & A_{c} &\to U_s A_c U_s^\dagger\,,
    & \phi_c &\to U_s\phi_c\,,\\
    &  & A_s &\to U_s A_s U_s^\dagger + \frac{i}{g}U_s\lc \partial\,, U_s^\dagger\rc\,,
    & \phi_s &\to U_s\phi_s\,.
    \end{aligned}
\end{equation}
Under collinear transformations, $A_s$ does not transform. However, it appears in the form of the soft-covariant derivative $D_s^\mu = \partial^\mu - ig A_s^\mu$ in the collinear transformation, like a background field.
Conversely, under a soft transformation, $A_s$ has the standard transformation \eqref{eq::SQCD::Full_Gluon_Transformation}, and $A_c$ has the covariant transformation of a (non-gauge) field in the adjoint representation.\footnote{From the soft perspective, any collinear field, regardless if gauge or matter, has the same transformation, depending only on its representation.} 
Note that the $x$-argument of the local transformations $U_c$ and $U_s$ has the same scaling as collinear and soft fields, respectively. This ensures that the scaling of the 
gauge fields \eqref{eq:gaugefieldlambdascaling} remains unaltered by the gauge symmetry of the effective theory.

The soft matter fields also 
do not transform under collinear gauge transformations.
To ensure these transformations, we make use of Wilson lines, and decompose the matter field as
\begin{equation}\label{eq::SQCD::ScalarSplit}
    \phi = \phi_c + WZ^\dagger \phi_s\,,
\end{equation}
introducing the Wilson line product
\begin{equation}\label{eq::SQCD::WZdefinition}
    WZ^\dagger = P\exp \lc ig \int_{-\infty}^0 ds\: n_+A(x+sn_+)\rc \Bar{P}\exp \lc -ig \int_{-\infty}^0 ds\: n_+A_s (x+sn_+)\rc\,.
\end{equation}
Since SCET is a non-local EFT, Wilson lines play a central role in the construction.  We will employ multiple other Wilson lines throughout this work. For more details on these, we refer to \cite{Beneke:2002ph,Beneke:2002ni}.

\subsection{Effective theory construction}\label{sec::SQCD::EFTDerivation}

After introducing the collinear and soft modes, we can now proceed to construct the effective theory.
This goes in four steps and follows the construction for QCD in \cite{Beneke:2002ni}:

\begin{enumerate}[(i)]
    \item First, we introduce the decompositions given in \eqref{eq::SQCD::GluonSplit}, \eqref{eq::SQCD::ScalarSplit} into the Lagrangian, thus describing the theory of collinear fluctuations in a soft background. This Lagrangian is not yet homogeneous in $\lambda$, that is, a given term does not necessarily have uniform scaling in $\lambda$. The inhomogeneity arises from the fact that in soft-collinear interactions 
$\int d^4x\,\psi_c(x)\psi_s(x)\ldots$, the $x$-argument 
of the soft field does not have the same counting as the 
collinear measure. Soft fields should only depend on the 
component 
\begin{equation}
x_-^\mu = \np x \,\frac{\nm^\mu}{2}\,,
\end{equation} 
which has the same scaling $\lambda^{-2}$ for both, collinear and soft fields, in soft-collinear interactions.
    The reason for this is that soft fields cannot resolve the structure of the collinear jet but only its direction. This is the position-space analogue of expanding in the small (relative to collinear) soft momentum components $k_\perp$ and $\np k$ in the soft-collinear interactions.
    \item Next, to render soft-collinear interactions homogeneous, we perform the multipole expansion of the soft fields $A_s(x) = A_s(x_-) + \dots$.
    However, the collinear fields, transforming as \eqref{eq::SQCD::FullGaugeTransformations}, do not respect this multipole expansion, as they transform with $U_s(x)$ and $D_s(x)$. Hence, the gauge transformations still mix different orders in $\lambda$.
\item To remedy this, we redefine the collinear fields $\phi_c \to \hat{\phi}_c$, so that the soft gauge transformation $U_s$ of these fields respects the multipole expansion and depends only on $x_-^\mu$.
    Coincidentally, this will make the transformation homogeneous in $\lambda$.
    Inserting these redefinitions into the Lagrangian, we find that the theory is equivalent to a theory formulated in terms of a new homogeneous background field $\nm A_s(x_-)$ rather than the full $A_s^\mu(x)$.
    \item Once the theory is expressed in terms of these new fields $\hat\phi_c$, $\hat A_c$, a very transparent structure of the soft-collinear interactions emerges.  There is a homogeneous soft-covariant derivative, and all subleading interactions are expressed only in terms of the soft field-strength tensor or its derivatives.
    At this point, we can expand each term in $\lambda$ and find homogeneous expressions.
\end{enumerate}
In the following sections we provide 
this construction in the sQCD case.

\subsubsection{Background-field Lagrangian}

Step (i): we insert the decompositions \eqref{eq::SQCD::GluonSplit}, \eqref{eq::SQCD::ScalarSplit} into the Lagrangian to obtain
\begin{eqnarray}
    \mathcal{L} &=& 
    \frac 12 \lc n_+D\phi_c\rc^\dagger n_-D \phi_c 
    + \frac 12 \lc n_-D \phi_c\rc^\dagger n_+D\phi_c \nonumber \\
    &&
    + \, \lc D_{\mu_\perp}\phi_c\rc^\dagger D^{\mu_\perp}\phi_c
    + \lc D_{s\mu} \phi_s\rc^\dagger D^{\mu}_s \phi_s \nonumber \\
    &&
    +\, \frac 12 \lc n_+D\phi_c\rc^\dagger n_-D WZ^\dagger \phi_s 
    + \frac 12 \lc \np D WZ^\dagger \phi_s\rc^\dagger \nm D\phi_c\nonumber \\
    &&
    +\, \frac 12 \lc n_-D\phi_c\rc^\dagger n_+D WZ^\dagger \phi_s 
    + \frac 12 \lc \nm D WZ^\dagger \phi_s\rc^\dagger \np D\phi_c\nonumber \\
    &&
    +\, \lc D_{\mu_\perp}\phi_c\rc^\dagger D^{\mu_\perp} WZ^\dagger \phi_s 
    + \lc D_{\mu_\perp} WZ^\dagger \phi_s\rc^\dagger D^{\mu_\perp}\phi_c\,,
\label{eq::SQCD::BackgroundFieldLagrangian}
\end{eqnarray}
where $D_\mu = \partial_\mu - ig A_{c\mu} - ig A_{s\mu}(x)$.
It can be easily checked that this Lagrangian is invariant under the background-field gauge symmetry \eqref{eq::SQCD::FullGaugeTransformations}.

This Lagrangian is not homogeneous in $\lambda$. It contains an inhomogeneous gauge-covariant derivative, due to its dependence on $A_{s\mu_\perp}$ and $\np A_{s}$, and the soft fields themselves are inhomogeneous in $\lambda$ in the soft-collinear interactions, as they still depend on $x^\mu$ rather than $x_-^\mu$.
Hence, we continue with step (ii) and perform the multipole expansion of all soft fields about $x_-^\mu$ as
\begin{equation}
\begin{aligned}
    \phi_s(x) &= \phi_s(x_-) 
    + x_\perp^\alpha\lc \partial_\alpha\phi_s\rc (x_-)\\
    &\quad
    + \frac 12  \nm x \lc \np\partial \phi_s\rc(x_-)
    + \frac 12 x_\perp^\alpha x_\perp^\beta \lc \partial_\alpha \partial_\beta \phi_s\rc (x_-) + \mathcal{O}(\lambda^3\phi_s)\,,
    \end{aligned}
\end{equation}
\begin{equation}
\begin{aligned}
    A_s(x) &= A_s(x_-) 
    + x_\perp^\alpha\lc \partial_\alpha A_s\rc (x_-)\\
    &\quad
    + \frac 12 \nm x \lc\np\partial A_s\rc(x_-)
    + \frac 12 x_\perp^\alpha x_\perp^\beta \lc \partial_\alpha \partial_\beta A_s\rc (x_-) + \mathcal{O}(\lambda^3 A_s)\,.
    \end{aligned}
\end{equation}
However, collinear fields still transform with $U_s(x)$, as seen in \eqref{eq::SQCD::FullGaugeTransformations}. 
This means that under a soft gauge transformation, different orders in $\lambda$ are mixed after multipole expansion, as these transformations depend on a gauge-parameter $\varepsilon_s(x)$ given by
\begin{equation}
    \varepsilon_s(x) = \varepsilon_s(x_-) + x_\perp^\alpha \lc \partial_\alpha\varepsilon_s \rc (x_-) + \dots\,,
\end{equation}
which generates an infinite tower of subleading terms.

To remedy this, in step (iii), we redefine the collinear fields so that they transform with homogeneous gauge transformations $U_s(x_-)$, generated by parameters $\varepsilon_s(x_-)$ which scale homogeneously in $\lambda$.
The all-order construction of these homogeneously-transforming collinear fields is presented in \cite{Beneke:2002ni}.
In short, we require the redefined fields $\hat{A}_c$ to transform as
\begin{equation}\label{eq::SQCD::HomGaugeTransformations}
    \begin{aligned}
    &\text{collinear:} & \hat{A}_c &\to U_c \hat{A}_c U_c^\dagger + \frac{i}{g} U_c\lc D_s(x_-)\,, U_c^\dagger \rc \,, \;& \hat{\phi}_c &\to U_c\hat{\phi}_c\,,\\
    &\text{soft:} & \hat{A}_c &\to U_s(x_-) \hat{A}_c U_s^\dagger(x_-)\,, & \hat{\phi}_c &\to U_s(x_-)\hat{\phi}_c\,,
    \end{aligned}
\end{equation}
where we introduced the homogeneous soft-covariant derivative
\begin{equation}
D_{s\mu} = \partial_\mu - ig\nm A_s(x_-)\, \frac{n_{+\mu}}{2}\,,
\label{eq:softder}
\end{equation}
which contains only the $\nm A_s$ component of the gauge field at $x_-^\mu$. From now on $D_s^\mu$ will refer to this definition unless otherwise mentioned. From the gauge-symmetry perspective, working with the homogenised collinear fields is equivalent to working with the usual collinear fields in a special background, where only $\nm A_s(x_-)$ is the dynamical background field. This new background field has the residual gauge symmetry depending only on the parameter $\varepsilon_s(x_-)$ at $x_-^\mu$.
This observation is made more transparent in the following.

\subsubsection{Wilson lines}\label{sec::SQCD::Wilsonlines}

We make use of two sets of Wilson lines to find the redefinition that connects the hatted collinear fields transforming as in \eqref{eq::SQCD::HomGaugeTransformations} to the original ones defined in \eqref{eq::SQCD::GluonSplit}, \eqref{eq::SQCD::ScalarSplit}.

The first one is the collinear Wilson line $W_c$. This semi-infinite Wilson line is defined as
\begin{equation}\label{eq::SQCD::collWilsonLine}
    W_c(x) = P\exp\lp ig\int_{-\infty}^0 ds\: \np \hat{A}_c(x+sn_+)\rp\,,
\end{equation}
depending only on the new collinear field $\hat{A}_c$ in the new soft background $\nm A_s$.\footnote{Note that $W_c$ defined in terms of the original field $A_c$ does not have a simple soft gauge transformation, as $\np A_s$ is non-vanishing. Instead, the Wilson line $W$, introduced in \eqref{eq::SQCD::WZdefinition} has a good transformation under the original gauge symmetry, as it contains both $A_c$ and the background field $A_s$. The Wilson line should always be defined in terms of $A_c$ plus the relevant soft background gluon field. It just happens in gauge theory (QCD) that the homogeneous soft background field does not have a $\np$ component, hence $W_c$ takes the form of a purely collinear object.}
Under the gauge symmetries, $W_c$ transforms as
\begin{equation}
    W_c(x) \xrightarrow{\mathrm{coll.}} U_c(x) W_c(x)\,, \qquad W_c(x) \xrightarrow{\mathrm{soft}} U_s(x_-)W_c(x) U_s^\dagger(x_-)\,.
\end{equation}
Note that in collinear light-cone gauge, defined by the condition $\np \hat{A}_c(x) = 0$, we have $W_c = 1$.
The interpretation of this Wilson line is quite simple. Given an arbitrary configuration $\hat{A}_c$, a gauge transformation with the  Wilson line $W_c^\dagger = W_c^{-1}$ brings the configuration into light-cone gauge. Hence, the collinear gauge-invariant collinear gluon field
\begin{equation}
\label{eq::SQCD::InvariantCollinearGluon}
g\mathcal{\hat{A}}_c^\mu = W_c^\dagger i \hat{D}^\mu W_c 
- i D_s^\mu\,,
\end{equation}
with $\hat{D}^\mu = \partial^\mu - ig \hat{A}_c^\mu - ig n_- A_s(x_-)\frac{n_+^\mu}{2}$. We note that the soft field is present 
only in $\nm\hat{\mathcal{A}}_c$, and further $\np \mathcal{\hat{A}}_c(x) = 0$, as can be seen from $i\np\hat{D} W_c=W_c i\np\partial$ 
and $W_c^\dagger W_c=1$. $\mathcal{\hat{A}}_c^\mu$ viewed as 
a function of $A_c$ can therefore be regarded as 
the collinear gluon field in a manifestly gauge-covariant 
formulation of light-cone gauge. Similarly, we define 
the collinear gauge-invariant collinear scalar field
\begin{equation}
\label{eq::SQCD::InvariantCollinearQuark}
\hat{\chi}_c = W_c^\dagger \hat{\phi}_c\,.
\end{equation}

To move the soft gauge transformation from $x$ to $x_-$, we make use of the finite-distance, straight Wilson line 
\begin{equation}\label{eq::SQCD::RWilsonLine}
    R(x) = P\exp\lp ig \int_0^1 ds\: (x-x_-)^\mu A_{s\mu}\lp x + s(x-x_-)\rp\rp\,.
\end{equation}
This object transforms as
\begin{align}
    R(x) \xrightarrow{\mathrm{coll.}} R(x)\,,\qquad R(x)\xrightarrow{\mathrm{soft}} U_s(x) R(x) U_s(x_-)^\dagger\,,
\label{eq:Rtrafo}
\end{align}
since the soft gluon field does not transform under the collinear gauge symmetry.
Similarly to $W_c$, we can interpret this Wilson line as a transformation that moves the fields to a certain gauge, here the fixed-line gauge $(x-x_-)^\mu A_{s\mu}(x) = 0$ \cite{Beneke:2002ni}, the light-cone generalisation of fixed-point, or Fock-Schwinger gauge $x^\mu A_\mu(x) = 0$.\footnote{This gauge appears quite naturally in the context of the multipole expansion. It allows us to express the gauge field in terms of the field-strength tensor. See Appendix \ref{sec::App::FixedPointGauge} for details.}

\subsection{Redefinition and new background field}

We are now able to relate the original fields, taken in collinear light-cone gauge, which implies $WZ^\dagger =1$, to the new, hatted fields. In order that the hatted fields transform as in \eqref{eq::SQCD::HomGaugeTransformations}, we proceed as follows \cite{Beneke:2002ni}. 
First, we multiply the hatted fields with $W_c^\dagger$ according to their collinear gauge transformation, to obtain manifestly collinear-gauge-invariant fields.
Then, we multiply them with $R$ according to their soft gauge transformation, to move the soft transformation matrix $U_s$ from $x_-$ to $x$. In this way, we find objects that are collinear gauge-invariant and transform with $U_s(x)$ according to \eqref{eq::SQCD::FullGaugeTransformations}, like the original fields.
In short, we have the redefinitions
\begin{eqnarray}
         \phi_c &=& R \underbrace{W_c^\dagger \hat{\phi}_c}_{\hat\chi_c}\,,\label{eq::SQCD::CollinearRedefinition}\\
        A_{\perp c} &=& R\biggl(\underbrace{W_c^\dagger \hat{A}_{c\perp} W_c + \frac{i}{g} W_c^\dagger \lc \partial_\perp\,, W_c\rc}_{=\mathcal{\hat{A}}_{\perp c}}\,\biggr) R^\dagger\,,
\\
        \nm A_c &=& R \biggl(\underbrace{ W_c^\dagger \nm \hat{A}_c W + \frac{i}{g} W_c^\dagger \lc \nm D_s(x_-)\,, W_c\rc}_{= \nm \mathcal{\hat{A}}_c}\,\biggr) R^\dagger\,. 
\label{eq::SQCD::CollinearRedefinition2}
\end{eqnarray}
Note that the original fields on the left-hand side are in collinear light-cone gauge, but not the hatted fields on the right. 
Since the new fields are defined with respect to the homogeneous background field $\nm A_s(x_-)$, it is useful to define the gauge-covariant combination
\begin{equation}\label{eq::SQCD::GaugeCovariantSoftGluon}
    \mathcal{A}_s(x) \equiv R^\dagger A_s(x) R + \frac{i}{g} R^\dagger \lc D_s\,, R\rc\,,
\end{equation}
where $D_s^\mu$ is defined in \eqref{eq:softder}. 
$\mathcal{A}^\mu_s(x)$ corresponds to the manifestly gauge-covariant part of the original field, that can be expressed in terms of the field-strength tensor and its covariant derivatives. Note that indeed $\mathcal{A}_s$ satisfies the 
fixed-line gauge condition
\begin{equation}
(x-x_-)\cdot\mathcal{A}_s(x)=0\,.
\end{equation}
The full-theory gauge potential $A_s(x)$ splits into the homogeneous background field $\nm A_s(x_-)$, which only appears inside $\nm D_s(x_-)$, and the gauge-covariant object $\mathcal{A}_s(x)$, which only depends on the field-strength tensor $F_{s\mu\nu}$.

\subsection{Fixed-line gauge}
\label{sec::SQCD::FixedLineGauge}

The preceding statements can be seen by employing fixed-line gauge.
In the following, we summarise some properties of this gauge. Further details on fixed-line gauge can be found in \cref{sec::App::FixedLineGauge}, supplemented by 
\cref{sec::App::FixedPointGauge}.

Fixed-line gauge is defined by the condition
\begin{equation}\label{eq::SQCD::FLgauge}
    (x-x_-)^\mu A_{s\mu}(x) = 0\,.
\end{equation}
Given a soft gauge field, a gauge transformation with 
$U=R^\dagger =R^{-1}$, where $R$ is defined in 
\eqref{eq::SQCD::RWilsonLine} will bring it into 
fixed-line gauge. 
Fixed-line gauge does not fix the gauge completely, 
however, but leaves $\nm A_s(x_-)$ undetermined. 
Hence, in this gauge, $A_s^\mu(x)$ can be expressed 
in terms of the homogeneous soft background field $\nm A_s(x_-)$, as well as subleading terms proportional to the field-strength tensor, as given in \eqref{eq::SQCD::FixedLineA+Identity} to \eqref{eq::SQCD::FixedLineA-Identity}.

In terms of the gauge-covariant formulation, 
$\mathcal{A}_s$ defined in \eqref{eq::SQCD::GaugeCovariantSoftGluon} can be expressed with the help of the identities \eqref{eq::app::FLA-R} to \eqref{eq::app::FLA+R} as 
\begin{align}\label{eq::SQCD::FLExpansionA-}
   \nm\mathcal{A}_s(x) &= \int_0^1 ds \:(x-x_-)^\mu \nm^\nu R^\dagger(y(s))  F_{s\mu\nu}(y(s))R(y(s))\,,\\
    \mathcal{A}_{s\,\nu_\perp}(x) &= \int_0^1 ds\: s(x-x_-)^\mu R^\dagger(y(s)) F_{s\mu\nu_\perp}(y(s))R(y(s))\,,\label{eq::SQCD::FLExpansionAp}\\
    \np\mathcal{A}_s(x) &= \int_0^1 ds\: s(x-x_-)^\mu \np^\nu R^\dagger(y(s))F_{s\mu\nu}(y(s))R(y(s))\label{eq::SQCD::FLExpansionA+}\,,
\end{align}
where $y(s) = x_- + s(x-x_-)$.
These identities give closed all-order expressions for the soft fields. Once expanded in $\lambda$, they generate an infinite tower of subleading terms \cite{Beneke:2002ni}.
We see that the redefined theory contains the homogeneous soft background field $\nm A_s(x_-)$, which only appears inside the covariant derivative $D_s$, as well as these subleading interactions, expressed purely in terms of $F_{s\mu\nu}$.
Once the integrals over $s$ and the Wilson line $R$ are expanded in powers of $\lambda$, these interactions will depend on $F_{s\mu\nu}$ and its (covariant) derivatives at $x_-$.

\subsection{All-order soft-collinear Lagrangian}

Now, in step (iv), we insert the redefinitions into the background-field Lagrangian \eqref{eq::SQCD::BackgroundFieldLagrangian}, i.e. we express the original fields $\phi_c$, $A_c$ in terms of $\hat{\phi}_c$, $\hat{A}_c$ as defined in \eqref{eq::SQCD::CollinearRedefinition} to \eqref{eq::SQCD::CollinearRedefinition2}.
To explain the construction, we first work with collinear matter fields $\phi_c$ only, setting $\phi_s =0$.\footnote{This discussion therefore covers the first three terms in \eqref{eq::SQCD::BackgroundFieldLagrangian}.} We reintroduce the soft matter fields in the end.

First, we express the collinear fields in terms of the manifestly gauge-invariant fields $\hat\chi_c, \mathcal{\hat A}_c$ defined in \eqref{eq::SQCD::InvariantCollinearQuark} and \eqref{eq::SQCD::InvariantCollinearGluon}, respectively.
Fixing collinear light-cone gauge for the original, unhatted fields, and setting $\phi_s=0$, the Lagrangian \eqref{eq::SQCD::BackgroundFieldLagrangian} simplifies to
\begin{equation}
    \mathcal{L} = 
    \frac 12 \lc n_+D_s(x)\phi_c\rc^\dagger n_-D \phi_c 
    + \frac 12 \lc n_-D \phi_c\rc^\dagger n_+D_s(x)\phi_c
    + (D_{\mu_\perp}\phi_c)^\dagger D^{\mu_\perp}\phi_c\,,
\end{equation}
where $D_\mu = \partial_\mu - ig A_{c\mu} - ig A_{s\mu}(x)$, and here $D_s(x)=\partial_\mu - ig A_{s\mu}(x)$ still contains the full soft gauge field.
We now insert the redefinitions \eqref{eq::SQCD::CollinearRedefinition} to \eqref{eq::SQCD::CollinearRedefinition2}, and express the Lagrangian in terms of the manifestly gauge-invariant $\hat{\chi}_c$ and $\mathcal{\hat{A}}_c$. We find
\begin{equation}
\begin{aligned}
    \mathcal{L} &= \frac{1}{2} \lc \lp \np\partial - ig\np \mathcal{A}_s(x)\rp\hat \chi_c\rc^\dagger \lp \nm D_s - ig \nm\mathcal{\hat{A}}_c - ig \nm \mathcal{A}_s(x)\rp\hat \chi_c + \text{h.c.}\\
    &\quad
    + \lc\lp \partial_{\mu_\perp} - ig \mathcal{\hat{A}}_{c\mu_\perp} - ig \mathcal{A}_{s\mu_\perp}(x)\rp\hat \chi_c\rc^\dagger \lp \partial^{\mu_\perp} - ig \mathcal{\hat{A}}_c^{\mu_\perp} - ig \mathcal{A}^{\mu_\perp}_s(x)\rp\hat \chi_c\,.
\end{aligned}
\end{equation}
Next, the identities \eqref{eq::SQCD::FLExpansionA-} to 
\eqref{eq::SQCD::FLExpansionA+} allow us to express $\mathcal{A}_s$ in terms of the field-strength tensor.
In addition, to simplify the notation, we introduce the Noether currents $j^{a}_\mu$ 
\begin{equation}
\begin{aligned}
    \np j^a &= i \hat \chi_c^\dagger t^a\np\overset{\leftrightarrow}{\partial}\hat \chi_c\,,\\
    j_{\mu_\perp}^a &= i \hat \chi_c^\dagger t^a \overset{\leftrightarrow}{\mathcal{D}}_{c\mu_\perp} \hat \chi_c\,,\\
    \nm j^a &= i \hat \chi_c^\dagger t^a\nm \overset{\leftrightarrow}{\mathcal{D}}\hat \chi_c\,,
\end{aligned}
\end{equation}
and the covariant derivatives
\begin{equation}
    \begin{aligned}
    \mathcal{D}_{c\mu_\perp} &= \partial_{\mu_\perp} - ig \mathcal{\hat{A}}_{c\mu_\perp}\,,\\
    \nm \mathcal{D} &= \nm D_s - ig \nm \mathcal{\hat{A}}_c\,.
    \end{aligned}
\end{equation}
The left-right arrow indicates 
\begin{equation}
   i \hat \chi_c^\dagger t^a \overset{\leftrightarrow}{\mathcal{D}}_{c\mu_\perp} \hat \chi_c = 
   i \lp \hat \chi_c^\dagger t^a \mathcal{D}_{c\mu_\perp} \hat \chi_c - \lc \mathcal{D}_{c\mu_\perp}\hat \chi_c^\dagger\rc t^a\hat \chi_c\rp\,,
\end{equation}
with $t^a$ the colour generators in the representation of the 
scalar field. This yields the Lagrangian 
\begin{equation}
    \mathcal{L} = \mathcal{L}^{(0)} + \mathcal{L}_\mathrm{sub}\,,
\end{equation}
where the leading-power terms are
\begin{equation}
    \begin{aligned}[b]
        \mathcal{L}^{(0)} &= \frac 12 \lc \np\partial\hat \chi_c\rc^\dagger \nm D_s\hat \chi_c + \frac 12 \lc \nm D_s \hat \chi_c\rc^\dagger \np\partial\hat \chi_c + \partial_{\mu_\perp}\hat{\chi}_c^\dagger \partial^{\mu_\perp}\hat \chi_c\\
    &\quad
    + \frac 12 g \nm\mathcal{\hat{A}}_c^a\np j^a
    + g \mathcal{\hat{A}}_{c\mu_\perp}^a j^{a\mu_\perp}
    + g^2 \mathcal{\hat{A}}_{c\mu_\perp}^a\mathcal{\hat{A}}_c^{b\mu_\perp}\hat \chi_c^\dagger t^a t^b\hat \chi_c\\
    &= \frac{1}{2}\lc \np \partial\hat \chi_c\rc^\dagger \nm \mathcal{D}\hat \chi_c 
        + \frac 12 \lc \nm \mathcal{D}\hat \chi_c\rc^\dagger \np\partial\hat \chi_c
        + \lc \mathcal{D}_{c\mu_\perp}\hat \chi_c\rc^\dagger \mathcal{D}_c^{\mu_\perp}\hat \chi_c\,.
    \end{aligned}
\end{equation}
The subleading terms are 
given exactly (to all orders) in $\lambda$ by
\begin{equation}
\begin{aligned}
    \mathcal{L}_\mathrm{sub} &= 
    \frac 12 g\np j^b \int_0^1 ds\: (x-x_-)^\mu \nm^\nu  R^{ab}(y(s)) F_{s\mu\nu}^a(y(s))\\
    &\quad
    + g j^{b\nu_\perp} \int_0^1 ds\: s(x-x_-)^\mu R^{ab}(y(s)) F_{s\mu\nu_\perp}^a(y(s))\\
    &\quad
    + \frac{1}{2}g \nm j^b \int_0^1 ds\: s(x-x_-)^\mu \np^\nu R^{ab}(y(s))F_{s\mu\nu}^a(y(s))\\
    &\quad
    + \frac{1}{2} g^2\hat \chi_c^\dagger \left\{t^a\,, t^b\right\} \hat \chi_c
    \int_0^1 ds\: (x-x_-)^\mu R^{da}(y(s)) n_-^\nu F_{s\mu\nu}^d(y(s))\\
    &\qquad
    \times\int_0^1 ds'\: s' (x-x_-)^\alpha n_+^\beta R^{eb}(y(s')) F_{s\alpha\beta}^e(y(s'))\\&\quad +\frac{1}{2} g^2\hat \chi_c^\dagger \left\{t^a\,, t^b\right\} \hat \chi_c \int_0^1 ds\: s(x-x_-)^\mu R^{da}(y(s)) F_{s\mu\nu}^d(y(s))\\&\qquad\times\eta_\perp^{\nu\beta}\int_0^1 ds'\: s' (x-x_-)^\alpha R^{eb}(y(s')) F_{s\alpha\beta}^e(y(s'))\,.
\end{aligned}
\end{equation}
Here, we introduced the adjoint $R$-Wilson line
\begin{equation}
    R^{ab}(x)t^b = R^\dagger(x) t^a R(x)\,.
\end{equation}
As will be seen below, $\mathcal{L}^{(0)}$ is homogeneous and contains the purely-collinear as well as the leading-power (eikonal) soft-collinear interactions. 
All other terms, part of $\mathcal{L}_{\mathrm{sub}}$, form the infinite tower of subleading soft-collinear interactions, which are given in terms of the field-strength tensor and its covariant derivatives, once the integrals are expanded in $\lambda$.

\subsection{Expansion in $\lambda$}

Note that $\mathcal{L}^{(0)}$ does not need any further expansion, and is already homogeneous in $\lambda$, scaling as $\mathcal{O}(\lambda^4)$.\footnote{Recall that the measure $d^4x$ scales as $\lambda^{-4}$, so the action scales as $\mathcal{O}(1)$. In the following, we give the scaling with respect to this leading Lagrangian.}
Expressing this in terms of the fields $\hat{\phi}_c,\hat{A}_c$, we find
\begin{align}
    \mathcal{L}^{(0)} &= \frac 12 \lc \np D_c\hat\phi_c\rc^\dagger \nm D\hat\phi_c + \frac 12 \lc \nm D \hat\phi_c\rc^\dagger \np D_c\hat\phi_c + \lc D_{c\mu_\perp}\hat\phi_c\rc^\dagger D_c^{\mu_\perp}\hat\phi_c\,,
\end{align}
where $D_c$ and $D$ are defined in terms of $\hat{A}_c$. We can now perform the $\lambda$ expansion of the subleading terms employing the expansions
\begin{align}
    \int_0^1 ds\: &(x-x_-)^\mu \nm^\nu R^\dagger(y(s)) g F_{s\mu\nu}(y(s))R(y(s)) = x_\perp^\mu \nm^\nu g F_{s\mu\nu} \nonumber\\
    &\quad + \frac 12 \nm x \np^\mu \nm^\nu g F_{s\mu\nu} 
    + \frac12 x_\perp^\mu x_{\perp\rho}\nm^\nu \lc D_s^\rho\,,g F_{s\mu\nu}\rc + \mathcal{O}(\lambda^5)\,,\\
    \int_0^1 ds\: &s(x-x_-)^\mu R^\dagger(y(s)) gF_{s\mu\nu_\perp}(y(s))R(y(s))
    = \frac 12 x_\perp^\mu g F_{s\mu\nu_\perp} + \mathcal{O}(\lambda^4)\,,\\
    \int_0^1 ds\: &s(x-x_-)^\mu \np^\nu R^\dagger(y(s))gF_{s\mu\nu}(y(s))R(y(s))
    = \mathcal{O}(\lambda^3)\,.
\end{align}
Keeping terms up to the second order in $\lambda$, we 
obtain the Lagrangian $\mathcal{L} = 
\mathcal{L}^{(0)} + \mathcal{L}_ \chi^{(1)} + \mathcal{L}_ \chi^{(2)}$, with successive orders given by
\begin{align}\label{eq::SQCD::L0final}
     \mathcal{L}^{(0)} &= \frac 12 \lc \np D_c\hat\phi_c\rc^\dagger \nm D\hat\phi_c + \frac 12 \lc \nm D \hat\phi_c\rc^\dagger \np D_c\hat\phi_c + \lc D_{c\mu_\perp}\hat\phi_c\rc^\dagger D_c^{\mu_\perp}\hat\phi_c\,,\\
     \mathcal{L}^{(1)}_\chi &= \frac 12 x_\perp^\mu \nm^\nu g F^a_{s\mu\nu} \np j^a\,,\\
     \mathcal{L}^{(2)}_\chi 
     &= \frac 14 \nm x \np^\mu \nm^\nu g F^a_{s\mu\nu} \np j^a
     + \frac 14 x_\perp^\mu x_{\perp\rho}\nm^\nu \mathrm{tr}\lp\lc D_s^\rho\,,g F_{s\mu\nu}\rc t^a\rp \np j^a
     + \frac 12 x_\perp^\mu g F_{s\mu\nu_\perp}^a j^{a\nu_\perp}\,.\label{eq::SQCD::L2final}
\end{align}
\vskip-0.2cm\noindent
While the expression in terms of the Noether current is convenient to see the structure of soft-collinear physics, for practical computations it can be advantageous to also express the Noether currents explicitly in terms of $\hat{\phi}_c$ and $\hat{A}_c$.
The subleading-power interaction Lagrangians are then given by 
\begin{align}
    \mathcal{L}_{\phi_c}^{(1)} &= \frac 12 \hat{\phi}_c^\dagger\lp x_\perp^\mu \nm^\nu W_c gF_{s\mu\nu} W_c^\dagger\rp i\np D_c\hat{\phi}_c + \mathrm{h.c.}\,,\\
    \mathcal{L}_{\phi_c}^{(2)} &= 
    \frac 14 \hat{\phi}_c^\dagger \lp \nm x \np^\mu \nm^\nu W_c g F_{s\mu\nu} W_c^\dagger \rp i\np D_c\hat{\phi}_c
    + \frac 14 \hat{\phi}_c^\dagger \lp x_\perp^\mu\nm^\nu x_{\perp\rho} W_c \lc D_s^\rho,\,g F_{s\mu\nu}\rc W_c^\dagger\rp in_+D_c\hat{\phi}_c \nn\\
    &\quad
    + \frac 12 \hat{\phi}_c^\dagger \lp x_\perp^\mu W_c g F_{s\mu\nu}W_c^\dagger \rp i D_{c\perp}^\nu \hat{\phi}_c + \mathrm{h.c.}
\end{align}

The terms containing soft matter fields can be treated analogously.
These terms give rise to two new contributions to the Lagrangian, denoted by  $\mathcal{L}_{ \phi_c\phi_s}^{(1)}$, $\mathcal{L}_{ \phi_c\phi_s}^{(2)}$, and are given by 
\begin{align}
    \mathcal{L}_{ \phi_c\phi_s}^{(1)} &= \frac 12 \lc \np D_c \hat{\phi}_c\rc^\dagger \nm \overset{\leftarrow}{D} W_c\phi_s
    + \lc D_{c}^{\mu_\perp}\hat{\phi}_c\rc^\dagger \overset{\leftarrow}{D}_{c\mu_{\perp}}W_c\phi_s
    + \frac 12 \lc \nm D \hat{\phi}_c\rc^{\dagger} \np\overset{\leftarrow}{D}_c W_c \phi_s
    +\mbox{ h.c.}\,,
    \\
    \mathcal{L}_{ \phi_c\phi_s}^{(2)} &= \lp \frac 12\lc i\np D_c\hat{\phi}_c\rc^\dagger \nm \overset{\leftarrow}{D} + \lc i D_{c}^{\mu_\perp} \hat{\phi}_c\rc^\dagger \overset{\leftarrow}{D}_{c\mu_\perp}
    + \frac12 \lc i \nm D \hat{\phi}_c\rc^\dagger \np\overset{\leftarrow}{D}_c
    \rp W_c\,
    x_{\perp}^\rho \lc D_{s\rho_\perp}\phi_s\rc\nonumber\\
    &\quad+ \lc D_{c\mu_\perp}\hat{\phi}_c\rc^\dagger W_c D_{s}^{\mu_\perp}\phi_s
    + \frac{1}{2}\phi^\dagger_s W_c^\dagger\lp \np D_c \nm D + D_{c\mu_\perp}D_c^{\mu_\perp}\rp W_c\phi_s + \mathrm{h.c.}
\end{align}
These Lagrangians represent the scalar analogues of the 
corresponding subleading-power SCET interactions for 
Dirac fermions given in \cite{Beneke:2002ni}.
We recall that the SCET Lagrangian is not affected by renormalisation, i.e. all coefficients are exact \cite{Beneke:2002ph} to all orders in $\alpha_s$. 

\subsection{$N$-jet operator basis}
\label{sec::SQCD::OperatorBasis}

To describe scattering processes in SCET, in addition to the Lagrangian, we require hard sources that generate energetic particles, the so-called $N$-jet operators.
These objects are necessary, as the Lagrangian only describes the interactions within one collinear sector and its interactions with the soft modes, as well as purely soft interactions.
Notably, there are no interactions between different collinear sectors in the Lagrangian.
Such processes generate hard vertices, which connect multiple particles belonging to different collinear sectors, and are encapsulated in the $N$-jet operators.
The operator basis of these objects in the position-space formalism has been worked out in \cite{Beneke:2017ztn,Beneke:2018rbh} and we state the most important results here.

A generic $N$-jet operator takes the form
\begin{equation}\label{eq::SQCD::Njet}
    \mathcal{J} = \int [dt]_N \: C(t_{i_1},t_{i_2},\dots) J_s(0) \prod_{i=1}^N J_i(t_{i_1},t_{i_2},\dots)\,,
\end{equation}
where $[dt]_N =\prod_{ik} dt_{i_k}$. These operators are non-local along their respective collinear direction $\nim$, 
indicated by the displacement $t_{i_k}$.
Here, $J_i$ denote the collinear and $J_s$ the soft building blocks, and $C(t_{i_1},t_{i_2},\dots)$ is the hard matching coefficient.
The hard scattering takes place at the space-time point $X=0$, where the soft and collinear building blocks are then evaluated.

The collinear and soft gauge symmetry of the effective theory severely constrains the operator basis. For $\mathcal{J}$ to be gauge-invariant, 
the collinear operators in a given sector $i$, must be invariant under their collinear gauge symmetry. Hence, we must have
\begin{equation}
\begin{aligned}
    J_i(x) &\xrightarrow{\mathrm{col.}} J_i(x)\,, \quad& J_i(x) &\xrightarrow{\mathrm{soft}} U_s(x_{i-})J_i(x)\,,
\end{aligned}
\end{equation}
where the soft transformation depends on the representation of the building block, but, crucially, for $J_i$ at $x$ is always evaluated at $x_{i-}$. 
This restricts the collinear building blocks to combinations of the manifestly gauge-invariant fields and their derivatives.
The gauge-invariant fields are constructed with the collinear Wilson line $W_{c_i}$.
These fields start at $\mathcal{O}(\lambda)$, since $\nip\hat{\mathcal{A}}_{c_i}=0$, with the combinations
\begin{equation}
    \hat{\chi}_{c_i} = W_{c_i}^\dagger\hat{\phi}_{c_i}\,,\qquad g\hat{\mathcal{A}}^\mu_{c_i\perp_i} = W_{c_i}^\dagger \lc i D^\mu_{c_i\perp_i}W_{c_i}\rc\,.   
\end{equation}
The first important observation is that additional $\mathcal{O}(1)$ objects, namely the derivatives $\nip \partial$, can be eliminated, since the collinear operators are already non-local along the $\nim^\mu$ direction. The derivative can thus be eliminated using integration by parts.
Crucially, this shows that at a given order in $\lambda$, there are only a finite number of operators. All objects that scale as $\mathcal{O}(1)$ are redundant by virtue of the non-locality or the collinear Wilson line. 
In summary, the elementary building blocks, denoted by $J^{A0}$, are given by 
\begin{equation}\label{eq::SQCD::BuildingBlocks}
    J_i^{A0}(t_i) \in \left\{ \hat{\chi}_{c_i}(t_i\nip)\,,\hat{\chi}_{c_i}^\dagger(t_i\nip)\,,\hat{\mathcal{A}}_{c_i\perp_i}(t_i\nip)\right\}\,.
\end{equation}

There are two possible ways to construct subleading operators from the building blocks $J^{A0}$:
\begin{enumerate}[(i)]
\item adding derivatives $i\partial_\perp \sim \mathcal{O}(\lambda)$ or $i\nim D_s \sim \mathcal{O}(\lambda^2)$ to the $A0$ currents. 
    Soft gauge-covariance requires $\nim D_s = \nim\partial - ig\nim A_s(x_{i-})$ instead of $\nim\partial$.
However, $\nim D_s$ can be eliminated by equation-of-motion relations \cite{Beneke:2017ztn}, so derivative operators are characterised entirely by the number of $\partial_\perp$ derivatives added.
    We denote these types of operators by $J^{An}$, where $n$ denotes the order in $\lambda$.\footnote{The 
$\mathcal{O}(\lambda^2)$ object $\nim\mathcal{A}_{c_i}$ 
can be eliminated using the gluon equations of motion.} 
\item adding more building blocks of the same collinear direction, as each block is itself of $\mathcal{O}(\lambda)$.
    An operator consisting of two (three) fields is denoted by $J^{Bn}$ ($J^{Cn}$).
\end{enumerate}

We can similarly restrict the soft building blocks that appear in the form of $J_s$.
Under gauge transformations, these operators transform as
\begin{equation}
    J_s(x) \xrightarrow{\mathrm{col.}} J_s(x)\,,\quad J_s(x) \xrightarrow{\mathrm{soft}} U_s(x)J_s(x)\,,
\end{equation}
where the soft transformation depends on the representation, as above.
The collinear transformation does not give any constraints, since soft fields do not transform.
Hence the only requirement is that $J_s$ is a soft gauge-covariant object, for example
\begin{equation}
    \phi_s(x)\sim\lambda^2\,,\quad F_{s\mu\nu}\sim\lambda^4\,,\quad i D_s^\mu\phi_s(x)\sim\lambda^4\,,
\end{equation}
where $D_s^\mu = i\partial^\mu - ig A_s^\mu(x)$ in the purely soft case. 
For $\mathcal{J}$ to be gauge-invariant, 
we require colour-neutrality, that is the soft transformations $U_s(0)$ of all collinear and soft building blocks 
together must combine to the trivial representation, 
$\prod_{\rm building\,blocks} U_s(0) = 1$.


\section{Perturbative gravity}
\label{sec::GR}

In this preparatory section, we present some relevant facts regarding the low-energy effective theory of quantum gravity for which we construct the soft-collinear effective Lagrangian -- the Einstein-Hilbert Lagrangian and its weak-field expansion.
   The quantisation of the Einstein-Hilbert action has a long history \cite{Arnowitt:1959eec,Arnowitt:1960es,Feynman:1963ax,Mandelstam:1968hz,Mandelstam:1968ud,Fradkin:1970pn,DeWitt:1967yk,DeWitt:1967ub,DeWitt:1967uc,tHooft:1974toh}.
   The approach of treating it as an effective field theory has been pioneered in \cite{Donoghue:1994dn}, and we adopt this approach as well.
   Further useful references for this point of view are given in \cite{Donoghue:2017pgk,Donoghue:1995cz}.

\subsection{Full theory}
    
Before constructing the effective theory, we have to specify 
what is the full theory.
            
The Einstein-Hilbert action is the simplest possible action for the metric field $g_{\mu\nu}$. It consists only of the scalar curvature and is given by
		\begin{equation}\label{eq::GR::EinsteinHilbertAction}
			S_{\mathrm{EH}} = -\frac{2}{\kappa^2}\int\de^4x\: \sqrt{-g} R\,,
		\end{equation}
		where $g$ is the metric determinant, $R = g^{\mu\nu}\Riem{\alpha}{\mu\alpha\nu}$ is the Ricci scalar and $\kappa^2 = 32\pi G_N$, with the Newton constant $G_N$.
		We use the convention
		\begin{equation}
		    \tensor{R}{^\mu_{\nu\alpha\beta}}= \partial_\alpha \chris{\mu}{\beta\nu}
		    - \partial_\beta\chris{\mu}{\alpha\nu} + \chris{\mu}{\alpha\lambda}\chris{\lambda}{\beta\nu} - \chris{\mu}{\beta\lambda}\chris{\lambda}{\alpha\nu}\,,
		\end{equation}
and the metric signature $(+,-,-,-)$. The equation of motion of this theory is the vacuum Einstein equation. As a quantum field theory, this action should be viewed as the first term of an effective field theory of gravity, which is an expansion in derivatives of $g_{\mu\nu}$. Then the full action is an infinite series \cite{Donoghue:1994dn}
		\begin{equation}\label{eq::GR::EinsteinHilbertEFT}
			S_{\mathrm{grav,EFT}} = -\int \de^4x\: \sqrt{-g} \left( \Lambda  + \frac{2}{\kappa^2} R - c_1 R^2 - c_2 R_{\mu\nu}R^{\mu\nu} + \dots\right)\,,
		\end{equation}
		where $\Lambda$ is the cosmological constant and $R_{\mu\nu} = \Riem{\alpha}{\mu\alpha\nu}$ is the Ricci tensor.
		
In the following, we construct the soft-collinear effective Lagrangian for the Einstein-Hilbert action. However, given the above, it is important to note that the construction applies to any other theory that modifies gravity at higher orders in $R$, as long as diffeomorphism invariance is still the fundamental symmetry and the leading term is the standard Einstein-Hilbert term.
		
For the matter theory, we consider for simplicity the 
minimally coupled real scalar field described by the action
        \begin{equation}\label{eq::GR::GeometricScalarField}
            S_\varphi = \int\de^4x \sqrt{-g}\: \lp\frac 12 g^{\mu\nu}\partial_\mu\varphi\partial_\nu\varphi -\frac{\lambda_\varphi}{4!}\varphi^4\rp\,. 
        \end{equation}

\subsection{Diffeomorphism invariance}
    
The symmetry group of gravitational theories is the diffeomorphism group of local coordinate changes
		\begin{equation}
			x^\mu \to y^{\mu}(x)\,.
		\end{equation}
	    In particular, we consider these in the form of a \emph{local translation}
		\begin{equation}\label{eq::GR::LocalTranslation}
			x^\mu \to y^\mu(x) = x^\mu + \varepsilon^\mu(x)\,,
		\end{equation}
defined with a vector field $\varepsilon^\mu(x)$. We can consider this diffeomorphism either as an active transformation, acting only on the fields, or as a passive coordinate transformation. The active point of view,
\begin{equation}
\varphi(x) \stackrel{!}{=} \varphi^\prime(y(x)) 
=  \varphi^\prime(x+\varepsilon(x))
\equiv T_\varepsilon \varphi^\prime(x),
\label{eq:active}
\end{equation} 
is more closely linked to the gauge-symmetry case, so we use this convention in the following. The last 
equation defines the translation operator $T_\varepsilon$. 
Its expansion in $\varepsilon$ is given by the 
Taylor expansion
\begin{equation}
T_\varepsilon f(x) = f(x) 
+ \varepsilon^\alpha(x)\partial_\alpha f(x) 
+ \frac 12 \varepsilon^\alpha(x) \varepsilon^\beta(x) \partial_\alpha\partial_\beta f(x)+ \mathcal{O}(\varepsilon^3)\,.
\label{eq:Tdef}
\end{equation}

Hence, under a diffeomorphism, a scalar field transforms as
\begin{equation}
\varphi(x) \to \varphi^\prime(x) = \lc U(x)\varphi(x)\rc\,,
\label{eq:scalargaugetrafofull}
    	\end{equation}
where due to \eqref{eq:active} $U(x)$ is the inverse 
translation, $U(x)=T_\varepsilon^{-1}$. In expanded form, 
the action of $U(x)$ is given by
\begin{eqnarray}
\lc U(x)\varphi(x)\rc &=& \varphi(x) 
    	    - \varepsilon^\alpha(x)\partial_\alpha\varphi(x) \nonumber \\
    	    &&+ \,\frac 12 \varepsilon^\alpha(x) \varepsilon^\beta(x) \partial_\alpha\partial_\beta \varphi(x)
    	    + \varepsilon^\alpha(x) \partial_\alpha\varepsilon^\beta(x) \partial_\beta \varphi(x) + \mathcal{O}(\varepsilon^3)\,.\qquad
\label{eq::GR::ScalarTransformExpanded}
    	\end{eqnarray}
We use the square-bracket to emphasise that $U(x)$ is a derivative-operator, and the derivatives act only on objects inside the square bracket, and that functions are evaluated at the position $x$ after the derivative is taken.\footnote{The definition \eqref{eq:Tdef} corresponds to the literal translation of 
$\varphi(x)$ given $x$, but note that $T_\varepsilon$ 
does not equal $\exp(\varepsilon^\alpha(x)\partial_\alpha)$ 
for space-time dependent translations. This entails that 
$T_\varepsilon^{-1}\not= T_{-\epsilon}$, but is instead 
given by the right-hand side of \eqref{eq::GR::ScalarTransformExpanded}.}
When the situation is unambiguous, we drop the square brackets and derivatives only act on the object directly to their right.
    	To linear order, this transformation corresponds to the Lie derivative of the scalar field, 
    	\begin{equation}
    	    \pounds_\varepsilon\varphi = -\varepsilon^\alpha\partial_\alpha\varphi\,.
    	\end{equation}
    	For a tensor field, we define the Jacobi matrices
    	\begin{equation}
    	    \tensor{U}{^\mu_\alpha}(x) = \frac{\partial y^\mu}{\partial x^\alpha}(x)\,,\quad \tensor{U}{_\mu^\alpha}(x) = \frac{\partial x^\alpha}{\partial y^\mu}(x)\,,
\label{eq::GR::JacobianDef}
\end{equation}
which are inverses with respect to one another.
A tensor field, e.g.~$\tensor{T}{_\mu^\nu}$, then transforms as
\begin{eqnarray}
&& \tensor{T}{_\mu^\nu}(x) \to \lc U(x) \tensor{U}{_\mu^\rho}(x)\tensor{U}{^\nu_\sigma}(x) \tensor{T}{_\rho^\sigma}(x)\rc
\nonumber\\
&&\hspace*{0.5cm} = \,\tensor{T}{_\mu^\nu}(x) 
    	    - \partial_\mu \varepsilon^\rho(x)\tensor{T}{_\rho^\nu}(x)
    	    + \partial_\sigma\varepsilon^\nu(x) \tensor{T}{_\mu^\sigma}(x)
    	    - \varepsilon^\alpha(x)\partial_\alpha \tensor{T}{_\mu^\nu}(x) 
    	    + \mathcal{O}(\varepsilon^2)\,.\qquad
\label{eq::GR::TensorTransformation}
\end{eqnarray}
The linear term agrees again with the Lie derivative of the tensor field $\tensor{T}{_\mu^\nu}$. 
The gauge transformations $U$, $\tensor{U}{_\mu^\alpha}$ and $\tensor{U}{^\mu_\alpha}$ satisfy a list of useful identities (cf. \cref{sec::app::UsefulIdentities}), which we employ in the following.

\subsection{Weak-field expansion}
\label{sec::WeakFieldExpansion}
   
The starting point for the effective field theory is the weak-field expansion of the actions \eqref{eq::GR::EinsteinHilbertAction}, \eqref{eq::GR::GeometricScalarField}. 
		While we do not necessarily consider weak fields later on,
		the expansion will be justified by the power-counting in the effective theory, as explained in \cref{sec::CG::PowerCountingGravity}.
		
		To perform the weak-field expansion, we define a background metric $\bar{g}_{\mu\nu}$ and perform a substitution
		\begin{equation}
			g_{\mu\nu}(x) = \bar{g}_{\mu\nu}(x) + \kappa h_{\mu\nu}(x)\,,
		\end{equation}
which we consider as an \emph{exact} expression defining 
$h_{\mu\nu}$.
		For now, we take as background metric the Minkowski metric
		\begin{equation}
			\bar{g}_{\mu\nu}(x) = \eta_{\mu\nu}\,,
		\end{equation}
but in \cref{sec::SG} we shall make use of a non-trivial background.
        In the weak-field expansion, we also define translations as
        \begin{equation}
            x^\mu \to x^\mu + \kappa\varepsilon^\mu(x)\,.
        \end{equation}
We multiply $\varepsilon^\mu$ with $\kappa$ so that $h_{\mu\nu}$ has a linear gauge transformation that does not contain $\kappa$ explicitly.\footnote{The leading-order transformation amounts to the linear shift
\begin{equation}
\label{eq:lineardiffs}
h_{\mu\nu} \to h_{\mu\nu} - \partial_\mu\varepsilon_\nu - \partial_\nu \varepsilon_\mu + \mathcal{O}(\varepsilon^2)\,,
\end{equation}
but the higher-order terms are non-linear in $h_{\mu\nu}$.
}
		We can now express the inverse metric $g^{\mu\nu}$ and all other functions of $g_{\mu\nu}$ as a power-series in $\kappa$. For the weak-field expansion, we assume $\kappa h_{\mu\nu}\ll1$ in the chosen reference frame.
		This allows us to truncate these expansions at a certain order in $h$, and also restricts the gauge transformations to those preserving $\kappa h_{\mu\nu}\ll1$.
		In practice, this means that we identify $\mathcal{O}(\varepsilon)=\mathcal{O}(h)$ and we truncate the gauge transformations at the desired order in $\varepsilon$.

Any action then turns into an infinite series in 
$h_{\mu\nu}$, resp. $\kappa$:
		\begin{equation}
		    S = \sum_{k=0}^\infty \kappa^k S^{(k)}
		\end{equation}
This action still has the full diffeomorphism invariance. 
However, once truncated at some order, the diffeomorphisms 
must also be truncated to consistent order. An immediate consequence of 
this feature is that we cannot have objects that are homogeneous in $h$ and gauge-invariant at the same time. 
Gauge-invariant objects, such as the Riemann tensor, are 
represented as series order-by-order in $h$ or $\kappa$. 
Working with a theory expanded in $h$, subleading terms 
in $\kappa$ appear in precise combinations to  
yield a gauge-invariant theory. This is a generic feature 
of non-linearly realised symmetries.

Starting from an extension \eqref{eq::GR::EinsteinHilbertEFT} of the Einstein-Hilbert action as the full theory, the higher-order terms $S^{(k)}$ take a different form, and will contain higher-derivative operators.
		Despite such modifications, the weak-field expansion proceeds the same way.

\subsubsection{Einstein-Hilbert Lagrangian}
		
    		In this expansion, the leading-order term of the Einstein-Hilbert action \eqref{eq::GR::EinsteinHilbertAction} is given by
    		\begin{equation}
    			S_\mathrm{EH}^{(0)} = \frac 12 \int\de^4x\: \Big\{
    			\partial_\alpha h_{\mu\nu} \partial^\alpha h^{\mu\nu} - 
    			\partial_\alpha h \partial^\alpha h - 
    			2 \partial_\mu h^{\mu\nu}(\partial_\alpha h^\alpha_\nu - \partial_\nu h)
    			\Bigr\}\,,
    		\end{equation}
    		where we introduced the trace $h = \eta^{\mu\nu}\tensor{h}{_{\mu\nu}}$.
    		
    		We quantise the fluctuation $h_{\mu\nu}$, and view this as a theory on Minkowski space.
    		This theory has a residual gauge symmetry inherited from the (truncated) diffeomorphism invariance \eqref{eq:lineardiffs}, and we fix it by working in a generalised de~Donder gauge, adding the gauge-fixing term
    		\begin{equation}
    		   S_\mathrm{gf} = b\int \de^4x\: \bigl(
    			\partial_\alpha h^\alpha_\mu - 
    			\frac{1}{2}\partial_\mu h\bigr)
    			\bigl(\partial_\beta h^{\beta\mu} - 
    			\frac{1}{2}\partial^\mu h\bigr)\,.
    		\end{equation}
The graviton propagator is then given by
    		\begin{equation} \label{eq::GR::GravitonPropagator}
    			D_{\mu\nu,\alpha\beta} = 
    			\langle 0\rvert T (h_{\mu\nu}(x) h_{\alpha\beta}(y))\lvert 0\rangle = 
    			i\int\frac{\de^4 p}{(2\pi)^4}\frac{e^{-ip\cdot(x-y)}}{p^2+i0}\bigl(P_{\mu\nu,\alpha\beta} +
    			\frac{1-b}{b}S_{\mu\nu,\alpha\beta})\,,
    		\end{equation} 
    		where
    		\begin{align}
    			P_{\mu\nu,\alpha\beta} &= \frac{1}{2}\,\bigl(\eta_{\mu\alpha} \eta_{\nu\beta} + 
    			\eta_{\mu\beta}\eta_{\nu\alpha} - 
    			\eta_{\mu\nu}\eta_{\alpha\beta})\,,\\
    			S_{\mu\nu,\alpha\beta} &= \frac{1}{2p^2}\,(\eta_{\mu\alpha}p_\nu p_\beta + 
    			\eta_{\mu\beta}p_\nu p_\alpha + 
    			p_\mu p_\alpha \eta_{\nu\beta} + 
    			p_\mu p_\beta \eta_{\nu\alpha})\,.
    		\end{align}
The trilinear $\mathcal{O}(\kappa)$ Lagrangian is given by
    		\begin{equation}\label{eq::GR::TrilinearLagrangian}
    		\begin{aligned}
    		    \mathcal{L}^{(1)} &= 
    		    \frac{1}{4}h^{\alpha\beta}(-2\partial_\mu h^{\mu}_\alpha\partial_\beta h
	            + 4 \partial_\mu h_{\alpha\nu}\partial_\beta h^{\mu\nu}
            	+ 2 \partial_\nu h^{\mu\nu}\partial_\mu h_{\alpha\beta}
            	- 2 \partial_\alpha h^{\mu}_\beta\partial_\mu h\\
            	&\qquad
            	+ \partial^\mu h_{\alpha\nu}\partial^\nu h_{\beta\mu}
            	- 2 \partial^\mu h_{\beta\nu}\partial_\mu h_\alpha^\nu)\\
            	&\quad + \frac{1}{8}\partial_\alpha (h^2) \partial^\alpha h
            	+ \frac{1}{4}\partial_\alpha(h^{\mu\nu}h_{\mu\nu})\partial^\alpha h
            	- \frac{1}{8}\partial_\alpha(h^2)\partial_\beta h^{\alpha\beta}
            	- \frac{3}{4}\partial_\alpha(h^{\mu\nu}h_{\mu\nu})\partial_\beta h^{\alpha\beta}\\
            	&\quad + \frac{1}{2}\partial_\alpha(h^{\mu\rho}h_\rho^\nu)\partial^\alpha h_{\mu\nu}
            	- \partial_\alpha(h^{\mu\rho}h_\rho^\nu)\partial_\mu h^\alpha_\nu
            	+ \frac{1}{2}\partial_\mu(h^{\mu\rho}h_{\rho}^\nu)\partial_\nu h 
            	- \frac{1}{8}h\partial_\alpha h \partial^\alpha h\\
            	&\quad - \frac{1}{8}h\partial_\alpha h_{\mu\nu}\partial^\alpha h^{\mu\nu}
            	+ \frac{1}{4} h \partial_\mu h_{\alpha\beta}\partial^\alpha h^{\mu\beta}\,.
    		\end{aligned}
    		\end{equation}
Due to the gauge-fixing term, there is also a ghost Lagrangian which follows from the standard Faddeev-Popov procedure. 
The soft-collinear ghost Lagrangian could be constructed in the same way as done for matter fields below, but we do not consider it in the following, as it provides no further insights.
        
\subsubsection{Minimally-coupled scalar field}
        
            For the minimally-coupled scalar field given in \eqref{eq::GR::GeometricScalarField}, we find the Lagrangian
			\begin{equation}
				\mathcal{L} = \mathcal{L}^{(0)} + \kappa\mathcal{L}^{(1)} + \kappa^2\mathcal{L}^{(2)} + \mathcal{O}(\kappa^3)\,,
			\end{equation}
			where
			\begin{align}\label{eq::GR::truncatedLagrangians1}
				\mathcal{L}^{(0)} &= \frac{1}{2}\partial_\mu\varphi\,\partial^\mu\varphi -
				\frac{\lambda_\varphi}{4!}\varphi^4\,,\\
				\label{eq::GR::truncatedLagrangians2}
				\mathcal{L}^{(1)} &= -\frac{1}{2}h_{\mu\nu}\left(
				\partial^\mu\varphi\,\partial^\nu\varphi - 
				\eta^{\mu\nu}\frac{1}{2}\partial_\alpha\varphi\,\partial^\alpha\varphi\right) -
				\frac{1}{2}h\frac{\lambda_\varphi}{4!}\varphi^4\,,\\
				\label{eq::GR::truncatedLagrangians3}
				\mathcal{L}^{(2)} &= 
				\frac{1}{2}\left(h^{\mu\alpha}h_{\alpha}^\nu - 
				\frac{1}{2}h h^{\mu\nu} +
				\frac{1}{8}(h^2 - 2 h^{\alpha\beta}h_{\alpha\beta})
				\eta^{\mu\nu}\right)
				\partial_\mu\varphi\,\partial_\nu\varphi\nonumber\\
				&\quad - 
				\frac{1}{8}(h^2 - 2 h^{\alpha\beta}h_{\alpha\beta})\frac{\lambda_\varphi}{4!}\varphi^4\,.
			\end{align}

\subsection{QCD vs. gravity}

Let us compare the gauge-theory case to gravity, to anticipate the difficulties and similarities of the SCET construction for gravity. 
\begin{enumerate}[(i)]
\item In gravity, the ``full theory'', which forms the basis of the SCET construction, is itself an effective theory of an unknown UV-complete theory of gravity. We must specify if we start from the pure Einstein-Hilbert theory \eqref{eq::GR::EinsteinHilbertAction} or include also higher-order terms \eqref{eq::GR::EinsteinHilbertEFT}. For simplicity, we adopt the Einstein-Hilbert theory.
 \item On top of this effective theory, we perform the weak-field expansion.
    This expansion causes a non-linear realisation of the diffeomorphism symmetry, which has to be truncated at a certain order in $\kappa$.
    \item The gauge charges, i.e. the generators of the local translations, are the momenta $P^\mu$. Unlike the gauge generators $t^a$ in QCD, these are also kinematic quantities and therefore have a non-trivial scaling in the SCET power-counting parameter $\lambda$.
    \item In QCD, the $\np A_c$ component of the collinear gauge field has $\mathcal{O}(1)$ scaling in $\lambda$ and must be controlled by the SCET gauge symmetry to all orders. In gravity, we shall see in Section~\ref{sec::CG::PowerCountingGravity} that $\np^\mu h_{\mu\nu_\perp}\sim\mathcal{O}(1)$ and 
there is even a  superleading component $\np^\mu \np^\nu h_{\mu\nu}\sim\mathcal{O}(\lambda^{-1})$ \cite{Beneke:2012xa}.
\end{enumerate}

These items have important implications for the form of the effective theory.
The first one is related to the systematics of renormalisation and loops in the theory, since the curvature expansion is equivalent to a loop expansion.
The natural size of the coupling constants $c_k$ of the $c_kR^{k+1}$ 
terms in \eqref{eq::GR::EinsteinHilbertEFT} is $c_k\sim \kappa^{2k-2}$, since this is the magnitude of 
the required counterterm. If we specify the full theory to 
consist of $R$ only, i.e.~the pure Einstein-Hilbert theory, we can expand in $h$ to all orders in 
$\kappa$, but cannot calculate loops. 
For the construction of the SCET Lagrangian presented below, 
this does not present any difficulty, since the SCET 
Lagrangian is not renormalised. However, if this Lagrangian 
were to be used for the systematic calculation of the soft-collinear expansion of a full-theory process including quantum corrections, one would have to include loop corrections to the matrix elements of SCET operators, together with the hard 
matching coefficients of these operators. The matching calculations involve loop corrections in the full theory, and hence an extension of the full theory beyond the Einstein-Hilbert Lagrangian. In this work, we focus on the principles that underlie the construction of the SCET gravity Lagrangian. These are governed by the diffeomorphism invariance of the full theory and independent of the precise form of the full theory. 
We therefore focus on the Einstein-Hilbert theory, 
expanding it to a fixed order in $\kappa$. 
{\em Then}, we expand this ``full theory'' to all orders in $\lambda$. 

The second and the third point are connected in the 
effective theory.
As mentioned before, in the effective theory we do not assume that the fields are necessarily weak. What we do assume, however, are the SCET kinematics. These kinematics then force us to perform an expansion analogous to the weak-field expansion, as will be explained in \cref{sec::CG}.

Crucially, since the gauge charges scale with $\lambda$, a gauge transformation links different orders of $\lambda$.\footnote{This is very similar to $n_\pm^\mu$ reparameterisation (RPI) transformations in QCD SCET \cite{Manohar:2002fd}. In fact, RPI transformations being related to the Lorentz symmetry of the full theory are related to the gauge transformations in SCET gravity, unlike the case of QCD.}
This implies that one cannot have objects that are simultaneously gauge-invariant and homogeneous in $\lambda$.
Thus, when we construct gauge-invariant objects, these can 
be classified by their leading $\lambda$-counting, but include an infinite tower of non-linear subleading terms which 
are fixed by the gauge symmetry of the effective theory. 
In practical computations, we work with these inhomogeneous but gauge-invariant building blocks and relate them to the original fields, analogous to QCD, where the building blocks are inhomogeneous in the strong coupling $g$ (but not in $\lambda$), due to the collinear Wilson line.

The fourth point is solved in a remarkably similar fashion to the QCD problem, where the $\np A_c$ component must be controlled. 
In QCD, the collinear Wilson line is employed to construct gauge-invariant operators, which satisfy the light-cone gauge condition $\np\mathcal{A}_c = 0$.
In gravity, one constructs analogues of collinear Wilson lines to define gauge-invariant graviton fields $\hinv_{\mu\nu}$, which satisfy the light-cone gauge condition $\mathfrak{h}_{\mu+}=0$ to all orders.
In this way, one controls the presence of the superleading $h_{++}\sim\mathcal{O}(\lambda^{-1})$ and leading $h_{+\mu_\perp}\sim\mathcal{O}(\lambda^0)$ components, and works with the transverse components only, which enter at $\mathcal{O}(\lambda)$, just like in QCD.

Besides these differences, the construction of the soft-collinear interactions and the $N$-jet operators is surprisingly similar to the gauge-theory case. One still performs the soft light-front multipole expansion and has to redefine the collinear fields to ensure that their gauge transformation respects this expansion.
The gauge condition analogous to fixed-line gauge is then a generalisation of the Riemann normal coordinates.
In these normal coordinates, the soft metric field and its derivative can be expressed in terms of a residual background field and the covariant Riemann tensor.
The structure of the Lagrangian then takes the same form -- a covariant derivative, and interactions described in terms of the Riemann tensor.
However, due to the previously mentioned points, all gauge-covariant objects are inhomogeneous in $\lambda$. To express this in terms of objects with a definite $\lambda$-counting, one has to perform the soft weak-field expansion.


\section{Collinear gravity}
\label{sec::CG}

We first consider the effective theory in the absence of soft fields, that is, 
the purely collinear sector of the minimally-coupled scalar field and
Einstein-Hilbert action. This theory is equivalent to the weak-field 
expansion of the full theory in light-cone gauge, but serves as a 
starting point for adding back the soft fields. Already here, we make
contact with some subtleties of gravity, namely the inhomogeneity inherent 
to the gauge transformations, due to the presence of infinitely many 
subleading collinear interaction terms from the expansion of the 
metric tensor around flat space, and due to the non-trivial scaling of 
the charge -- the momentum. Furthermore, we introduce the analogues of the Wilson lines used in QCD.
    
After stating the power-counting and motivating the collinear expansion, we proceed to write down the purely collinear Lagrangian.
    We briefly discuss the remaining symmetry, but overall the situation is very analogous to the weak-field expansion.
    In the collinear sector, we encounter a component of the graviton that gives rise to power-enhancement and must be controlled in the operator basis.
    We introduce a redefinition that yields gauge-invariant objects, which satisfy the light-cone gauge condition.
    This ensures that no power-enhanced operators can appear in the 
SCET Lagrangian and $N$-jet operators.

\subsection{Power-counting}
\label{sec::CG::PowerCountingGravity}
    
Like in QCD SCET, we introduce light-like reference vectors $n_{\pm}^\mu$ and assign the collinear scaling \eqref{eq::SQCD::MomentumPowerCounting} to momenta, given by
$(\np p\,, p_\perp\,,\nm p) \sim (1\,,\lambda\,,\lambda^2)\,.$
In the following, we make use of the short-hand notation
		\begin{equation}
			n_+^\mu n_+^\nu h_{\mu\nu} = h_{++}\,,
		\end{equation}
similarly for the other components. Since we adopt the active point of view, where gauge transformations, which correspond to a change of coordinates, act as purely internal transformations on the fields, these reference vectors do not transform under diffeomorphisms.\footnote{Hence a component $\np^\mu A_\mu = A_+$ is not gauge-invariant despite the index contraction, as it is not contracted with the metric tensor.}
        
Due to the different scaling of the collinear momentum components, the components of the graviton $h_{\mu\nu}$ also scale differently. Their scaling is 
determined by the two-point function \eqref{eq::GR::GravitonPropagator}, which gives \cite{Beneke:2012xa} 
		\begin{equation}\label{eq::CG::GravitonScaling}
		\begin{aligned}
			h_{++}&\sim \lambda^{-1}\,, & h_{+\mu_\perp} &\sim 1\,, & h_{+-}&\sim \lambda\,,\\
			h_{\mu_\perp\nu_\perp}&\sim \lambda\,, & h_{\mu_\perp-}&\sim \lambda^2\,, & h_{--}&\sim \lambda^{3}\,, 
			\end{aligned}
		\end{equation}
as well as  $h \sim\lambda$ for the trace. Contracting $h_{\mu\nu}$ with a 
vector of the same collinear sector always yields power-suppression by 
one order in $\lambda$, since
		\begin{equation} 
			h^{\mu\nu} V_\nu = 
			\frac{1}{2}(h^{\mu}_+V_- + 
			h^{\mu}_-V_+) + 
			h^{\mu}_{\perp} V_\perp \sim \lambda V^\mu\,.
\label{eq:collinearcontraction}
		\end{equation}

Two important conclusions immediately follow from these scalings. First, 
there is the component $h_{++}$ that scales with an \emph{inverse power} 
of $\lambda$. This component must be controlled to all orders to obtain 
a sensible $\lambda$ expansion as otherwise any subleading term 
can contribute at lower orders through couplings to $h_{++}$ \cite{Beneke:2012xa}.
Even though any contraction $h^{\mu\nu}V_\nu$ is suppressed, this only holds for contractions within the same collinear sector. With two sectors $i$ and $j$, the leading-order contribution of such a contraction is
		\begin{equation}
			h^{\mu\nu}_i V_{j\nu} = \frac{n_{i-}^\mu}{2}\frac{n_{i-}n_{j-}}{4}h_{+_i +_i} V_{+_j} + \mathcal{O}(\lambda^0) \sim \mathcal{O}(\lambda^{-1})\,,
		\end{equation}
which clearly poses a problem. For example, it could lead to collinear singularities, which are absent in gravity \cite{Weinberg:1965nx,Akhoury:2011kq}. Thus we have to show that $h_{++}$ and also $h_{+\mu_\perp}\sim 1$  are not valid building blocks in the gauge-invariant $N$-jet 
operator basis. To achieve this, we perform a field redefinition that is equivalent to working in light-cone gauge. This procedure is analogous to the introduction of the collinear Wilson line that controls $\np A_c$ in the QCD case, as presented in \cref{sec::SQCD::Wilsonlines}. 
		
Second, within a collinear sector, the expansion in $\lambda$ agrees with 
the expansion in $h_{\mu\nu}$ or $\kappa$ in the full theory. This follows, 
since any additional power of $\kappa h_{\mu\nu}$ also adds one index 
contraction and hence a power of $\lambda$ according to 
\eqref{eq:collinearcontraction}. This greatly simplifies the construction. 
In essence, one only needs to write the weak-field theory expanded to the 
desired order in $h_{\mu\nu}$ in light-cone gauge. 
In the following, we set $\kappa=1$ to simplify the notation.

\subsection{Gauge transformations}
		
In this section we consider the gauge transformations of the collinear 
fields in more detail. In SCET gravity these transformations must be 
truncated at a certain order in $\lambda$, in order to keep the 
expansion valid. In the purely collinear sector, this is equivalent to 
truncating them at a certain order in $\varepsilon$ as in the 
weak-field expansion, since  collinear contractions are suppressed 
in $\lambda$, and these contractions agree with the expansion in $h$.
		
To derive the $\lambda$-scaling of the parameter $\varepsilon$, we require that the infinitesimal gauge transformation
		\begin{equation}
			h_{\mu\nu} \to h_{\mu\nu} - \partial_\mu\varepsilon_\nu - \partial_\nu\varepsilon_\mu \sim \mathcal{O}(h_{\mu\nu})
		\end{equation}
is homogeneous in $\lambda$. Therefore, we must assign the scaling
		\begin{equation}
			n_+\varepsilon \sim \frac{1}{\lambda}\,,\quad n_-\varepsilon\sim\lambda\,,\quad \varepsilon^{\mu_\perp}\sim 1
		\end{equation}
 to $\varepsilon^\mu(x)$. Looking at the coordinate transformation, and keeping in mind that the collinear position coordinate scales as
		\begin{equation}
			n_+x \sim \frac{1}{\lambda^2}\,,\quad n_-x \sim 1\,,\quad x_\perp^\alpha \sim \frac{1}{\lambda}\,,
		\end{equation}
we see that $\varepsilon^\mu$ scales as $\lambda x^\mu$, consistent with 
the notion of a \emph{small} translation.
		However, unlike in the QCD case, in gravity the gauge-parameter $\varepsilon^\mu$ has a non-trivial scaling in $\lambda$. This new feature of gravity will complicate the subsequent constructions.
		For the scalar field, the transformation up to $\mathcal{O}(\lambda^2)$ is given by \eqref{eq::GR::ScalarTransformExpanded}, where $\mathcal{O}(\varepsilon^3) = \mathcal{O}(\lambda^3)$, and the transformation of $h_{\mu\nu}$ takes the form
		\begin{equation}
		    \begin{aligned}
		        h^\prime_{\mu\nu} &= h_{\mu\nu} - \partial_\mu\varepsilon_\nu - \partial_\nu \varepsilon_\mu -\partial_\mu\varepsilon^\alpha h_{\alpha\nu}
		        -\partial_\nu\varepsilon^\alpha h_{\alpha\mu}
		        -\varepsilon^\alpha\partial_\alpha h_{\mu\nu}\\
		        &\quad
		        +  \partial_\mu\varepsilon^\alpha \partial_\alpha\varepsilon_\nu 
		        +  \partial_\nu\varepsilon^\alpha \partial_\alpha\varepsilon_\mu 
		        +  \partial_\mu\varepsilon_\alpha \partial_\nu\varepsilon^\alpha
		        + \varepsilon^\alpha\partial_\alpha(\partial_\mu\varepsilon_\nu + \partial_\mu\varepsilon_\nu) + \mathcal{O}(\lambda^3)\,.
		    \end{aligned}
		\end{equation}
Hence, the gauge symmetry of the collinear sector (in the absence of soft fields) is fully equivalent to the corresponding gauge symmetry of the truncated full theory, and it also mixes different orders in $\lambda$, implying that the field redefinitions and gauge-invariant objects cannot be homogeneous, but will be defined order by order in $\lambda$.
		
This can be understood easily. The inhomogeneous gauge symmetry ensures that subleading terms in the Lagrangian are related to the leading ones, in order to combine into a geometric quantity, defined again as a series in $h_{\mu\nu}$, and is thus inhomogeneous in $\lambda$. To facilitate the notation, it is thus possible to identify from which geometric object certain terms in the expansion stem. These objects then have a leading term, and associated with it an infinite tower of subleading terms.
However, this tower is fully constrained by (collinear) gauge invariance.

\subsection{Collinear Wilson line}
\label{sec::CG::CollinearWilsonLine}

As stated before, we must ensure that the $\lambda$-enhanced component 
$h_{++}$ is not a valid building block for the collinear $N$-jet operators. 
Further, we have to ensure that also $h_{+\mu_\perp}\sim\mathcal{O}(1)$ is 
controlled, in order to have only a finite number of building blocks at a 
given order in $\lambda$. In QCD, this is achieved by introducing the collinear Wilson line $W_c$ as in \cref{sec::SQCD::Wilsonlines}.
	    This Wilson line is inhomogeneous in the gauge coupling $g$, hence we can already anticipate that the gravitational analogue will be inhomogeneous in $\kappa$, and therefore also in $\lambda$. This is consistent with the inhomogeneous gauge symmetry.
	    
There are two equivalent points of view of this Wilson line. 
The first one is that we use it to redefine the graviton  
field  $h_{\mu\nu}$ to render it manifestly gauge-invariant, ensuring that after this redefinition $h_{\mu\nu}$ satisfies the collinear light-cone gauge condition $h_{+\mu} = 0$. 
This leads to what we may call covariant light-cone gauge, covariant in 
the sense that we construct an object $\hinv_{\mu\nu}$, which depends 
on $h_{\mu\nu}$, but is manifestly gauge-invariant, and 
satisfies $\hinv_{\mu+}=0$. From the second point of view, we express the 
fields in a special coordinate system, as suggested and explained in \cite{Donnelly:2015hta}. Here, one can use the geodesic equation to solve for the gauge-fixing parameter.
This procedure was also used in \cite{Chakraborty:2019tem}.
We will present both ways, as  
each has its advantages. The second construction is 
relegated to Appendix~\ref{sec::GeometricConstruction} including 
the demonstration that it is equivalent to the first.
	   
The collinear field redefinition works as follows: a scalar field 
transforms under a diffeomorphism as
		\begin{equation}
			\varphi(x) \to \varphi^\prime(x) = U(x)\varphi(x)\,,
		\end{equation}
see \eqref{eq:scalargaugetrafofull}. 
In order to compensate the transformation with $U(x)$, we consider 
the translation operator $W_c^{-1}=T_{\theta_c[h]}$, which corresponds to 
the coordinate transformation
\begin{equation}
y=x+\theta_c[h]\,,
\label{eq:thetachtranslation}
\end{equation}
with the graviton field-dependent parameter 
$\theta_c[h]\equiv\theta_c[h_{\mu\nu}(x)]$ chosen such that, given 
$h_{\mu\nu}$,  in the 
new coordinate system $h'_{+\mu}(x)=0$, which corresponds to 
light-cone gauge. Then, under an {\em arbitrary} gauge transformation 
$U(x)$, $W_c^{-1}$ transforms as\cite{Donnelly:2015hta}\footnote{We denote this object as $W_c^{-1}$ to emphasise that it corresponds to $W_c^\dagger$, and behaves inversely to the gauge transformation $U(x)$.}
\begin{equation}\label{eq::CG::Wilsontransformation}
W_c^{-1} \to W_c^{-1} U^{-1}(x)\,,
\end{equation}
where we assume that gauge transformations vanish at infinity, more precisely $x-\infty \,n_+$.
The transformation \eqref{eq::CG::Wilsontransformation} can be explicitly verified with $\theta_c$ as computed below.
In the second, geometric approach, this identity becomes evident.
By expressing the field in the new coordinates, the relevant gauge transformations are the diffeomorphisms at $x=-\infty$.
As the gauge transformations are taken to vanish there, the resulting expressions are manifestly gauge-invariant.
In other words, $W_c$ really acts as the analogue of the Wilson line \eqref{eq::SQCD::collWilsonLine} and moves the gauge transformation from point $x$ to $x=-\infty$, where it vanishes.
This is the essence of \eqref{eq::CG::Wilsontransformation}.
We can proceed to define the gauge-invariant scalar field $\chi_c$ as
		\begin{equation}
			\chi_c = \lc W^{-1}_c\varphi\rc \,,
		\end{equation}
which obviously satisfies $\chi_c\to\chi_c^\prime=\chi_c$.
Note that by construction $W_c^{-1}$ depends on the graviton field 
$h_{\mu\nu}$, but we suppress this dependence in the following. 
For a tensor field we further make use of the Jacobi matrices $\jac{\mu}{\alpha}$, $\jacin{\alpha}{\mu}$,
which we denote as $\tensor{W}{^\mu_\alpha}$ and $\tensor{W}{_{\mu}^\alpha}$, respectively. 
	
To derive $W_c$, i.e.~the parameter $\theta_c[h]$ in 
\eqref{eq:thetachtranslation}, we consider the gauge-invariant metric 
tensor\footnote{			Alternatively, we can compute this parameter directly from the geodesic equation. The details can be found in \cref{sec::GeometricConstruction}.}

\begin{equation}
\label{eq::ginv}
\eta_{\mu\nu} + \mathfrak{h}_{\mu\nu}(x) = \tensor{W}{^\alpha_\mu}
\tensor{W}{^\beta_\nu}[W_c^{-1}g_{\alpha\beta}(x)]\,,
			\end{equation}
where $g_{\alpha\beta}(x) = \eta_{\alpha\beta}+h_{\alpha\beta}(x)$.
As this transformation is inhomogeneous in $\lambda$, we expand the 
parameter $\theta_c[h]$ in $\lambda$ (dropping the argument to simplify 
notation),
		\begin{equation}
                \theta_c^\mu = \theta_c^{\mu(0)} + \theta_c^{\mu(1)} + \dots\,,
            \end{equation}
        where the superscript denotes the suppression in $\lambda$ relative to the leading term $\theta_c^{(0)}$, and calculate 
            \begin{equation}\label{eq::CG::WilsonLineDefinition}
                W_c^{-1} = T_{\theta_c} = 1 + \theta_c^\alpha\partial_\alpha + \frac 12 \theta_c^\alpha \theta_c^\beta \partial_\alpha\partial_\beta + \mathcal{O}(\theta_c^3)\,.
            \end{equation}
Expanding \eqref{eq::ginv}, we find
			\begin{equation}
				\hinv_{\mu\nu} = h_{\mu\nu} + 
				\partial_\mu \theta_{c\nu} +
				\partial_\nu \theta_{c\mu} + 
				\partial_\mu \theta_c^\alpha h_{\alpha\nu} + 
				\partial_\nu \theta_c^\alpha h_{\alpha\mu} + 
				\theta_c^\alpha\partial_\alpha h_{\mu\nu} + 
				\partial_\mu\theta_c^\alpha \partial_\nu \theta_{c\alpha} + 
				\mathcal{O}(\theta_c^3,\theta_c^2h)\,.
			\end{equation} 
We now impose the condition $\hinv_{\mu+}=0$ and solve iteratively for 
$\theta_c$. Solving first $\hinv_{++}=0$, and then inserting the solution 
into $\hinv_{\mu+} = 0$, yields
\begin{align}\label{eq::CG::colltheta}
				\theta^{(0)}_{c\mu} &= -\frac{1}{\delp}\biggl(h_{\mu+} - 
				\frac 12 \frac{\partial_\mu}{\delp} h_{++}\biggr)\,,\\
				\theta^{(1)}_{c+} &= -\frac 12 \frac{1}{\delp}\biggl( 
				2 \delp \theta_c^{(0)\alpha} h_{\alpha+} + 
				\theta_c^{(0)\alpha}\partial_\alpha h_{++} + 
				\delp \theta_c^{(0)\alpha}\delp \theta^{(0)}_{c\alpha}\biggr)\,,\\
				\theta^{(1)}_{c\mu} &= -\frac{1}{\delp}\biggl(
				\partial_\mu \theta^{(1)}_{c+} + 
			    \partial_\mu \theta_c^{(0)\alpha} h_{\alpha+} + 
				\delp \theta_c^{(0)\alpha} h_{\alpha\mu} + 
				\theta_c^{(0)\alpha}\partial_\alpha h_{\mu+}\nonumber\\
				&\quad + 
				\partial_\mu \theta_c^{(0)\alpha}\delp \theta^{(0)}_{c\alpha}\biggr)\,.\label{eq::CG::colltheta2}
			\end{align}
Here we defined the action of the inverse of 
$i\np\partial$ on functions $f(x^\mu)$ as 
\begin{equation}
\frac{1}{i\np\partial+i\epsilon}f(x^\mu) = -i 
\int_{-\infty}^0 \!ds \,f(x^\mu+s \np^\mu)\,.
\label{eq:inverserder}
\end{equation}
Looking at the linear terms in $\hinv_{\mu\nu}$, one finds
			\begin{equation}\label{eq::CG::collinvgraviton1}
				\hinv_{\mu\nu} = h_{\mu\nu} - 
				\frac{\partial_\mu}{\delp}\left( h_{\nu+} - \frac 12 \frac{\partial_\nu}{\delp} h_{++}\right) -
				\frac{\partial_\nu}{\delp}\left( h_{\mu+} - \frac 12 \frac{\partial_\mu}{\delp}h_{++}\right) +
				\mathcal{O}(\lambda h_{\mu\nu})\,,
			\end{equation}
			which is strikingly similar to the gauge-invariant collinear gluon field \eqref{eq::SQCD::InvariantCollinearGluon},
			\begin{equation}\label{eq::gaugeInvAPerp}
				\mathcal{A}_{c\mu_\perp} = \frac{1}{g} W^\dagger_c \brac{i D_{c\mu_\perp} W_c} = A_{c\mu_\perp} - \frac{\partial_{\mu_\perp}}{\delp}A_{c+} + \mathcal{O}(g A_{c\mu_\perp})\,. 
			\end{equation}
However, the graviton field is defined via an infinite series order-by-order in $\lambda$, whereas $\mathcal{A}_{c\mu}$ has a definite power-counting in $\lambda$, but is an infinite series in the gauge coupling $g$. This shows that in SCET gravity, the kinematic SCET expansion in $\lambda$ is closely linked to the coupling expansion in $\kappa$.
						
In this notation, the analogy to QCD is immediate. In QCD, we fixed light-cone gauge by using the inverse gauge transformation, denoted by $W_c^\dagger$, and redefining each field according to its collinear gauge symmetry.
			Now, in gravity, we define the object $W_c^{-1}$ which acts inversely compared to a gauge transformation $U(x)$, but the redefinition is again according to the gauge transformation.
    		
\subsection{Manifestly gauge-invariant Lagrangian}
    	
As already stated, the purely collinear theory is equivalent to the weak-field expansion \eqref{eq::GR::truncatedLagrangians1}, \eqref{eq::GR::truncatedLagrangians2} in light-cone gauge.
    	Alternatively, we can choose to express the original fields $h_{\mu\nu}$ and $\varphi$ in terms of the manifestly gauge-invariant fields $\hinv_{\mu\nu}$ and $\chi_c$.
    	
    	Inserting the redefinitions for $h_{\mu\nu}$ and $\varphi$, we find the manifestly gauge-invariant matter Lagrangian in covariant light-cone gauge
\begin{align}\label{eq::ManifestlyInvariantCollLag}
\mathcal{L}^{(0)} &= \frac{1}{2}\partial_\mu \chi_c \partial^\mu \chi_c - 
			\frac{\lambda_\varphi}{4!} \chi_c^4\,,\\
\mathcal{L}^{(1)} &= -
			\frac 12 \mathfrak{h}_{\mu\nu}(\partial^\mu \chi_c\partial^\nu \chi_c - 
			\eta^{\mu\nu}\frac{1}{2}\partial_\alpha \chi_c\partial^\alpha \chi_c) - 
\frac{1}{2}\mathfrak{h}\frac{\lambda_\varphi}{4!}\chi_c^4\,,\\
\mathcal{L}^{(2)} &=  
\frac{1}{2}\biggl(\mathfrak{h}^{\mu\alpha}\mathfrak{h}^\nu_\alpha - 
			\frac{1}{2}\mathfrak{h}\mathfrak{h}^{\mu\nu} + 
			\frac{1}{8}(\mathfrak{h}^2 - 2 \mathfrak{h}^{\alpha\beta}\mathfrak{h}_{\alpha\beta})\eta^{\mu\nu}\biggr)\partial_\mu \chi_c\partial_\nu \chi_c\nonumber\\
			&\quad - 
			\frac{1}{8}(\mathfrak{h}^2 - 2 \mathfrak{h}^{\alpha\beta}\mathfrak{h}_{\alpha\beta})\frac{\lambda_\varphi}{4!} \chi_c^4
		\end{align} 
up to $\mathcal{O}(\lambda^2)$. This corresponds to the weak-field expanded Lagrangian of the full theory derived before, but with $h_{\mu\nu}\to\mathfrak{h}_{\mu\nu}$.
		Due to the light-cone gauge condition $\mathfrak{h}_{\mu+}=0$,
		we immediately see that there are no leading-power collinear gravitational interactions and thus no analogue of the $\mathcal{O}(1)$ derivative  $\np D_c$  in QCD.
To derive the Einstein-Hilbert action, 
we use the equations of motion up to (relative) $\mathcal{O}(\lambda)$,
\begin{eqnarray}\label{eq::CG::EOMh}
&& \mathfrak{h} =\frac{1}{2}\biggl(\mathfrak{h}_{\alpha_\perp \beta_\perp}\mathfrak{h}^{\alpha_\perp  \beta_\perp}-\frac{1}{\partial^2_+}(\partial_+\mathfrak{h}_{\alpha_\perp  \beta_\perp}\partial_+\mathfrak{h}^{\alpha_\perp  \beta_\perp})\biggr)\,,\\
&& \mathfrak{h}_{- \mu_\perp} =-2\frac{\partial^{ \alpha_\perp}}{\partial_+}\mathfrak{h}_{ \mu_\perp   \alpha_\perp}+\biggl(
        -\frac{\partial_{ \mu_\perp}}{\partial^3_+}(\partial_+\mathfrak{h}_{  \alpha_\perp  \beta_\perp}\partial_+\mathfrak{h}^{  \alpha_\perp  \beta_\perp})\label{eq::CG::EOMhperp}
        \\
        &&\hspace*{1cm}
        +\,\frac{1}{\partial^2_+}\Bigl(-2\partial^2_+\mathfrak{h}_{ \mu_\perp   \alpha_\perp}\frac{\partial_{\beta_\perp}}{\partial_+}\mathfrak{h}^{  \alpha_\perp  \beta_\perp}
        +2\mathfrak{h}^{  \alpha_\perp  \beta_\perp}\partial_+\partial_{ \alpha_\perp}\mathfrak{h}_{ \mu_\perp  \beta_\perp}
        +\partial_+\mathfrak{h}_{ \alpha_\perp  \beta_\perp}\partial_{ \mu_\perp}\mathfrak{h}^{  \alpha_\perp  \beta_\perp}\Bigr)\biggr)\,,\nn
        \end{eqnarray}
		to eliminate the subleading components $\mathfrak{h}_{\mu_\perp-}$ and $\mathfrak{h}_{--}$.
		We then obtain the Lagrangian expressed in terms of only the physical, transverse modes $\mathfrak{h}_{\mu_\perp\nu_\perp}$: 
		\begin{align}
        \mathcal{L}^{(0)}_{\textrm{EH}}&=\,\frac{1}{2}\partial_{\mu}\ch_{\alpha_\perp\beta_\perp}\partial^{\mu}\ch^{\alpha_\perp\beta_\perp}\,,\\
        \mathcal{L}^{(1)}_{\textrm{EH}}&=\,\frac{1}{2}\biggl(\ch_{\alpha_\perp\beta_\perp}\partial^2_+\ch^{\alpha_\perp\beta_\perp}\frac{\partial_{\rho_\perp}\partial_{\sigma_\perp}}{\partial^2_+}\ch^{\rho_\perp\sigma_\perp}
        - 2\ch_{\alpha_\perp\beta_\perp}\partial_+\partial^{\rho_\perp}\ch^{\alpha_\perp\beta_\perp}\frac{\partial^{\sigma_\perp}}{\partial_+}\ch_{\rho_\perp\sigma_\perp}
        \nonumber\\
        &\quad\phantom{\frac\kappa 2\biggl[} + \ch_{\alpha_\perp\beta_\perp}\ch_{\rho_\perp\sigma_\perp}\partial^{\rho_\perp}\partial^{\sigma_\perp}\ch^{\alpha_\perp\beta_\perp}
        - 2\ch_{\alpha_\perp\sigma_\perp}\ch_{\beta_\perp\rho_\perp}\partial^{\rho_\perp}\partial^{\sigma_\perp}\ch^{\alpha_\perp\beta_\perp}\nonumber\\
        &\quad\phantom{\frac\kappa 2\biggl[}
        - 4\ch_{\alpha_\perp\sigma_\perp}\partial_+\ch^{\alpha_\perp\beta_\perp}\frac{\partial^{\rho_\perp}\partial^{\sigma_\perp}}{\partial_+}\ch_{\rho_\perp\beta_\perp}\biggr)\,.
        \end{align}
		
Note that $\mathfrak{h}_{\mu\nu} = \mathfrak{h}^{(1)}_{\mu\nu} + \mathfrak{h}^{(2)}_{\mu\nu} + \dots$ expressed in terms of $h_{\mu\nu}$ is not homogeneous in $\lambda$, and we have
\begin{align}
			\mathfrak{h}^{(1)}_{\mu\nu} &= h_{\mu\nu} + 
			\partial_\mu\theta_{c\nu}^{(0)} + \partial_\nu \theta_{c\mu}^{(0)}\,,\\
			\mathfrak{h}^{(2)}_{\mu\nu} &= \partial_\mu \theta^{(1)}_{c\nu} + 
			\partial_\nu \theta^{(1)}_{c\mu} + 
			\partial_\mu \theta_c^{(0)\alpha}h_{\alpha\nu} + 
			\partial_\nu \theta_c^{(0)\alpha}h_{\alpha\mu} + 
			\theta_c^{(0)\alpha}\partial_\alpha h_{\mu\nu} + 
			\partial_\mu\theta^{(0)}_{c\alpha} \partial_\nu \theta_c^{(0)\alpha}\,,
		\end{align}
		where $\theta_c^{(i)}$ is determined from $h_{\mu\nu}$ according to \eqref{eq::CG::colltheta} -- \eqref{eq::CG::colltheta2}.
		The object $\hinv_{\mu\nu}^{(1)}$ is gauge-invariant under the linear transformation, but only the full $\mathfrak{h}_{\mu\nu}$ is gauge-invariant under the all-order in $\lambda$ non-linear transformation.
		However, the Lagrangian is homogeneous in the sense that each term has a leading-order piece and an infinite series of subleading terms, which are truncated at the desired order in $\lambda$.
		If we now choose to fix light-cone gauge, all these subleading terms vanish, as their purpose is to ensure that the non-linear and subleading gauge transformations are taken into account, rendering $\hinv_{\mu\nu} $ manifestly invariant.
		Hence, in light-cone gauge these terms vanish, and the Lagrangian is manifestly homogeneous in $\lambda$.
		
For computations, we can thus either fix light-cone gauge and work with the manifestly invariant Lagrangian, or use the original Lagrangian and choose another convenient gauge. For the sources, which we discuss later in more detail, this construction controls the power-enhanced $h_{++}$ and $\mathcal{O}(1)$ components $h_{+\mu_\perp}$ to all orders in $\lambda$, since only $\mathfrak{h}_{\mu\nu}$ can appear in the sources.
		This allows us to write down a finite number of building blocks that can contribute to the $N$-jet operators at a given order in $\lambda$.
		
We conclude this section with a short summary of the most important insights:
\begin{itemize}
\item The collinear expansion agrees with the weak-field expansion.
\item The collinear ``Wilson line'' $W_c^{-1}$ is the exact analogue of the 
literal Wilson line $W_c^\dagger$, which appears in the QCD case. In addition to the derivation in \cite{Donnelly:2015hta,Chakraborty:2019tem}, we presented a more intuitive approach using explicit gauge-fixing.
		    \item By working with the manifestly gauge-invariant fields $\mathfrak{h}_{\mu\nu}$, which correspond to the graviton field in light-cone gauge, we can control the dangerous modes in the source operators and have manifest $\lambda$-counting.
\end{itemize}
Also note that the situation in the purely collinear theory is completely analogous to the QCD scenario presented before. The collinear theory is the full theory---perturbative gravity---in light-cone gauge. Any noteworthy differences are already present in the full theory, and are mainly features of the non-linearly realised diffeomorphism symmetry.


\section{Soft-collinear gravity}\label{sec::SG}

In this section, we derive the soft-collinear Lagrangian.
While the actual implementations and details differ quite drastically from the gauge-theory case, the overall concepts and intuition behind the construction are the same.
Thus, we follow the scheme outlined in \cref{sec::SQCD} as closely as possible, using the same terminology and notation, but pointing out all relevant differences.
The main distinction is the gauge symmetry, and once we understand its implications, the generalisation of the SCET construction is straightforward.

The outline of this section is as follows:
First, we introduce the field content and its power counting, then we proceed with the discussion of the soft and collinear gauge transformations. 
We again treat the soft field as a background field with collinear fluctuations on top of it.
We point out the subtleties, discussing in detail the inhomogeneous nature of the theory, before we proceed with the actual construction of soft-collinear gravity, which follows closely the one presented in \cref{sec::SQCD}.
We emphasise the effective gauge symmetry, which guides the construction just as before.

In step (iii) in QCD, we made use of fixed-line gauge. In gravity, we introduce the analogue of this gauge, which we denote as fixed-line normal coordinates (FLNC), the light-cone generalisation of Riemann normal coordinates (RNC).
We explain this part in detail, as it is crucial for the subsequent construction of soft-collinear gravity.
The final Lagrangian will turn out to be structurally very similar to the gauge-theory case.
To conclude, we discuss the minimal $N$-jet operator basis, in analogy to \cref{sec::SQCD::OperatorBasis}.

\subsection{Power-counting, field content and gauge symmetry}\label{sec::SG::GaugeTransform}

We employ the usual SCET power counting introduced in \cref{sec::SQCD::PowerCounting}.
As field content, we consider a real scalar field $\varphi$ and the metric tensor $g_{\mu\nu}$.
We split the metric tensor as
\begin{equation}\label{eq::SG::MetricTensorSplit}
    g_{\mu\nu} = \eta_{\mu\nu} + h_{\mu\nu} + s_{\mu\nu} \equiv g_{s\mu\nu} + h_{\mu\nu}\,,
\end{equation}
introducing the collinear graviton $h_{\mu\nu}$ and the soft graviton $s_{\mu\nu}$, and also include a collinear and a soft scalar field, denoted by $\varphi_c$, $\varphi_s$, respectively.
Ultimately, we perform an expansion around Minkowski space. However, it is convenient to keep the soft metric tensor $g_{s\mu\nu} \equiv \eta_{\mu\nu} + s_{\mu\nu}$, as this makes many points in the derivation more transparent and we can use geometric notions to see the all-order structure.
The scaling of the collinear graviton \eqref{eq::CG::GravitonScaling} and the scalar fields \eqref{eq::SQCD::ScalarScaling} is the same as before, and the soft graviton field has a homogeneous $\lambda$-scaling as
\begin{equation}
    s_{\mu\nu}\sim\lambda^2\,.
\end{equation}

In analogy to \cref{sec::SQCD::GaugeSymmetry}, where we introduced a different gauge transformation for the soft and collinear gluon fields, we can use the split \eqref{eq::SG::MetricTensorSplit} to let the two fields transform differently, thereby realising two symmetries.

The original fields transform with the full diffeomorphisms \eqref{eq::GR::TensorTransformation} as
\begin{equation}
\begin{aligned}
    g_{\mu\nu} &\to U \lc\tensor{U}{_\mu^\alpha} \tensor{U}{_\nu^\beta}g_{\alpha\beta}\rc\\
    &= g_{\mu\nu} - \nabla_\mu\varepsilon_\nu - \nabla_\nu\varepsilon_\mu + \mathcal{O}(\varepsilon^2)\,,\\
    \varphi &\to \lc U\varphi\rc \\
    &= \varphi - \varepsilon^\alpha\nabla_\alpha \varphi + \mathcal{O}(\varepsilon^2)\,,
\end{aligned}
\end{equation}
where $\nabla$ is the covariant derivative with respect to the full metric tensor $g_{\mu\nu}$.
Once we perform the soft-collinear expansion, we have to truncate this transformation at some order in $\lambda$, but we keep the symbols of the full transformation to keep the notation concise.

We now introduce the analogous notions of a ``collinear'' and a ``soft'' gauge symmetry as in \cref{sec::SQCD::GaugeSymmetry}, adopting the point of view that the collinear fluctuation lives in a soft background.

For the ``collinear'' symmetry, only the fluctuations $h_{\mu\nu}$ and $\varphi_c$ transform, and we describe the transformation by the parameter $\varepsilon_c^\mu$. We have to ensure that for the full fields, e.g. $g_{s\mu\nu} + h_{\mu\nu}$, this is still the usual diffeomorphism transformation.
	This leads to
	\begin{equation}\label{eq::SG::CollinearSymmetry}
	\begin{aligned}
	h_{\mu\nu} &\to \lc U_c\left( \tensor{U}{_{c\mu}^\alpha}\tensor{U}{_{c\nu}^\beta}(g_{s\alpha\beta} + h_{\alpha\beta})\right)\rc - g_{s\mu\nu}\,,\\
	\varphi_c &\to \lc U_c\varphi_c\rc\,,\\
	g_{s\mu\nu} &\to g_{s\mu\nu}\,,\\
	\varphi_s &\to \varphi_s\,.
	\end{aligned}
	\end{equation}
	It is evident that the sum $g_s + h$ has the  diffeomorphism transformation of the full metric tensor.
	This is completely analogous to 
	QCD in \eqref{eq::SQCD::FullGaugeTransformations}, and the translation $\lc U_c\varphi_c\rc$ corresponds to the large gauge 
	transformation $[U_c\phi_c]$ in gauge theory. 
	It is also conceptually the same as background field 
	gauge in QCD. The only difference is that in the usual background 
	field method, the matter (quark) fields do not have a non-vanishing 
	background, so we would set $\varphi_s=0$ and make the split only 
	for the gauge field. Here we also want a soft matter background 
	for later quantisation.
	
	For the ``soft'' gauge transformations, we want the background metric $g_{s\mu\nu}$ and the ``soft'' scalar field to have the standard diffeomorphism transformation.
	This implies
	\begin{equation}\label{eq::SG::SoftSymmetry}
	\begin{aligned}
	h_{\mu\nu} &\to \lc U_s \left( \tensor{U}{_{s\mu}^\alpha}\tensor{U}{_{s\nu}^\beta}h_{\alpha\beta}\right)\rc\,,\\
	\varphi_c &\to \lc U_s\varphi_c\rc\,,\\
	g_{s\mu\nu} &\to \lc U_s \left( \tensor{U}{_{s\mu}^\alpha}\tensor{U}{_{s\nu}^\beta} g_{s\alpha\beta}\right)\rc\,,\\
	\varphi_s &\to \lc U_s \varphi_s\rc\,.
	\end{aligned}
	\end{equation}
	$h_{\mu\nu}$ transforms like an ordinary tensor field and so does the background metric $g_{s\mu\nu}$. 
	Comparing this to the QCD case \eqref{eq::SQCD::FullGaugeTransformations}, we recognise the analogous transformations, namely, the collinear fluctuation $h_{\mu\nu}$ transforms like a tensor field, i.e. like an ordinary matter field, under the soft gauge transformation.
	The soft fluctuation, $s_{\mu\nu}$, inherits the gauge transformation
	\begin{equation}
	    s_{\mu\nu} \to \lc U_s\tensor{U}{_{s\mu}^\alpha} \tensor{U}{_{s\nu}^\beta}(\eta_{\alpha\beta} + s_{\alpha\beta})\rc - \eta_{\mu\nu}\,,
	\end{equation}
	which is the standard gauge transformation of a metric perturbation, just as the soft gluon $A_s$ has the standard gauge transformation of a gluon \eqref{eq::SQCD::FullGaugeTransformations}.
	
	Similar to QCD \eqref{eq::SQCD::ScalarSplit}, we relate the full theory scalar fields to the effective theory fields by introducing the ``Wilson line'' $WZ^{-1}$
	\begin{equation}\label{eq::SG::ScalarFieldSplit}
	    \varphi = \varphi_c + \lc WZ^{-1} \varphi_s\rc\,.
	\end{equation}
	Here, $W^{-1}$ and $Z^{-1}$ are defined as in \eqref{eq::CG::WilsonLineDefinition}, but with the full graviton and the soft graviton field, respectively, that is
	\begin{equation}
	    W \equiv T^{-1}_{\theta[g_s+h]}\,,\quad Z^{-1} \equiv T_{\theta[g_s]}\,,
	\end{equation}
	where $\theta[g_{\mu\nu}]$ denotes the parameter $\theta$ that fixes light-cone gauge for the metric tensor $g_{\mu\nu}$.
	In other words, to compute $W$, we fix light-cone gauge $h_{{\rm full},\,\mu+}= 0$ for the full-theory fluctuation $h_{{\rm full},\,\mu\nu} = h_{\mu\nu} + s_{\mu\nu}$, and to compute $Z$ we fix light-cone gauge for only the soft fluctuation, $s_{\mu+}\equiv0$ exactly as was done in the QCD construction \eqref{eq::SQCD::WZdefinition}.

\subsection{Effective theory construction}

We can now proceed with the construction of the effective Lagrangian, following \cref{sec::SQCD::EFTDerivation} as closely as possible, but pointing out in which aspects gravity differs.
The plan of the derivation is as follows:
\begin{enumerate}[(i)]
    \item Starting from the full theory, we perform the weak-field expansion
    \begin{align}
        g_{\mu\nu}(x) = g_{s\mu\nu}(x) + h_{\mu\nu}(x)\,,
    \end{align}
    where $h_{\mu\nu}$ is the collinear graviton and $g_{s\mu\nu}$ is a dynamical soft background.
    This expansion \emph{duplicates} the gauge symmetry into two copies, and we call them ``collinear'' \eqref{eq::SG::CollinearSymmetry}, if with respect to $h_{\mu\nu}$, and ``soft'' \eqref{eq::SG::SoftSymmetry} if it is associated to $g_{s\mu\nu}$.
    \item We multipole-expand the soft fields in soft-collinear interactions about $x_-^\mu = n_-x \frac{n_+^\mu}{2}$ as
    \begin{align}
        \varphi_s(x) = \varphi_s(x_-) + (x-x_-)^\alpha\partial_\alpha\varphi_s(x_-) + \dots\,,
    \end{align}
    to ensure homogeneous $\lambda$ scaling.
    \item We require a field redefinition of the collinear fields to ensure soft gauge transformations respect the multipole expansion. This is achieved by the analogue of the $R$ Wilson line.
    In the gravitational case, the residual soft gauge transformations are not homogeneous in $\lambda$, due to the $\lambda$-scaling of the charges, but they respect the multipole expansion.
    To find these field redefinitions, we construct the analogue of fixed-line gauge in the gravitational setting.
    Just as in gauge theory, we can then identify the residual soft background fields that are linked to the residual gauge transformation.
    \item We insert the redefintions and express the theory in terms of the new soft background field $\hat{g}_{s\mu\nu}$. We can then perform the $\lambda$ expansion, and sort the new background field into a covariant derivative.
    This Lagrangian, expressed in terms of covariant derivatives, then looks strikingly similar to the gauge theory: we have a soft-covariant derivative and all subleading interactions are expressed in terms of the field-strength tensor, i.e. the Riemann tensor.
\end{enumerate}
Points (i) -- (iv) are indeed the exact same steps as in the gauge-theory case. In gravity, we perform an additional step in (iv) to indicate the close link of soft gravity to soft gauge theory by introducing a soft-covariant derivative, which is usually not present for scalar fields coupled to gravity.
Due to the nature of gravity, we cannot give closed expressions in most cases.
However, it is always possible to determine the relevant quantities to the desired order in $\lambda$, and often there is even a geometric interpretation for the appearing quantities.

\subsection{Background-field Lagrangian}\label{sec::SG::BG}
	
In step (i), to find the Lagrangian in the background field formalism, we simply need to introduce the split \eqref{eq::SG::MetricTensorSplit} in the full theory.
We start from the action
\begin{align}\label{eq::MinimalCoupledScalarAction}
    S = \int\de^4x\: \sqrt{-g}\lp \frac 12 g^{\mu\nu}\partial_\mu\varphi\partial_\nu\varphi - \frac{\lambda_\varphi}{4!}\varphi^4\rp
\end{align}
and insert the expansion of $g^{\mu\nu}$ in $h_{\mu\nu}$.
We write
\begin{equation}\label{eq::SG::BFLagrangian}
    \mathcal{L} = \mathcal{L}_\varphi + \mathcal{L}_{\varphi h} + \mathcal{L}_{\varphi hh} + \mathcal{O}(h^3)\,,
\end{equation}
where the individual terms are given by
\begin{align}\label{eq::SG::BFLagrangian0}
    \mathcal{L}_{\varphi} &= \frac 12 \sqrt{-g_s} g_s^{\mu\nu}\partial_\mu \varphi \partial_\nu \varphi
    -\sqrt{-g_s}\frac{\lambda_\varphi}{4!}\varphi^4
    \,,\\
    \mathcal{L}_{\varphi h} &= -\frac 12 \sqrt{-g_s} \left(g_s^{\mu\alpha} g_s^{\nu\beta} h_{\alpha\beta} - \frac 12 g_s^{\alpha\beta} h_{\alpha\beta}g_s^{\mu\nu} \right)\partial_\mu \varphi\partial_\nu \varphi
    -\sqrt{g_s}\lp\frac{1}{2}g_{s}^{\alpha\beta} h_{\alpha\beta}\rp\frac{\lambda_\varphi}{4!}\varphi^4
    \,,\label{eq::SG::BFLagrangian1}\\
    \mathcal{L}_{\varphi hh} &= \frac 12 \sqrt{-g_s} \left(g_s^{\mu\alpha} g_s^{\nu\beta} g_s^{\rho\sigma} h_{\alpha\rho}h_{\beta\sigma} - \frac 12 g_s^{\alpha\beta} h_{\alpha\beta} g_s^{\mu\rho} g_s^{\nu\sigma} h_{\rho\sigma}
    \right.
    \nn\\
    &\qquad\phantom{\sqrt{-g_s} }  \left.
    + \frac 18 g_s^{\mu\nu}(g_s^{\alpha\beta} h_{\alpha\beta})^2
    - \frac 14 g_s^{\mu\nu}g_s^{\rho\alpha} g_s^{\sigma\beta}  h_{\rho\sigma} h_{\alpha\beta}\right) \partial_\mu \varphi\partial_\nu \varphi\nn\\
    &\quad - \sqrt{-g_s}\lp 
    \frac 18 (g_s^{\alpha\beta}h_{\alpha\beta})^2
    -\frac 14 g_s^{\rho\alpha}g_s^{\sigma\beta} h_{\rho\sigma}h_{\alpha\beta}\rp
    \frac{\lambda_\varphi}{4!}\varphi^4
    \,,\label{eq::SG::BFLagrangian2}
\end{align}
where $\varphi = \varphi_c + WZ^{-1}\varphi_s$ and the subscript indicates the order in $h$.
In this formulation, gauge-invariance under the soft transformation is manifest, as each object transforms like a proper tensor, and indices are contracted with $g_{s\mu\nu}$ to yield a scalar. The soft metric determinant $\sqrt{-g_s}$ appears to render the action manifestly gauge-invariant.
Collinear gauge-invariance is not obvious, as the gauge transformation connects different orders in $h$.
Nonetheless, the Lagrangian is invariant order-by-order.
However, note that this Lagrangian is not yet homogeneous in $\lambda$.

For the purely gravitational part, we start from the 
Einstein-Hilbert action
\begin{equation}
    S_{\rm EH} = -2 \int d^4x\: \sqrt{-g} R\,,
\end{equation}
where $R$ denotes the Ricci scalar.
We split the metric tensor $g_{\mu\nu}$ as in \eqref{eq::SG::MetricTensorSplit}, which is equivalent to perturbing the determinant and the Ricci scalar about the background $g_{s\mu\nu}$.
This results in
\begin{equation}
    \mathcal{L}_{\rm EH} = \mathcal{L}_{s} + \mathcal{L}_{hh} + \mathcal{L}_{hhh} + \mathcal{O}(h^4)\,.
\end{equation}
There is no $\mathcal{L}_{h}$ by the equation of motion.
The terms in $\mathcal{L}_s$ form the purely-soft gravitational theory. (The leading terms of the purely collinear theory are found in $\mathcal{L}_{hh}$, once one expands $g_{s\mu\nu}= \eta_{\mu\nu}+ s_{\mu\nu}$ in $s_{\mu\nu}$.)
Explicitly, $\mathcal{L}_{hh}$ reads
    \begin{align}
    \mathcal{L}_{hh} &= \sqrt{-{g}_s}\,\biggl(
    \frac 12 \nabla_\mu h_{\alpha\beta} \nabla^\mu h^{\alpha\beta} 
    - \frac 12 \nabla_\mu h \nabla^\mu h
    + \nabla_\alpha h^{\alpha\beta}\nabla_\beta h
    - \nabla_\alpha h^{\alpha\beta}\nabla_\mu h^\mu_\beta \nn\\
    &\quad - 4 R_{\alpha\beta} h^{\alpha\mu} h^\beta_\mu
    + 2 R_{\alpha\beta\mu\nu} h^{\alpha\mu}h^{\beta\nu} 
    + R_{\alpha\beta} h h^{\alpha\beta}
    - \frac{1}{4}R(h^2 - 2 h^{\alpha\beta}h_{\alpha\beta})
    \biggr)\,.\label{eq::SG::BilinearEH}
\end{align}
Here, $\nabla_\mu$ is the covariant derivative with respect to $g_{s\mu\nu}$, $R_{\alpha\beta}$ and $R_{\alpha\beta\mu\nu}$ are the purely-soft Ricci and Riemann tensor, and the background metric is also used to raise and lower indices, i.e.
$h^{\mu\nu} = g_s^{\mu\alpha} g_s^{\nu\beta} h_{\alpha\beta}$ and $h=g_{s}^{\mu\nu} h_{\mu\nu}$.
Once we perform the $\lambda$ expansion, these terms will simplify considerably, as e.g. $R_{\alpha\beta}\sim\lambda^6$.
One can also derive the explicit version of the trilinear background-field Lagrangian $\mathcal{L}_{hhh}$. However, for our purposes, namely obtaining the soft-collinear Einstein-Hilbert theory to $\mathcal{O}(\lambda)$, we may use of the full-theory trilinear result \eqref{eq::GR::TrilinearLagrangian} with minimal substitutions, as explained later in \cref{sec::SG::SCLagrangian}.

\subsection{Inhomogeneities in $\lambda$}

From the background field Lagrangian \eqref{eq::SG::BFLagrangian}, one can see that there are two sources of inhomogeneity in $\lambda$.
First, due to the inhomogeneity of the charges (the momentum $P^\mu$ has non-trivial and inhomogeneous scaling), and second due to the fact that soft fields rather than on $x^\mu$, depend only on the light-cone coordinate $x_-^\mu = n_+x \frac{n_-^\mu}{2}$.
	
The first type of inhomogeneity already appears in the full theory as well as the purely collinear one, as discussed in \cref{sec::GR,sec::CG}.
Once the metric field $g_{\mu\nu} = \eta_{\mu\nu} + \kappa h_{\mu\nu}$ is expanded, the theory turns into an infinite series in $\kappa$. In turn, the full diffeomorphism symmetry also gets expanded in $\kappa$.
	This is inherited in the purely collinear theory, where the kinematic $\lambda$ expansion agrees with the $\kappa$ expansion.
	The reason for this is that the gauge charges, that is the collinear momenta $P^\mu$, have a non-trivial and inhomogeneous $\lambda$-scaling 
	$(n_+P, n_-P, P_\perp)\sim(1,\lambda^2,\lambda)$. In turn, the coordinate $x^\mu$ also acquires an inhomogeneous scaling and correspondingly, the gauge-parameters $\varepsilon_c^\mu(x)$, which are linked to translations $x \to x + \varepsilon_c$.
	In contrast, in QCD, the charge and the gauge parameter $\varepsilon(x)$ do not scale with $\lambda$.
	
	However, this is a feature of the full theory, and simply means that any object that is a series in $\kappa$ does not have a homogeneous scaling in $\lambda$, whereas any object in QCD that is a series in the strong coupling $g$, is homogeneous in $\lambda$.
	Such inhomogeneous objects have a well-defined leading scaling, and then generate an infinite number of subleading terms, which are constrained by the inhomogeneous gauge symmetry.
	A good comparison for this are the Wilson lines. In QCD, they are infinite series in the coupling $g$, but each term in the series scales as $\mathcal{O}(\lambda^0)$. Consequently, the Wilson line is an $\mathcal{O}(\lambda^0)$ object.
	In gravity, it is an infinite series in $\kappa$, which, since $\kappa$ is a dimensionful constant, is then of course also inhomogeneous in $\lambda$.
	If we ignored this different scaling, both constructions would look very similar (up to the fact that the gravitational case has no simple closed form as an exponential).
	Hence, it seems that this kind of inhomogeneity is unavoidable when dealing with gravity, as it just links different linear terms to their non-linear completion, related by higher powers in $\kappa$.
	It is just an artifact of the perturbative expansion leading to a non-linearly realised symmetry.
	
	The second type of inhomogeneity, on the other hand, is also present in QCD.
	When we multiply soft and collinear fields, we have to ensure that the soft fields $g_{s\mu\nu}(x)$, $\varphi_s(x)$ are only evaluated at the coordinate $x^\mu_-$. This means that in the Lagrangian, they have to appear as $g_{s\mu\nu}(x_-)$, $\varphi_s(x_-)$.
	Hence, in step (ii) of the EFT construction, we perform the light-front multipole expansion \cite{Beneke:2002ph}, defined as
	\begin{equation}
	\begin{aligned}
         g_{s\mu\nu}(x) &= g_{s\mu\nu}(x_-) + x_\perp^\alpha\lc\partial_\alpha g_{s\mu\nu}\rc(x_-)\\
        &\quad
        + \frac 12 \nm x \lc\np \partial g_{s\mu\nu}\rc(x_-)
        +\frac 12 x_\perp^\alpha x_\perp^\beta \lc\partial_\alpha\partial_\beta g_{s\mu\nu}\rc(x_-) + \mathcal{O}(\lambda^3 g_{s\mu\nu})\,,
    \end{aligned}
    \end{equation}
    and similarly for $\varphi_s$. Any soft field, once it appears with a collinear field, generates another infinite series in $\lambda$ because of this expansion.
	Notably, these terms are all of the same order in $\kappa$, and hence not related to the non-linearities of general relativity.
	Moreover, the soft gauge transformations of collinear fields also have to respect this multipole expansion. Therefore, in step (iii), we redefine the collinear fields accordingly.
	In QCD, this is achieved by the introduction of the $R$ Wilson line \cite{Beneke:2002ni}, see \cref{sec::SQCD::FixedLineGauge}, and leads to a completely homogeneous gauge transformation in $\lambda$.
	In gravity, as will be seen, one can construct the analogue of the $R$ Wilson line.
	
	In summary, gravitational theories are inherently non-linear and thus inhomogeneous in the power-counting parameter $\lambda$.
	This is a feature of the full theory and of no importance for soft-collinear physics.
	The multipole expansion, however, is the same as for gauge theory, and one can perform analogous redefinitions to bring these terms into a more transparent form.
	In the following, we show how this is done in gravity.
	
	\subsection{Multipole expansion and normal coordinates}
	
	Due to the presence of inherently inhomogeneous gauge transformations, one cannot use homogeneous gauge symmetry as the guiding principle of the construction. 
	We recall that in QCD, the soft sector is expressed in terms of a number of suitable building blocks that respect the multipole expansion, namely the new soft-covariant derivative $D_s$, depending only on $\nm A_s(x_-)$, and the field-strength tensor $F_{s\mu\nu}$, see \eqref{eq::SQCD::L0final} -- \eqref{eq::SQCD::L2final}.
	The homogeneous gauge symmetry is then a consequence of the homogeneous residual background field.
	For gravity, we want to find an analogous split of the soft background field into a new residual background field, the corresponding covariant derivative, and subleading terms with expansions similar to \eqref{eq::SQCD::FLExpansionA-} -- \eqref{eq::SQCD::FLExpansionA+}.
    As these form the heart of the construction, the fixed-line gauge for gravity is motivated and introduced in detail.
	
	How do we achieve this goal? In QCD, the redefinition of the matter field \eqref{eq::SQCD::CollinearRedefinition} is equivalent to evaluating the soft field in fixed-line gauge, which is a partial gauge-fixing and a generalisation of fixed-point gauge. 
	Both gauge conditions share many similarities.
	Hence, we first discuss the analogue of the simpler case, fixed-point gauge.
	Once this is understood and the notation is set up, the generalisation to the analogue of fixed-line gauge is presented.
	In the end, we can identify the new background field as well as the gauge-covariant subleading terms, which 
are given in terms of the Riemann tensor.

\subsubsection{Illustration: fixed-point gauge and Riemann normal coordinates}
	
Before tackling the more complicated fixed-line scenario, we consider the simpler case of the multipole expansion about $x=0$ and the Riemann normal coordinates (RNC).
	This construction is well-known. Nevertheless, we repeat
	the derivation and explanation here from a slightly different angle, which facilitates the understanding of the subsequent SCET generalisation.
	
	In the following, the theory consists only of a metric field $g_{\mu\nu}(x)$ (neither soft nor collinear) and a scalar field $\varphi(x)$.
	In gravity, the gauge conditions corresponding to fixed-point gauge are the well-known Riemann normal coordinates.
	Here, the multipole expansion of the metric tensor $g_{\mu\nu}(x)$ is expressed as
	\begin{align}\label{eq::SG::MetricRNC}
    \tilde{g}_{\mu\nu}(x) &= g_{\mu\nu}(0) - \frac 16 x^\alpha x^\beta (\tensor{R}{^\rho_{\alpha\nu\beta}}(0)g_{\rho\mu}(0) + \tensor{R}{^\rho_{\alpha\mu\beta}}(0)g_{\rho\nu}(0)) + \mathcal{O}(x^3)\,,
    \end{align}
    where the higher-order terms in $x$ contain only derivatives of the Riemann tensor.
    Notably, the first derivative term is absent.
    In addition, one can diagonalise $g_{\mu\nu}(0)$ via an additional transformation to obtain the well-known result
    \begin{equation}\label{eq::SG::MetricRNCDiagonalised}
        \check{g}_{\mu\nu}(x) = \eta_{\mu\nu} - \frac 13 x^\alpha x^\beta R_{\alpha\mu\beta\nu}(0) + \mathcal{O}(x^3)\,,
    \end{equation}
    where the metric tensor is expressed purely in terms of the Riemann tensor and its derivatives.
    
    Already here, the situation is slightly different to the gauge-theory case:
    In QCD in fixed-point gauge, $A_{\mu}(0)$ vanishes and the first (and higher) derivatives $x^\alpha\lc\partial_\alpha A_{\mu}\rc(0)$ are expressed in terms of the field-strength tensor (and its derivatives).
    In gravity, on the other hand, the first derivative $x^\alpha\lc\partial_\alpha g_{\mu\nu}\rc(0)$ vanishes, and only the second (and higher) derivatives appear in terms of the Riemann tensor (and its derivatives). The leading term $g_{\mu\nu}(0)$ is unaffected by the RNC but can be diagonalised.
    
    It is instructive for the following to discuss RNC in detail. RNC can be fixed in different ways.
    One approach is to impose the gauge condition
    \begin{equation}\label{eq::SG::RNCgaugecondition}
    x^\mu x^\nu \chris{\alpha}{\mu\nu}(x) = 0
    \end{equation}
    on the Christoffel symbols, stressing the similarity to fixed-point gauge $x^\mu A_\mu(x) = 0$.
    Another approach is to construct a reference frame where geodesics in the vicinity of the origin are parametrised as straight lines.
    Both approaches lead to the same result.
    To set up the notation for the following section, we derive and verify \eqref{eq::SG::MetricRNC} as well as \eqref{eq::SG::MetricRNCDiagonalised} in our notation.
    In the subsequent text, if no argument of a field is given, it is understood to be evaluated at $x=0$, e.g $\chris{\mu}{\alpha\beta} \equiv \chris{\mu}{\alpha\beta}(0)$.
    
    In Riemann normal coordinates, a geodesic $y^\mu(s)$ passing through the origin at $s=0$ and through the point $x$ close to the origin at $s=1$ is parametrised as 
    \begin{equation}\label{eq::SG::geodesic}
        y^\mu(s) = sx^\mu\,.
    \end{equation}
    In a generic reference frame, this geodesic is given by
    \begin{equation}\label{eq::SG::RNCDerivationy}
        y^\mu(s) = sx^\mu + v^\mu(s)\,,
    \end{equation}
    where $v^\mu(s)$ denotes the displacement of both reference frames, starting at $\mathcal{O}(s^2)$ and satisfying $v^\mu(0)= 0$. Note that $v^\mu(s)$ is also $x$-dependent, but we suppress this dependence in the following.
    The geodesic is a solution of the equation
    \begin{equation}\label{eq::SG::RNCGeodesicequation}
        \frac{d^2 y^\mu(s)}{ds^2} + \chris{\mu}{\alpha\beta}(y(s)) \frac{d y^\alpha(s)}{ds} \frac{d y^\beta(s)}{ds} = 0\,.
    \end{equation}
    Inserting \eqref{eq::SG::RNCDerivationy} into this equation, one solves  
iteratively for $v^\mu(s)$ in the weak-field expansion. At 
leading order, expanding
    \begin{equation}
        \chris{\mu}{\alpha\beta}(sx + v(s)) = \chris{\mu}{\alpha\beta}(sx) + v^\rho(s)\lc\partial_\rho \chris{\mu}{\alpha\beta}\rc(sx) + \mathcal{O}(s^2)\,,
    \end{equation}
one finds
    \begin{equation}
        \frac{d v^{(0)\mu}(s)}{ds^2} = -\chris{\mu}{\alpha\beta}(sx)x^\alpha x^\beta\,,
    \end{equation}
    which is solved by
    \begin{equation}
        v^{(0)\mu}(s) = -\int_0^s ds^\prime\: \int_0^{s^\prime} d{s^{\prime\prime}}\: x^\alpha x^\beta \,\chris{\mu}{\alpha\beta}(xs^{\prime\prime})\,.
    \end{equation}
    To evaluate the integral, $\chris{\mu}{\alpha\beta}(xs^{\prime\prime})$ has to be Taylor-expanded about $x=0$. This leads to an infinite series in $s$, which is controlled by the expansion and the integral.
    One now reinserts $v^{(0)}(s)$ into \eqref{eq::SG::RNCGeodesicequation} and computes the next correction of $\mathcal{O}(s)$.
    This can be done to any desired order, but there is no closed expression for the result.
    
    From the parameter $v^\mu(s)$, one can determine the relation between a generic reference frame $x$ and the RNC frame $\tilde{x}$ from \eqref{eq::SG::RNCDerivationy} by setting $s=1$ and identifying the left-hand side of \eqref{eq::SG::RNCDerivationy} with the generic coordinate, and $x^\mu$ in the right-hand side as the RNC, consistent with the notion of $v^{\mu}$ as the displacement of both reference frames. One obtains
    \begin{equation}\label{eq::SG::RNCDerivationx}
        x^\mu = \tilde{x}^\mu - \frac{1}{2} \tilde{x}^\alpha \tilde{x}^\beta \chris{\mu}{\alpha\beta} + \frac 16 \tilde{x}^\alpha \tilde{x}^\beta \tilde{x}^\nu \lp 2\chris{\mu }{\alpha\tau}\chris{\tau}{\beta\nu} - \lc\partial_\nu\chris{\mu}{\alpha\beta}\rc\rp + \mathcal{O}(\tilde{x}^4)\,.
    \end{equation}
    To obtain the standard form of the RNC, i.e. $\tilde{x}$ expressed in terms of $x$, one inverts \eqref{eq::SG::RNCDerivationx} to find
    \begin{equation}\label{eq::SG::RNCDerivationInversex}
        \tilde{x}^\mu = x^\mu + \frac 12 x^\alpha x^\beta \chris{\mu}{\alpha\beta} + \frac 16 x^\alpha x^\beta x^\nu \lp \chris{\mu }{\alpha\tau}\chris{\tau}{\beta\nu} + \lc\partial_\nu\chris{\mu}{\alpha\beta}\rc\rp + \mathcal{O}(x^4)\,.
    \end{equation}
    One can now proceed to check \eqref{eq::SG::MetricRNC} using the explicit coordinate transformation \eqref{eq::SG::RNCDerivationInversex}, e.g. by computing
    \begin{equation}\label{eq::SG::tileg}
        \tilde{g}_{\mu\nu}(\tilde{x}) = \frac{\partial x^\alpha}{\partial \tilde{x}^\mu}(x) \frac{\partial x^\beta}{\partial \tilde{x}^\nu}(x) g_{\alpha\beta}(x)\,.
    \end{equation}
    In this check, $g_{\mu\nu}(x)$ on the right-hand side is understood to be multipole-expanded about $x=0$ as
    \begin{equation}\label{eq::SG::RNCMetricMultipole}
        g_{\mu\nu}(x) = g_{\mu\nu} + x^\alpha\lc\partial_\alpha g_{\mu\nu}\rc + \frac 12 x^\alpha x^\beta \lc\partial_\alpha\partial_\beta g_{\mu\nu}\rc + \mathcal{O}(x^3)\,.
    \end{equation}
    From metric compatibility $\nabla_\alpha g_{\mu\nu}=0$, one finds the useful relations
    \begin{align}\label{eq::SG::Metriccompat}
    \partial_\alpha g_{\mu\nu} &= \chris{\lambda}{\alpha\mu}g_{\lambda\nu} + \chris{\lambda}{\alpha\nu}g_{\lambda\mu}\,,\\
    \partial_\beta \partial_\alpha g_{\mu\nu} &= \lc \partial_\beta \chris{\lambda}{\alpha\mu}\rc g_{\lambda\nu} + \chris{\lambda}{\alpha\mu}\chris{\rho}{\beta\lambda}g_{\rho\nu} + \chris{\lambda}{\alpha\mu}\chris{\rho}{\beta\nu}g_{\rho\lambda} + \lp \mu\leftrightarrow\nu\rp\,.
    \end{align}
    With these explicit results, it is straightforward to evaluate \eqref{eq::SG::tileg} and verify \eqref{eq::SG::MetricRNC}.
    
    Equipped with the coordinate transformations \eqref{eq::SG::RNCDerivationx}, \eqref{eq::SG::RNCDerivationInversex}, we now construct the analogue of the $R$ Wilson line for the RNC scenario. Recall from \cref{sec::App::FixedPointGauge} that for gauge theory and fixed-point gauge, $V$ of \eqref{eq::App::FPWilsonLine}, the analogue of the $R$ Wilson line, is used to transport the soft gauge transformation of collinear fields from $x$ to $x=0$. This Wilson line takes the form of an inverse gauge transformation and moves the soft field to fixed-point gauge, where it can then be expressed in terms of the field-strength tensor.
    In fixed-point gauge, there is no left-over background field, as the gauge-fixing is complete, and there is no residual gauge transformation, only the global transformation $U(0)$.
    Formally, however, the construction is similar and thus instructive for the more complicated fixed-line scenario.
    
    In gravity, we start from the other direction: We have the explicit coordinate transformations \eqref{eq::SG::RNCDerivationx}, \eqref{eq::SG::RNCDerivationInversex} and can construct the $R$ Wilson line.
    Then, we can apply the Wilson line and investigate the residual gauge transformation, which will also be a global transformation.
    Finally, this allows us to derive the necessary redefinition of the collinear fields.
    
    Instead of doing the coordinate transformation \eqref{eq::SG::RNCDerivationInversex}, we perform an (inverse) active transformation of the metric tensor $g_{\mu\nu}(x)$, taken in RNC, to move it to the generic coordinates $x$. We denote this redefined metric tensor by $\tilde{g}_{\mu\nu}(x)$.
    In this object $\tilde{g}_{\mu\nu}(x)$, the generic coordinate $x$ then takes the form of the RNC coordinate, i.e. we obtain the expansion \eqref{eq::SG::MetricRNC}, as explained in the following.
    
    From \cref{sec::CG::CollinearWilsonLine}, we understand that the analogue of the $R$ Wilson line \eqref{eq::SQCD::RWilsonLine} is a translation operator
    \begin{equation}\label{eq::SG::RWilsonRNC}
            R^{-1}_{\rm RNC}(x) \equiv T_{\theta_{\rm RNC}(x)} =  1 + \theta_{\rm RNC}^\alpha(x)\partial_\alpha + \frac 12 \theta_{\rm RNC}^\alpha(x)\,\theta_{\rm RNC}^\beta(x) \partial_\alpha\partial_\beta + \mathcal{O}(\theta_{\rm RNC}^3)\,,
    \end{equation}
    where $\theta_{\rm RNC}^\alpha(x)$ has to be determined from the RNC coordinates \eqref{eq::SG::RNCDerivationx}, where we write
    \begin{equation}
        x^\mu = \lc T_{\theta_{\rm RNC}} \tilde{x}^\mu\rc\,.
    \end{equation}
    By comparison with \eqref{eq::SG::RNCDerivationx}, we find
    \begin{equation}\label{eq::SG::ThetaRNC}
        \theta_{\rm RNC}^\mu(x) \equiv -\frac 12 x^\alpha x^\beta \chris{\mu}{\alpha\beta} + \frac 16 x^\alpha x^\beta x^\nu \lp 2 \chris{\mu }{\alpha\tau}\chris{\tau}{\beta\nu} - \lc\partial_\nu\chris{\mu}{\alpha\beta}\rc\rp + \mathcal{O}(x^4)\,.
    \end{equation}
    From this translation, one can derive the Jacobi-matrices
    \begin{equation}\label{eq::SG::RNCJacobiDef}
    \tensor{R}{_\mu^\alpha}(x) = \frac{\partial \tilde{x}^\alpha}{\partial x^\mu}(x)\,,\quad 
    \tensor{R}{^\mu_\alpha}(x) = \frac{\partial x^\mu}{\partial \tilde{x}^\alpha}(x)\,.
    \end{equation}
    The gauge-invariant metric tensor is then defined as
    \begin{equation}\label{eq::SG::MetricRNCDefinition}
        \tilde{g}_{\mu\nu}(x) \equiv \tensor{R}{^\alpha_\mu}(x)\tensor{R}{^\beta_\nu}(x) \lc R^{-1}_{\rm RNC}(x) g_{\alpha\beta}(x)\rc\,.
    \end{equation}
    One should understand this definition as follows: For a generic metric tensor $g_{\mu\nu}$ not in RNC, we perform an inverse active transformation using $R^{-1}(x)$ and the Jacobians to define an object that satisfies the RNC condition.
    If we fix the coordinates $x$ to be RNC, we have $R(x)=1$ and we see that the metric tensor in RNC and the expression \eqref{eq::SG::MetricRNCDefinition} agree. To evaluate \eqref{eq::SG::MetricRNCDefinition}, only $\tensor{R}{^\alpha_\mu}(x)$ is necessary. Using \eqref{eq::SG::ThetaRNC}, one finds
\begin{eqnarray}
\tensor{R}{^\alpha_\mu}(x) &=& \delta^\alpha_\mu - x^\rho \chris{\alpha}{\rho\mu} + 
        x^\rho x^\sigma\lp  \frac{1}{3} \chris{\alpha}{\mu\lambda}\chris{\lambda}{\rho\sigma} + \frac{2}{3} \chris{\alpha}{\rho\lambda}\chris{\lambda}{\sigma\mu} - \frac 16 \lc\partial_\mu\chris{\alpha}{\rho\sigma}\rc - \frac 13 \lc \partial_\rho \chris{\alpha}{\sigma\mu}\rc\rp
\nonumber\\ &&+ \,\mathcal{O}(x^3)\,.
\label{eq::SG::RNCRJacobi}   
\end{eqnarray}
With \eqref{eq::SG::Metriccompat}, one verifies that \eqref{eq::SG::MetricRNCDefinition} agrees with \eqref{eq::SG::MetricRNC}.
    
    We learn that the metric tensor defined in \eqref{eq::SG::MetricRNCDefinition} is manifestly gauge-invariant, and expressed in terms of $g_{\mu\nu}(0)$ and the Riemann tensor.
    This object has a residual symmetry respecting the multipole expansion about $x=0$, namely \emph{global} transformations $\tensor{A}{_\mu^\alpha}\in \mathrm{GL}(1,3)$ and \emph{global} translations $\varepsilon^\mu$\footnote{The Riemann normal coordinates, as constructed, are defined with respect to the origin at $x=0$. However, due to our active point of view, we can still translate the fields to a different space-time coordinate $x+\varepsilon$. Viewed as a coordinate transformation, this is equivalent to shifting the RNC to the new origin.}, i.e.
    \begin{equation}\label{eq::SG::RNCResidualGauge}
        x^\mu \to \tensor{A}{_\alpha^\mu} x^\alpha + \varepsilon^\mu\,. 
    \end{equation}
    In turn, applying $R(x)$ to the matter fields $\varphi(x)$, one can infer that it moves the gauge transformation (local translation) $\varepsilon^\mu(x)$ to $x=0$, where it is given by the linear transformation \eqref{eq::SG::RNCResidualGauge}.
    In contrast to QCD, the homogeneous gauge transformation is not just given by parameter $\varepsilon(0)$, but contains an \emph{additional} linear transformation.
    This is required in order to have non-trivial tensor transformations.
    
    Observe that the RNC as defined in \eqref{eq::SG::RNCDerivationx} are trivial at the linear level, and the non-trivial transformation begins only at $\mathcal{O}(x^2)$.
    We can modify the linear transformation by choosing different initial conditions when solving the geodesic equation \eqref{eq::SG::RNCGeodesicequation}.
    One convenient choice of coordinates is to perform on top of \eqref{eq::SG::RNCDerivationInversex} an additional coordinate transformation such that $g_{\mu\nu}$ equals the Minkowski metric $\eta_{\mu\nu}$ at the origin.
    To find the relevant parameters, we first write
    \begin{equation}\label{eq::SG::Definitionm}
        g_{\mu\nu}(0) = \tensor{e}{_\mu^\alpha}\tensor{e}{_\nu^\beta}\eta_{\alpha\beta}\,,
    \end{equation}
    introducing the global ``vierbein'' $\tensor{e}{_\mu^\alpha}$, whose existence is guaranteed since $g_{\mu\nu}(0)$ is a symmetric matrix.
    We call $\tensor{e}{_\mu^\alpha}$ a ``vierbein'' in analogy to the curved-space definition, but we do not introduce any local inertial reference frames.
    In the weak-field expansion $g_{\mu\nu} = \eta_{\mu\nu} + h_{\mu\nu}$, these $\tensor{e}{_\mu^\alpha}$ are given in terms of $h_{\mu\nu}$ by
    \begin{equation}
        \tensor{e}{_\mu^\alpha} = \delta_\mu^\alpha + \frac 12 h_{\mu}^\alpha - \frac{1}{8}h_{\mu\beta}h^{\beta\alpha} + \mathcal{O}(h^3)\,.
    \end{equation}
    Using this matrix $\tensor{e}{_\mu^\alpha}$, \eqref{eq::SG::RNCDerivationInversex} is modified as $\check{x}^\mu = \tensor{e}{_\rho^\mu}\tilde{x}^\rho$, and 
    \begin{equation}\label{eq::SG::RNCDiagx}
        \check{x}^\mu = \tensor{e}{_\rho^\mu}\lp x^\rho + \frac 12 x^\alpha x^\beta \chris{\rho}{\alpha\beta} + \frac 16 x^\alpha x^\beta x^\nu \lp \chris{\rho}{\alpha\tau}\chris{\tau}{\beta\nu} + \lc\partial_\nu\chris{\rho}{\alpha\beta}\rc\rp\rp + \mathcal{O}(x^4)\,.
    \end{equation}
    Computing the metric tensor in these coordinates, i.e. using the translation
    \begin{equation}
        x^\mu \equiv \check{x}^\mu + \check{\theta}_{\rm RNC}^\mu(\check{x})\,,
    \end{equation}
    and the definitions \eqref{eq::SG::RNCJacobiDef} one obtains
    \begin{equation}
        \check{g}_{\mu\nu}(\check{x}) = \frac{\partial x^\alpha}{\partial \check{x}^\mu}(x) \frac{\partial x^\beta}{\partial \check{x}^\nu}(x) g_{\alpha\beta}(x)\,,
    \end{equation}
    and one immediately reproduces \eqref{eq::SG::MetricRNCDiagonalised}.
    To find the parameter $\check{\theta}_{\rm RNC}^\mu(x)$ one again needs to invert \eqref{eq::SG::RNCDiagx}.
    For this, one requires the inverse matrix $\tensor{E}{^\mu_\alpha}$, defined via
    \begin{equation}\label{eq::SG::RNCInverseM}
        \tensor{E}{^\mu_\beta}\tensor{e}{_\alpha^\beta} = \delta^\mu_\alpha\,.
    \end{equation}
    The parameter $\check{\theta}_{\rm RNC}^\mu(x)$ is then given by
    \begin{equation}\label{eq::SG::ThetaRNCDiag}
    \begin{aligned}
    \check{\theta}_{\rm RNC}^\mu(x) &= (\tensor{E}{^\mu_\rho} - \delta^\mu_\rho) x^\rho - \frac 12 x^\rho x^\sigma \tensor{E}{^\alpha_\rho}\tensor{E}{^\beta_\sigma}\chris{\mu}{\alpha\beta}\\
    &\quad+ \frac 16 x^\rho x^\sigma x^\lambda \tensor{E}{^\alpha_\rho}\tensor{E}{^\beta_\sigma}\tensor{E}{^\nu_\lambda}(2\chris{\mu}{\alpha\tau}\chris{\tau}{\beta\nu} - \brac{\partial_\nu\chris{\mu}{\alpha\beta}}) + \mathcal{O}(x^4)\,.
\end{aligned}
\end{equation}
Evaluating the metric tensor in RNC \eqref{eq::SG::RNCDiagx} with the parameter $\check{\theta}_{\rm RNC}^\mu(x)$, one obtains an object satisfying the diagonalised expansion \eqref{eq::SG::MetricRNCDiagonalised}.

The residual gauge transformation, i.e. the global transformations \eqref{eq::SG::RNCResidualGauge}, are also affected by this linear transformation. They change into global translations and global Lorentz transformations, the background symmetry of the Minkowski metric tensor $\eta_{\mu\nu}$.

To summarise, Riemann normal coordinates \eqref{eq::SG::RNCDerivationInversex} allow us to express the derivatives of the metric tensor in terms of the Riemann tensor. These coordinates completely fix the gauge, and the residual transformations are the (global) symmetry transformations of the zeroth-order term $g_{\mu\nu}(0)$.
By modifying the RNC by an additional global transformation \eqref{eq::SG::Definitionm}, it is possible to change this residual symmetry to global Poincar\'e transformations.
In turn, redefining matter fields with the $R$ Wilson line \eqref{eq::SG::RWilsonRNC} with the corresponding parameter \eqref{eq::SG::ThetaRNCDiag} changes the matter gauge transformations from local translations to global Poincar\'e transformations, consistent with the idea of moving the gauge transformation from $x$ to $x=0$.
The key point is that already in this simple setting, the residual gauge transformation of gravity is more complicated than the gauge-theory analogue, and it is easier to start from the gauge condition, RNC, then work backwards.

\subsubsection{Fixed-line normal coordinates}
\label{sec::SG::FLNC}

We return to the SCET construction and extend the discussion of the previous section to the light-front multipole expansion around $x_-^\mu = \np x \frac{\nm^\mu}{2}$ by generalising the RNC to their fixed-line version, in analogy to gauge theory.
Recall that already in gauge theory, as explained in \cref{sec::SQCD::FixedLineGauge}, the fixed-line generalisation is only a partial gauge-fixing, and there is a left-over, unconstrained component $\nm A_s(x_-)$, which serves as a new background field in the effective theory.
Hence, we anticipate that the generalisation in gravity will also only be a partial gauge-fixing, and a residual, unconstrained background field will arise.

Instead of \eqref{eq::SG::RNCMetricMultipole}, the soft metric tensor is expanded as
\begin{align}
    g_{s\mu\nu}(x) &= g_{s\mu\nu}(x_-) + x_\perp^\alpha\lc\partial_\alpha g_{s\mu\nu}\rc(x_-) +
    \frac 12 \nm x \lc\np\partial g_{s\mu\nu}\rc(x_-)\nn\\
    &\quad
    + \frac 12 x_\perp^\alpha x_\perp^\beta \lc\partial_\alpha\partial_\beta g_{s\mu\nu}\rc(x_-)
    + \mathcal{O}(\lambda^3 g_{s\mu\nu})\,,
\end{align}
and we wish to find the analogue to \eqref{eq::SG::MetricRNCDiagonalised}, i.e. express the second and higher derivatives via the Riemann tensor.
We denote this generalisation of Riemann normal coordinates by fixed-line normal coordinates (FLNC).

The derivation of these coordinates is analogous to the one of the RNC presented in the previous section.
The generalisation of the gauge condition \eqref{eq::SG::RNCgaugecondition} reads
\begin{equation}
    (x-x_-)^\alpha (x-x_-)^\beta \chris{\mu}{\alpha\beta}(x) = 0\,,
\end{equation}
which notably does not restrict the $\chris{\mu}{--}$ components. It is convenient to again consider the geometric construction of required parameter $\theta_{\rm FLNC}(x)$.
To declutter the notation, {\em we suppress $x_-^\mu$ in the arguments of soft fields in the following}, e.g $g_{s\mu\nu}(x_-)\equiv g_{s\mu\nu}$.

Instead of parametrising the geodesic as \eqref{eq::SG::geodesic}, one chooses 
\begin{equation}\label{eq::SG::FLgeodesic}
    y^\mu(s) = x_-^\mu + s(x-x_-)^\mu + v^\mu(s)\,,
\end{equation}
and constructs the RNC only transverse to $x_-^\mu$.
The derivation then follows the RNC case, which uses the geodesic equation \eqref{eq::SG::RNCGeodesicequation} to solve for $v^\mu(s)$ iteratively.
The fixed-line generalisation of \eqref{eq::SG::RNCDerivationInversex} is given by
    \begin{align}\label{eq::SG::FLNCDerivationInversex}
        \tilde{x}^\mu &= x^\mu + \frac 12 (x-x_-)^\alpha (x-x_-)^\beta \chris{\mu}{\alpha\beta}\\
        &\quad+ \frac 16 (x-x_-)^\alpha (x-x_-)^\beta (x-x_-)^\nu \lp \chris{\mu }{\alpha\tau}\chris{\tau}{\beta\nu} + \lc\partial_\nu\chris{\mu}{\alpha\beta}\rc\rp + \mathcal{O}((x-x_-)^4)\,.\nn
    \end{align}
As discussed before, these coordinates can still be modified by an additional ``linear'' transformation $(x-x_-)^\mu \to \tensor{A}{_\alpha^\mu}(x_-)\, (x-x_-)^\alpha$ to change $g_{s\mu\nu}(x_-)$.
Motivated by the RNC result, we introduce the ``vierbein'' $\tensor{e}{_\mu^\alpha}(x_-)$ via\footnote{Note that \eqref{eq::SG::MetricTensorVierbeinFLNC} defines the ``vierbein'' only up to an $x_-$-dependent Lorentz transformation. This local Lorentz transformation carries Greek indices and is linked to the gravitational gauge symmetry, i.e. diffeomorphisms. We do not introduce local inertial frames at each point $x$.}
\begin{equation}\label{eq::SG::MetricTensorVierbeinFLNC}
    g_{s\mu\nu}(x_-) \equiv \tensor{e}{_\mu^\alpha}(x_-)\tensor{e}{_\nu^\beta}(x_-) \eta_{\alpha\beta}\,.
\end{equation}
This ``vierbein'' is the analogue of $\tensor{e}{_\mu^\alpha}$ defined in \eqref{eq::SG::Definitionm}, and its weak-field expansion is given by
\begin{equation}
    \tensor{e}{_\mu^\alpha} = \delta_\mu^\alpha + \frac 12 s_\mu^\alpha - \frac 18 s_{\mu\beta}s^{\beta\alpha} + \mathcal{O}(s^3)\,.
\end{equation}
Next, we take the coordinate $\tilde{x}^\mu$ given in \eqref{eq::SG::FLNCDerivationInversex}, and perform a linear transformation in the transverse components $\tilde{x}_\perp$, $\nm\tilde{x}$ using the ``vierbein''.
This yields a new coordinate system, denoted by $\check{x}$, given as
\begin{equation}\label{eq::SG::DiagFLNCRelationtonondiag}
    \check{x}_- = \tilde{x}_-\,,\quad \check{x}^\mu_\perp = \tensor{e}{_{\alpha}^{\mu_\perp}} \tilde{x}^\alpha\,,\quad \nm \check{x} = n_{-\rho}\tensor{e}{_\alpha^{\rho}} \tilde{x}^\alpha\,,
\end{equation}
where the $\tilde{x}_-$-coordinate is unchanged.
Evaluating \eqref{eq::SG::DiagFLNCRelationtonondiag} and using \eqref{eq::SG::FLNCDerivationInversex} yields $\check{x}$ in terms of the original coordinate $x$. We are interested in the inverse relation to obtain the parameter $\theta_{\rm FLNC}$.
The original coordinate $x$ is expressed in terms of $\check{x}$ as 
\begin{align}\label{eq::SG::FLNCInversex}
    x^\mu &= \check{x}^\mu + (\tensor{E}{^\mu_\alpha} - \delta^\mu_\alpha)(\check{x}-\check{x}_-)^\alpha - \frac 12 (\check{x}-\check{x}_-)^\rho (\check{x}-\check{x}_-)^\sigma \tensor{E}{^\alpha_\rho}\tensor{E}{^\beta_\sigma}\chris{\mu}{\alpha\beta}\\
    &\quad+ \frac 16 (\check{x}-\check{x}_-)^\rho (\check{x}-\check{x}_-)^\sigma (\check{x}-\check{x}_-)^\kappa \tensor{E}{^\alpha_\rho}\tensor{E}{^\beta_\sigma}\tensor{E}{^\nu_\kappa}(2\chris{\mu}{\alpha\lambda}\chris{\lambda}{\beta\nu} - \brac{\partial_\nu\chris{\mu}{\alpha\beta}})
    + \mathcal{O}(\check{x}^3)\nn\,,
\end{align}
where we introduced the inverse ``vierbein'' $\tensor{E}{^\mu_\alpha}(x_-)$ via
\begin{equation}
    \tensor{E}{^\mu_\alpha}(x_-)\tensor{e}{_\nu^\alpha}(x_-) = \delta^\mu_\nu\,.
\end{equation}
Its weak-field expansion is given by
\begin{equation}
    \tensor{E}{^\mu_\alpha} = \delta^\mu_\alpha - \frac 12 s^\mu_\alpha + \frac 38 s^{\mu\beta}s_{\beta\alpha} + \mathcal{O}(s^3)\,.
\end{equation}
From \eqref{eq::SG::FLNCInversex} one can now read off the parameter $\theta^\mu_{\rm FLNC}(x)$ to be 
\begin{eqnarray}
\label{eq::SG::FLNCLLParameter}
    \theta^{\mu}_{\rm FLNC}(x) &=& (\tensor{E}{^\mu_\rho} - \delta^\mu_\rho)(x-x_-)^\rho - \frac 12 (x-x_-)^\rho (x-x_-)^\sigma \tensor{E}{^\alpha_\rho}\tensor{E}{^\beta_\sigma}\chris{\mu}{\alpha\beta}\\
    &&\hspace*{-1.5cm}+ \,\frac 16 (x-x_-)^\rho (x-x_-)^\sigma (x-x_-)^\lambda \tensor{E}{^\alpha_\rho}\tensor{E}{^\beta_\sigma}\tensor{E}{^\nu_\lambda}(2\chris{\mu}{\alpha\tau}\chris{\tau}{\beta\nu} - \brac{\partial_\nu\chris{\mu}{\alpha\beta}}) + \mathcal{O}(x^4)\,.\nn
\end{eqnarray}
Note the formal similarity to \eqref{eq::SG::ThetaRNCDiag}.
However, the FLNC parameter \eqref{eq::SG::FLNCLLParameter} differs from \eqref{eq::SG::ThetaRNCDiag} in two ways: first, each soft field depends on $x^\mu_-$, that is, lives on the light-cone, as opposed to constant fields at $x=0$ in \eqref{eq::SG::ThetaRNCDiag}. Second, the expansion in $(x-x_-)$ only contains $x_\perp$ and $\nm x$, but not $x_-$.
For these two reasons, this results in an incomplete gauge-fixing of $g_{s\mu-}(x)$, as we explicitly check below.

With the parameter \eqref{eq::SG::FLNCLLParameter} we define the 
fixed-line generalisation of the $R$ Wilson line \eqref{eq::SG::RWilsonRNC} as the translation
\begin{equation}\label{eq::SG::RFLNCDef}
    R^{-1}_{\rm FLNC}(x) = T_{\theta_{\rm FLNC}(x)}\,,
\end{equation}
and from this the Jacobian $\tensor{R}{^\alpha_\mu}$ \eqref{eq::SG::RNCJacobiDef}.
Then, analogous to the static case \eqref{eq::SG::MetricRNCDefinition}, we define the metric tensor $\check{g}_{s\mu\nu}(x)$ in fixed-line gauge as
\begin{equation}\label{eq::SG::FLNCMetricDefinition}
\check{g}_{s\mu\nu}(x) \equiv \tensor{R}{^\alpha_\mu}(x)\tensor{R}{^\beta_\nu}(x) \lc R^{-1}_{\rm FLNC}(x) g_{s\alpha\beta}(x)\rc\,,    
\end{equation}
where the Jacobians $\tensor{R}{^\alpha_\mu}$ are constructed from $R^{-1}_{\rm FLNC}(x)$.

We now proceed to evaluate the expression \eqref{eq::SG::FLNCMetricDefinition} explicitly. 
First, we consider the transverse components $\check{g}_{s\mu_\perp\nu_\perp}(x)$.
The Jacobian $\tensor{R}{^\alpha_{\mu_\perp}}(x)$ \eqref{eq::SG::RNCJacobiDef} is given by
\begin{align}
    \tensor{R}{^\alpha_{\mu_\perp}}(x) &= \tensor{E}{^\alpha_{\mu_\perp}} - (x-x_-)^\rho \tensor{E}{^\kappa_{\mu_\perp}}\tensor{E}{^\lambda_\rho}\chris{\alpha}{\kappa\lambda}\nn\\
    &\quad+ \frac 16 (x-x_-)^\rho(x-x_-)^\sigma \biggl( 
    2 \tensor{E}{^\kappa_{\mu_\perp}}\tensor{E}{^\lambda_\rho}\tensor{E}{^\nu_\sigma}(2 \chris{\alpha}{\nu\tau}\chris{\tau}{\kappa\lambda} - \brac{\partial_\nu\chris{\alpha}{\kappa\lambda}})\nn\\
    &\quad+ \tensor{E}{^\kappa_{\rho}}\tensor{E}{^\lambda_\sigma}\tensor{E}{^\nu_\mu_\perp}(
    2\chris{\alpha}{\nu\tau}\chris{\tau}{\kappa\lambda} - \brac{\partial_\nu\chris{\alpha}{\kappa\lambda}})
    \biggr) + \mathcal{O}(x^3)\,.
\end{align}
Evaluating \eqref{eq::SG::FLNCMetricDefinition}, we find
\begin{equation}\label{eq::SG::gppresult}
    \check{g}_{s\mu_\perp\nu_\perp}(x) =
        \eta_{\mu_\perp\nu_\perp} - \frac 13 (x-x_-)^\rho (x-x_-)^\sigma \tensor{E}{^\alpha_\rho}\tensor{E}{^\beta_\sigma}\tensor{E}{^\kappa_{\mu_\perp}}\tensor{E}{^\lambda_{\nu_\perp}}R_{\alpha\kappa \beta\lambda} + \mathcal{O}(x^3)\,.
\end{equation}
The transverse components of the metric tensor in FLNC take a form similar to the result from standard RNC \eqref{eq::SG::MetricRNCDiagonalised}. The same is also true for the components $\check{g}_{\mu_\perp+}$ and $\check{g}_{++}$.
Next, we consider
\begin{equation}
    \check{g}_{s\mu_\perp-}(x) = \tensor{R}{^\alpha_{\mu_\perp}}(x)\tensor{R}{^\beta_-}(x) \lc R^{-1}_{\rm FLNC}(x) g_{s\alpha\beta}(x)\rc\,.
\end{equation}
The combination $\tensor{R}{^\alpha_{\mu_\perp}}(x)\tensor{R}{^\beta_-}(x)$ evaluates to
   \begin{align}
        \tensor{R}{^\alpha_{\mu_\perp}}(x)&\tensor{R}{^\beta_-}(x) =\tensor{E}{^\alpha_{\mu_\perp}}n_-^\beta - y^\rho\left( \tensor{E}{^\kappa_{\mu_\perp}}\tensor{E}{^\lambda_\rho}\chris{\alpha}{\kappa\lambda}\nm^\beta - \tensor{E}{^\alpha_{\mu_\perp}}\partial_-\tensor{E}{^\beta_\rho}\right)\nn\\
        &+ y^\rho y^\sigma \left(
        \frac 16 \nm^\beta\left( 
    2 \tensor{E}{^\kappa_{\mu_\perp}}\tensor{E}{^\lambda_\rho}\tensor{E}{^\nu_\sigma}
    +\tensor{E}{^\kappa_\rho}\tensor{E}{^\lambda_\sigma}\tensor{E}{^\nu_{\mu_\perp}}\right)
    \left(2\chris{\alpha}{\nu\tau}\chris{\tau}{\kappa\lambda} - \brac{\partial_\nu\chris{\alpha}{\kappa\lambda}}\right)\right.\nn\\
    & \left.- \partial_-\tensor{E}{^\beta_\rho}\tensor{E}{^\kappa_{\mu_\perp}}\tensor{E}{^\lambda_\sigma}\chris{\alpha}{\kappa\lambda} - \frac 12 \tensor{E}{^\alpha_{\mu_\perp}}\partial_-(\tensor{E}{^\kappa_\rho}\tensor{E}{^\lambda_\sigma}\chris{\beta}{\kappa\lambda})\right) + \mathcal{O}(x^3)\,,
    \label{eq::SG::gperp-1}
    \end{align}
    where $y^\rho \equiv (x-x_-)^\rho$.
    The remaining factor $\lc R^{-1}_{\rm FLNC}(x) g_{s\alpha\beta}(x)\rc$ is given by
    \begin{align}
        \lc R^{-1}_{\rm FLNC}(x) g_{s\alpha\beta}(x) \rc &= g_{s\alpha\beta} + y^\rho \tensor{E}{^\kappa_\rho}\lc\partial_\kappa g_{s\alpha\beta}\rc + \frac 12 y^\rho y^\sigma \tensor{E}{^\kappa_\rho}\tensor{E}{^\lambda_\sigma}\lc\partial_\kappa\partial_\lambda g_{s\alpha\beta}\rc\nn\\
        &\quad- \frac 12 y^\rho y^\sigma \tensor{E}{^\kappa_\rho}\tensor{E}{^\lambda_\sigma}\chris{\tau}{\kappa\lambda}\lc\partial_\tau g_{s\alpha\beta}\rc + \dots\,.\label{eq::SG::gperp-2}
    \end{align}
Combining \eqref{eq::SG::gperp-1} and \eqref{eq::SG::gperp-2}, one finds that the $\mathcal{O}(x^0)$ term
    \begin{equation}
        \check{g}_{s\mu_\perp-}^{(0)}(x) = \tensor{E}{^\alpha_{\mu_\perp}}g_{s\alpha-} = \tensor{E}{^\alpha_{\mu_\perp}}\tensor{e}{_\alpha^\rho}\tensor{e}{_-^\sigma}\eta_{\rho\sigma} = \tensor{e}{_{-{\mu_\perp}}}
\label{eq::SG::gperp-result0}
    \end{equation}
is the ``vierbein'' $\tensor{e}{_\mu^\alpha}(x_-)$ introduced in \eqref{eq::SG::MetricTensorVierbeinFLNC}.
    At $\mathcal{O}(x)$, we use \eqref{eq::SG::Metriccompat} to rewrite the derivative of the metric in the first line of the following equation, and find
\begin{align}
        \check{g}_{s\mu_\perp-}^{(1)}&(x) = y^\rho\left( \tensor{E}{^\alpha_{\mu_\perp}}\tensor{E}{^\kappa_\rho}\brac{\partial_\kappa g_{s\alpha-}} - \tensor{E}{^\kappa_{\mu_\perp}}\tensor{E}{^\lambda_\rho}\chris{\alpha}{\kappa\lambda}g_{s\alpha-} + \tensor{E}{^\alpha_{\mu_\perp}}\lc\partial_-\tensor{E}{^\beta_\rho}\rc g_{s\alpha\beta}\right)\nn\\
        &= y^\rho \left( \tensor{E}{^\alpha_{\mu_\perp}}\tensor{E}{^\kappa_\rho}(\chris{\tau}{\kappa\alpha}g_{s\tau-} + \chris{\tau}{\kappa-}g_{s\tau\alpha}) - \tensor{E}{^\kappa_{\mu_\perp}}\tensor{E}{^\lambda_\rho}\chris{\alpha}{\kappa\lambda}g_{s\alpha-} + \tensor{E}{^\alpha_{\mu_\perp}}\lc \partial_-\tensor{E}{^\beta_\rho}\rc g_{s\alpha\beta}\right)\nn\\
        &= y^\rho\left( \tensor{E}{^\alpha_{\mu_\perp}}\lc \partial_-\tensor{E}{^\beta_\rho}\rc g_{s\alpha\beta} + \tensor{E}{^\alpha_{\mu_\perp}}\tensor{E}{^\kappa_\rho}\chris{\beta}{\kappa-}g_{s\beta\alpha}\right)\nn\\
        &= - y^\alpha \lc\Omega_{-}\rc_{\alpha\mu_\perp}\,,\label{eq::SG::gperp-result1}
        \end{align}
where we introduced the ``spin-connection''\footnote{We call this object ``spin-connection'' as it is defined like the standard spin-connection, but from the ``vierbein'' $\tensor{e}{_\mu^\alpha}$ \eqref{eq::SG::MetricTensorVierbeinFLNC}. Thus, it does not have an independent local Lorentz transformation but inherits one when performing a diffeomorphism.}
        \begin{equation}\label{eq::SG::SpinConnectionDefinition}
            \lc\Omega_\mu\rc^{\alpha\beta} =
            \tensor{e}{_\nu^\alpha}\brac{\partial_\mu\tensor{E}{^{\nu \beta}}}
            + 
            \tensor{e}{_\nu^\alpha}\chris{\nu}{\sigma\mu}\tensor{E}{^{\sigma \beta}}
        \end{equation}
in the last line. At second order, after some algebra, we find
        \begin{align}
            \check{g}_{s\mu_\perp-}^{(2)}(x) &= -\frac 23y^\alpha y^\beta \tensor{E}{^\kappa_\alpha}\tensor{E}{^\lambda_\beta}\tensor{E}{^\rho_{\mu_\perp}}n_-^\nu R_{\rho\kappa\nu\lambda}\,.\label{eq::SG::gperp-result2}
        \end{align}
Combining \eqref{eq::SG::gperp-result0}, \eqref{eq::SG::gperp-result1} and \eqref{eq::SG::gperp-result2}, we obtain
        \begin{align}\label{eq::SG::gperp-result}
            \check{g}_{s\mu_\perp-}(x) &= e_{-\mu_\perp} - y^\alpha \lc\Omega_{-}\rc_{\alpha \mu_\perp} - \frac 23 y^\alpha y^\beta \tensor{E}{^\kappa_\alpha}\tensor{E}{^\lambda_\beta}\tensor{E}{^\rho_{\mu_\perp}}n_-^\nu R_{\rho\kappa\nu\lambda} + \mathcal{O}(x^3)\,,
        \end{align}
and similarly for $\check{g}_{s+-}$,
        \begin{align}\label{eq::SG::g+-result}
            \check{g}_{s+-}(x) &= e_{-+} - y^\alpha \lc\Omega_{-}\rc_{\alpha +} - \frac 23 y^\alpha y^\beta \tensor{E}{^\kappa_\alpha}\tensor{E}{^\lambda_\beta}\tensor{E}{^\rho_+}n_-^\nu R_{\rho\kappa\nu\lambda} + \mathcal{O}(x^3)\,.
        \end{align}
Finally, for $\check{g}_{s--}(x)$, some algebra results in
        \begin{align}
            \check{g}_{s--}(x) &= g_{s--} - 2 y^\rho \lc\Omega_{-}\rc_{\rho \alpha}\tensor{e}{_-^\alpha} - y^\rho y^\sigma \tensor{E}{^\kappa_\rho} \tensor{E}{^\lambda_\sigma} n_-^\mu n_-^\nu R_{\mu\kappa\nu\lambda}\nn\\
            &\quad+ y^\rho y^\sigma(\brac{\partial_- \tensor{E}{^\mu_\rho}}\brac{\partial_-\tensor{E}{^\nu_\sigma}}g_{s\mu\nu} + 2 \brac{\partial_- \tensor{E}{^\mu_\rho}}\tensor{E}{^\kappa_\sigma}\chris{\lambda}{\kappa-}g_{s\lambda\mu}\nn\\
            &\qquad+ \tensor{E}{^\kappa_\rho}\tensor{E}{^\lambda_\sigma}\chris{\alpha}{\lambda-}\chris{\beta}{\kappa-}g_{s\alpha\beta}) + \mathcal{O}(x^3)\,,
        \end{align}
and expressing the metric as $g_{s--} = \tensor{e}{_-^\alpha}\tensor{e}{_-^\beta}\eta_{\alpha\beta}$, yields
\begin{eqnarray}
\check{g}_{s--}(x) &=& (\tensor{e}{_-^\alpha} - y^\rho\tensor{\lc\Omega_-\rc}{_{\rho}^\alpha})(\tensor{e}{_-^\beta} - y^\sigma\tensor{\lc\Omega_-\rc}{_{\sigma}^\beta})\eta_{\alpha\beta} - y^\alpha y^\beta \tensor{E}{^\kappa_\alpha}\tensor{E}{^\lambda_\beta}n_-^\mu n_-^\nu R_{\mu\kappa\nu\lambda} 
\nn\\
&& + \,\mathcal{O}(x^3)\,.
\label{eq::SG::g--}
\end{eqnarray}
        
        Let us compare the metric components \eqref{eq::SG::gppresult}, \eqref{eq::SG::gperp-result}, \eqref{eq::SG::g--} in fixed-line normal coordinates to the corresponding ones \eqref{eq::SG::MetricRNCDiagonalised} in Riemann normal coordinates.
        The first observation is that the transverse components \eqref{eq::SG::gppresult} behave like in standard RNC.
        Here, the metric tensor $\check{g}_{s\mu_\perp\nu_\perp}$ is expressed in terms of the Minkowski metric $\eta_{\mu_\perp\nu_\perp}$ and the Riemann tensor.
        The same holds true for $\check{g}_{s\mu_\perp\!+}$.
        The gauge theory case is analogous. Here, the $A_{s\perp}$ and $\np A_s$ fields satisfy the same expansion as the gluon field in fixed-point gauge.
        Only $\nm A_s$ has a different one, where the background field $\nm A_s(x_-)$ appears, cf. \eqref{eq::app::FLA-R} -- \eqref{eq::app::FLA+R}.
        
        Like in the gauge theory case, we can now split $\check{g}_{s\mu\nu}$ into the residual background field, which we denote by $\hat{g}_{s\mu\nu}$, and the gauge-covariant part that contains only the Riemann tensor, i.e. the analogue of \eqref{eq::SQCD::GaugeCovariantSoftGluon}, which we denote by $\mathfrak{g}_{s\mu\nu}$.
        We write 
        \begin{equation}\label{eq::SG::BGMetricSplitRiemann}
            \check{g}_{s\mu\nu}(x) \equiv \hat{g}_{s\mu\nu}(x) + \mathfrak{g}_{s\mu\nu}(x)\,,
        \end{equation} 
and deduce from \eqref{eq::SG::gppresult}, \eqref{eq::SG::gperp-result} \eqref{eq::SG::g+-result}, \eqref{eq::SG::g--}:
\begin{align}
\label{eq::SG::ResidualMetric1}
            &\hat{g}_{s+-}(x) = e_{-+} - (x-x_-)^\alpha \lc\Omega_{-}\rc_{\alpha +}\,,\\[0.15cm]
&\hat{g}_{s\mu_\perp-}(x) = e_{-\mu_\perp} - (x-x_-)^\alpha  \lc\Omega_{-}\rc_{\alpha\mu_\perp}\,,
\label{eq::SG::ResidualMetric1b}\\
        	&\hat{g}_{s--}(x) = \lp\tensor{e}{_-^\alpha} - (x-x_-)^\rho \tensor{\lc\Omega_-\rc}{_{\rho}^\alpha}\rp\label{eq::SG::ResidualMetric2}
        	\lp \tensor{e}{_-^\beta} - (x-x_-)^\sigma \tensor{\lc\Omega_-\rc}{_{\sigma}^\beta}\rp \eta_{\alpha\beta}\,,\\
        	&\hat{g}_{s\mu_\perp\nu_\perp}(x) = \eta_{\mu_\perp\nu_\perp}\,.\label{eq::SG::ResidualMetric3}
        	\end{align}
        Thus, the residual soft background field has two contributions, $\tensor{e}{_-^\alpha}(x_-)$ and $\tensor{\lc\Omega_-\rc}{_{\alpha\beta}}(x_-)$. 
        In the weak-field expansion, these two quantities are given by
        \begin{align}\label{eq::SG::VierbeinWFExpansion}
            \tensor{e}{_-^\alpha} &= \delta_-^\alpha + \frac 12 s_-^\alpha - \frac{1}{8} s_{-\beta}s^{\beta\alpha} + \mathcal{O}(s^3)\,,\\
            \tensor{\lc\Omega_-\rc}{_{\alpha\beta}} &= -\frac 12 \lp \lc \partial_\alpha s_{\beta-}\rc - \lc\partial_\beta s_{\alpha-}\rc\rp + \mathcal{O}(s^2)\,,\label{eq::SG::SpinConnectionWFExpansion}
        \end{align}
and one sees that the linear term in $\tensor{e}{_-^\alpha}(x_-)$ is proportional to $s_-^\alpha(x_-)$, whereas the linear term in $\tensor{\lc\Omega_-\rc}{_{\alpha\beta}}(x_-)$ is proportional to the derivative of $s_{\mu\nu}(x)$, evaluated at $x_-$ {\em after} taking the derivative.
        In the effective theory, all soft fields depend only on the light-cone variable $x_-$, and therefore these two fields must be treated as independent objects, since the full $x$-dependence of $s_{\mu\nu}(x)$ can no longer be accessed -- the derivatives required to compute $\lc\Omega_{-}\rc_{\alpha\beta}$ are not obtainable from $s_{\mu\nu}(x_-)$.
        
        Apart from these two fields, only the Riemann tensor $R_{\mu\nu\alpha\beta}(x_-)$ and its derivatives appear in the gauge-covariant part $\mathfrak{g}_{s\mu\nu}$. Note how similar to the QCD case the result is. There, the soft gluon field is expressed in terms of the homogeneous background field $\nm A_s(x_-)$ and its field-strength tensor $F_{\mu\nu}^s$ via the covariant expression $\mathcal{A}_s$ \eqref{eq::SQCD::GaugeCovariantSoftGluon}.
        
This is one of the most important insights for the gravitational construction.
        In both theories, QCD and gravity, properly taking into account the multipole expansion gives rise to non-vanishing soft background fields, which form an essential part of the soft-collinear physics.
        The fixed-line gauge condition in QCD corresponds to fixed-line normal coordinates in gravity.
In the gravitational case, the non-vanishing background field contains two independent objects, one related to the ``vierbein'' and one related to the ``spin-connection''.
        While soft-collinear QCD is thus covariant with respect to $\nm A_s(x_-)$, soft-collinear gravity must respect the symmetries of the background $\hat{g}_{s\mu\nu}$, i.e.~of the two fields $\tensor{e}{_-^\alpha}(x_-)$ and $\lc\Omega_{-}\rc_{\alpha\beta}(x_-)$.
        In the following, we show how these two residual background fields can be arranged into a soft-covariant derivative, and how all building blocks in soft-collinear gravity can be rendered covariant with respect to this emergent background and its symmetries.
	
\subsubsection{Soft-covariant derivative and emergent gauge symmetry}
	
	In this section, we explain how the soft-covariant derivative arises in the effective theory and determine the residual gauge symmetry of the soft background.
	
	Consider again the split \eqref{eq::SG::BGMetricSplitRiemann} of the fixed-line metric tensor into the background field $\hat{g}_{s\mu\nu}$ and the covariant part $\mathfrak{g}_{s\mu\nu}$.
	In the effective theory, only the metric $\hat{g}_{s\mu\nu}$ is used to raise and lower indices, and any tensor must be covariant with respect to its symmetry.
	For scalar fields, we construct tensors using the derivative $\partial_\mu$, and from this a manifestly-scalar object is defined as
	\begin{equation}\label{eq::SG::derivativecalc1}
	    \hat{g}_s^{\mu\nu}\partial_\mu\varphi_c\,\partial_\nu\varphi_c\,.
	\end{equation}
	Next, observe from \eqref{eq::SG::ResidualMetric1} -- \eqref{eq::SG::ResidualMetric3} that only $\hat{g}_s^{\mu-}$ is non-trivial.
	In \eqref{eq::SG::derivativecalc1}, this implies for the scalar combination
	\begin{equation}
	    \hat{g}_s^{\mu\nu}\partial_\mu\varphi_c\,\partial_\nu\varphi_c = \hat{g}_s^{\mu-}\partial_\mu\varphi_c\,\np\partial\varphi_c + \eta_\perp^{\mu\nu}\partial_{\mu_\perp}\varphi_c\,\partial_{\nu_\perp}\varphi_c\,.
	\end{equation}
	We thus identify the soft-covariant $\nm D_s$ derivative as
	\begin{align}\label{eq::SG::softcovariantderivative}
            \nm D_s &\equiv \hat{g}_s^{\mu-}\partial_\mu\\
            &=
            \partial_- - \frac 12 s_{-\mu} \partial^\mu + \frac 18 s_{+-}s_{--}\partial_+ + \frac{1}{16} s_{-\alpha_\perp}s^{\alpha_\perp}_-\partial_+
            - \frac 14 \tensor{\lc\Omega_-\rc}{_{\mu\nu}}J^{\mu\nu} + \mathcal{O}(\lambda^3)\nn\,,
    \end{align}
where $J^{\mu\nu} = (x-x_-)^\mu \partial^\nu - (x-x_-)^\nu \partial^\mu \equiv (x-x_-)^{[\mu}\partial^{\nu]}$ denotes the angular momentum (Lorentz generator) operator.
    Let us stress again that, unlike QCD, the soft-covariant derivative in gravity contains two independent fields, $\tensor{e}{_{-\mu}}(x_-)$ and $\lc\Omega_{-}\rc_{\alpha\beta}(x_-)$, of $\mathcal{O}(\lambda^2)$ and $\mathcal{O}(\lambda^4)$, respectively. These two fields couple at linear order to the momentum and angular momentum of the scalar field.

    Next, we investigate the residual gauge transformations of $\hat{g}_{s\mu\nu}(x)$ explicitly, and verify that the derivative defined above in \eqref{eq::SG::softcovariantderivative} is indeed covariant.
    The residual transformations are simply the symmetry transformations of the background $\hat{g}_{s\mu\nu}(x)$, and correspond to $U_s(x_-)$ in the gauge theory case \eqref{eq::SQCD::HomGaugeTransformations}.
    We compute explicitly the infinitesimal transformations.
For this, one first performs the weak-field expansion
\begin{equation}\label{eq::SG::WFExpansionHatBG}
    \hat{g}_{s\mu\nu}(x) = \eta_{\mu\nu} + \hat{s}_{\mu\nu}(x)
\end{equation}
to separate the constant background $\eta_{\mu\nu}$ from the ($\lambda$-inhomogeneous) fluctuation $\hat{s}_{\mu\nu}$ and
expresses $\hat{s}_{\mu\nu}$ in terms of the original soft fluctuation $s_{\mu\nu}(x)$.
Then, one uses the gauge transformation
\begin{equation}\label{eq::SG::SoftFluctuationTransformation}
    s_{\mu\nu}(x) \to s_{\mu\nu}(x) - \partial_\mu \varepsilon_\nu(x) - \partial_\nu \varepsilon_\mu(x) + \mathcal{O}(s^2)\,,
\end{equation}
to determine the transformations of $\hat{s}_{\mu\nu}$.
For example, one obtains for $\hat{s}_{\mu_\perp-}(x)$ 
(dropping non-linear terms) from 
\eqref{eq::SG::ResidualMetric3}, \eqref{eq::SG::VierbeinWFExpansion},
\eqref{eq::SG::SpinConnectionWFExpansion}
	\begin{align}
	\hat{s}_{\mu_\perp-}(x) &= \frac 12 s_{\mu_\perp-} + \frac 12 x_\perp^\alpha \lc \partial_{[\alpha}s_{\mu_\perp]-}\rc + \frac 14 n_-x \lc\partial_{[+}s_{\mu_\perp]-}\rc\,.
	\end{align} 
We can rewrite the gauge transformation \eqref{eq::SG::SoftFluctuationTransformation} for this case as
\begin{align}
	\frac 12 s_{\mu_\perp-} &\to \frac 12 s_{\mu_\perp-} - \frac 12 \partial_{\mu_\perp}\varepsilon_- - \frac 12 \partial_-\varepsilon_{\mu_\perp}\nn\\
	&= \frac 12 s_{\mu_\perp-} - \partial_{-}\varepsilon_{\mu_\perp} + \frac 12 \left(\partial_-\varepsilon_{\mu_\perp}-\partial_{\mu_\perp}\varepsilon_-\right)\nn \\
	&= \frac 12 s_{\mu_\perp-} - \partial_{-}\varepsilon_{\mu_\perp} + \omega_{-\mu_\perp}\,,
	\end{align}
	where we introduced 
	\begin{equation}
	    \omega_{\mu\nu} \equiv \frac 12 \left(\partial_\mu\varepsilon_\nu - \partial_\nu\varepsilon_\mu\right)\,.
	\end{equation}
Computing this for the other components and accounting for the contribution from $\lc\Omega_{-}\rc_{\alpha\beta}$, we find the infinitesimal gauge transformation of $\hat{s}_{\mu-}$ to be
\begin{equation}\label{eq::SG::InfinitesimalGaugeTransS}
    \hat{s}_{\mu-}(x) \to \hat{s}_{\mu-}(x) - \partial_{-}\varepsilon_{\mu} + \omega_{-\mu_\perp} - \partial_- \left((x-x_-)^\alpha \omega_{\mu\alpha}\right) + \dots\,.
\end{equation}

Hence, the infinitesimal transformation can be defined in terms of the parameters $\varepsilon_-(x_-)$ and $\omega_{\mu\nu}(x_-)$ of the coordinate transformation
	\begin{align}
		x^\mu &\to x^\mu + \varepsilon^\mu(x_-) + \tensor{\omega}{^{\mu}_{\nu}}(x_-)(x-x_-)^\nu\,,\label{eq::SG::infinitesimalresidualtransformation}
	\end{align} 
	consisting of a translation with parameter $\varepsilon^\mu(x_-)$ and a rotation (Lorentz transformation) about $(x-x_-)$ with parameter $\tensor{\omega}{^{\mu}_{\nu}}(x_-)$. This infinitesimal transformation corresponds to a local Poincar\'e transformation that lives on the light-cone $x_-^\mu = \np x \frac{\nm^\mu}{2}$.
	It is ``homogeneous'' in the sense that the parameter only depends on $x_-$, respecting the multipole expansion. The generators no longer contain this $x_-$, as the angular momentum is defined about $(x-x_-)$.
	However, due to the scaling of the generators, i.e. the momentum $P^\mu$ and the angular momentum $J^{\mu\nu}$, the transformation itself is not homogeneous in $\lambda$, that is, the components do not all have the same $\lambda$-scaling.
	
	From the all-order expression for $\hat{g}_{s\mu\nu}$ \eqref{eq::SG::ResidualMetric1} -- \eqref{eq::SG::ResidualMetric3}, we can compute the gauge transformation to any desired order in $\lambda$, capturing also the non-linear terms, both in $\varepsilon_\mu(x)$ as well as in the soft fluctuation $s_{\mu\nu}(x)$.
	To find these non-linear terms, one starts from \eqref{eq::SG::ResidualMetric1} -- \eqref{eq::SG::ResidualMetric3}, and performs the expansion in $s_{\mu\nu}$ for the appearing background fields $\tensor{e}{_-^\alpha}(x_-)$ and $\lc\Omega_{-}\rc_{\alpha\beta}(x_-)$ to the desired order in $\lambda$.
	Then, using the full diffeomorphism transformation of the fluctuation $s_{\mu\nu}(x)$, one computes the transformations of $\tensor{e}{_-^\alpha}$ and $\lc\Omega_{-}\rc_{\alpha\beta}$, and finally deduces the combined transformation of $\hat{g}_{s\mu\nu}$.
	
	We proceed to investigate the behaviour of derivatives under this symmetry.
	Under the infinitesimal gauge transformation \eqref{eq::SG::infinitesimalresidualtransformation},
	\begin{align}
	    \varphi_c^\prime(x) &=
	    \varphi(x) - \varepsilon^\alpha\partial_\alpha\varphi(x) - \omega_{\alpha\beta}\,(x-x_-)^\beta\partial^\alpha \varphi(x) + \mathcal{O}(\varepsilon^2)\,.
	\end{align}
	Applying a derivative, we find for the $\partial_\perp$ and $\np\partial$ derivatives of the field the infinitesimal transformations
	\begin{align}\label{eq::partialperptrans}
	    \partial_{\mu_\perp} \varphi(x) &\to T_{\varepsilon+\omega}^{-1}\lc\partial_{\mu_\perp}\varphi(x)\rc - \omega_{\mu_\perp\alpha}\partial^\alpha \varphi(x) + \mathcal{O}(\varepsilon^2)\,,\\
	    \partial_{+} \varphi(x) &\to T_{\varepsilon+\omega}^{-1}\lc\partial_{+}\varphi(x)\rc - \omega_{+\alpha}\partial^\alpha \varphi(x) + \mathcal{O}(\varepsilon^2)\,,\label{eq::partial+trans}
	\end{align}
	where $T_{\varepsilon+\omega}^{-1}$ denotes the translation
	\begin{equation}
	    T_{\varepsilon+\omega}^{-1} = 1 - \varepsilon^\alpha \partial_\alpha - \omega_{\alpha\beta}\,(x-x_-)^\beta\partial^\alpha + \mathcal{O}(\varepsilon^2)\,.
	\end{equation}
	These transformations are already the correct transformations for derivatives under an infinitesimal Lorentz transformation.
	The $\nm\partial$ component, however, does not have the right transformation. This derivative also acts on the gauge parameters and one finds instead
	\begin{equation}
	    \nm\partial\varphi(x) \to T_{\varepsilon+\omega}^{-1}\lc \nm\partial\varphi(x)\rc - \nm\partial \varepsilon^\alpha \partial_\alpha\varphi(x) - \nm\partial \omega_{\alpha\beta}\,(x-x_-)^\beta\partial^\alpha \varphi(x) + \mathcal{O}(\varepsilon^2)\,.
	\end{equation}
	However, from the explicit result for the soft-covariant derivative defined in \eqref{eq::SG::softcovariantderivative}, one can compute the gauge transformation of the linear terms
	\begin{equation}
	    \nm D_s = \partial_- - \frac 12 s_{-\mu} \partial^\mu
            - \frac 14 \tensor{\lc\Omega_-\rc}{_{\mu\nu}}J^{\mu\nu} + \dots
	\end{equation}
	as follows:
	using the transformation of $\hat{s}_{\mu\nu}$ determined in \eqref{eq::SG::InfinitesimalGaugeTransS}, one finds
	\begin{align}
	    \frac 12 s_{-\mu} &\to \frac 12 s_{-\mu} - \partial_-\varepsilon_\mu + \omega_{-\mu} + \dots\,,\\
	    \lc\Omega_{-}\rc_{\alpha\beta} &\to \lc\Omega_{-}\rc_{\alpha\beta} - \partial_- \omega_{\alpha\beta} + \dots\,,
	\end{align}
	and one can immediately see that 
	\begin{equation}
	    \nm D_s\varphi(x) \to T_{\varepsilon+\omega}^{-1}\lc \nm D_s\varphi(x)\rc - \omega_{-\alpha}D_s^\alpha\varphi(x)\,,\label{eq::covariantinftrans}
	\end{equation}
	where $D_s^\alpha = \frac{\nm^\alpha}{2}\np \partial + \partial^\alpha_\perp + \frac{\np^\alpha}{2} \nm D_s$. Thus, the covariant derivative as defined in \eqref{eq::SG::softcovariantderivative} has a covariant transformation under the residual gauge transformation.
	This serves as an explicit check that the geometric intuition used to define $\nm D_s$ is correct.
    
    In addition to this soft-covariant derivative $\nm D_s$, we also require the presence of the residual soft metric determinant $\sqrt{-\hat{g}_s}$ together with the measure $d^4x$, as usual, to render the action manifestly soft background-field gauge-invariant.
    Besides these two objects, the soft metric field can only appear in the Riemann tensor or its derivatives, as a result of fixed-line normal coordinates, and in the definition of manifestly gauge-invariant collinear building blocks, like $\mathfrak{h}^{\mu\nu}$, as explained below.
    This shows that we have achieved the gauge-invariant soft-collinear factorisation we set out to find in the beginning.
Before we therefore move to the Lagrangian itself, let us summarise the most important points:
    \begin{itemize}
        \item The analogue of fixed-point gauge in gravity are the Riemann normal coordinates. The light-cone generalisation, called fixed-line normal coordinates \eqref{eq::SG::FLNCLLParameter}, can be obtained in a similar fashion.
        \item 
        These coordinates only partially fix the gauge. The soft metric tensor $g_{s\mu\nu}$ contains a residual background field $\hat{g}_{s\mu\nu}$ \eqref{eq::SG::ResidualMetric1}~--~\eqref{eq::SG::ResidualMetric3} which can be expressed in terms of the leading $\tensor{e}{_{-\mu}}(x_-)$ and subleading $\lc\Omega_{-}\rc_{\alpha\beta}(x_-)$ fields, defined in \eqref{eq::SG::VierbeinWFExpansion} and \eqref{eq::SG::SpinConnectionWFExpansion}.
        These fields are the gravitational counterpart of $\nm A_s(x_-)$ in QCD, and are considered to be independent fields in the effective theory.
        \item The remaining sub-subleading and all higher-order terms are expressed in terms of the Riemann tensor, and can be computed in a systematic fashion to the desired order in $\lambda$.
        These terms are the analogues of \eqref{eq::SQCD::FLExpansionA-}~--~\eqref{eq::SQCD::FLExpansionA+}.
        \item From the residual background field, we can obtain a residual ``homogeneous'' gauge transformation, which corresponds to local Poincar\'e transformations living on the classical trajectory $x_-^\mu$. This emergent light-cone gauge symmetry controls the soft objects that can appear in the effective Lagrangian.
        \item The specific form of this background field $\hat{g}_{s\mu\nu}$ and its residual gauge symmetry allows us to define a soft-covariant derivative $\nm D_s$ \eqref{eq::SG::softcovariantderivative}.
        The other derivatives transform already correctly under the gauge symmetry.
        \item The soft graviton field then only appears inside the determinant $\sqrt{-\hat{g}_s}$, in the soft-covariant derivative $\nm D_s$, inside the Riemann tensor, or in the definition of manifestly gauge-invariant collinear objects, as shown below.
    \end{itemize}
    With all these notions understood, we can proceed with the derivation of the soft-collinear gravity Lagrangian, which is now remarkably similar to the QCD construction.

\subsection{Soft-collinear Lagrangian}\label{sec::SG::SCLagrangian}

Recall that we derived the background-field Lagrangian \eqref{eq::SG::BFLagrangian}, and we need to perform the multipole expansion in step (ii) as well as the collinear field redefinition in step (iii). For these redefinitions, we need the analogues of the $W_c$ and $R$ Wilson lines.

The new, hatted collinear fields are defined with respect to the homogeneous soft background field $\hat{g}_{s\mu\nu}$ given in \eqref{eq::SG::ResidualMetric1} -- \eqref{eq::SG::ResidualMetric3}, and its (non-linear) inverse $\hat{g}_s^{\mu\nu}$.
The QCD analogues are the hatted fields defined in \eqref{eq::SQCD::CollinearRedefinition} -- \eqref{eq::SQCD::CollinearRedefinition2}, which are covariant with respect to the soft background $\nm A_s(x_-)$. The most important feature of 
gravity is that the 
background field consists of two independent parts, one related to $g_{s-\mu}(x_-)$ and one to its first derivative. 
While they do not have simple closed expressions in terms 
of the soft metric fluctuation, we can construct them to all orders, as explained in \cref{sec::SG::FLNC}.

\subsubsection{``Wilson lines'' and collinear field redefinitions}

Recall that for gauge theory, 
we made use of two Wilson lines.
The first one, $W_c$, defined in \eqref{eq::SQCD::collWilsonLine}, moves the collinear fields to collinear light-cone gauge.
The second one, denoted by $R$ in \eqref{eq::SQCD::RWilsonLine} is needed to obtain the homogeneous gauge transformation of the hatted collinear fields. 
In gravity, we use the analogues of both of these Wilson lines.

The analogue of $W_c$, which we also denote as $W_c$, was already introduced in the purely collinear theory in \cref{sec::CG::CollinearWilsonLine} to control the enhanced and unsuppressed components of $h_{\mu\nu}$, namely $h_{++}$ and $h_{+\mu_\perp}$.
It can be obtained by fixing collinear light-cone gauge.
However, in the soft-collinear theory, there is one major difference to the purely collinear and even the QCD case.
Namely, because the homogeneous background field contains a non-vanishing $\hat{g}_{s+-}$ component \eqref{eq::SG::ResidualMetric1}, the collinear gauge transformation of $h_{+-}$ depends on this soft background, so the gauge transformation of $h_{\mu+}$ is \emph{not independent} of the soft background, unlike in QCD \eqref{eq::SQCD::HomGaugeTransformations}, where $\np A_c$ has a purely collinear transformation.
Thus, the soft field and soft-covariant derivative appear in $W_c$, and it is no longer a purely collinear object.
This ``Wilson line'' is defined as
\begin{equation}\label{eq::SG::CollWilsonLine}
    W_c^{-1} = T_{\theta_{\rm LC}} = 1 + \theta_{\rm LC}^\alpha\partial_\alpha + \mathcal{O}(\lambda^2)\,,
\end{equation}
with parameter $\theta_{\rm LC}$ defined as in \eqref{eq::CG::colltheta} but of the form \eqref{eq::app::v1}, namely
\begin{equation}\label{eq::SG::ThetaLCGauge}
	   \theta_{\rm LC}^\mu = - \frac{1}{(\delp)^2}\hat{\Gamma}^\mu_{++} + \frac{1}{(\delp)^2}\left( 2 \hat{\Gamma}^\mu_{\tau+}
	   \frac{1}{\delp}\hat{\Gamma}^\tau_{++}
	   + \partial_\nu \hat{\Gamma}^\mu_{++}\frac{1}{(\delp)^2}\hat{\Gamma}^\nu_{++}\right) + \mathcal{O}(\lambda^2)\,.
\end{equation}
Here, $\tensor{\hat{\Gamma}}{^\mu_{\alpha\beta}}$ is the Christoffel symbol constructed from $\hat{g}_{s\mu\nu}+h_{\mu\nu}$, i.e.~as a weak-field expansion around the homogeneous background $\hat{g}_{s\mu\nu}$.
The parameter $\theta_{\rm LC}^\mu$ corresponds to \eqref{eq::CG::colltheta} computed with respect to the full metric field $\hat{g}_s+h$, but expressed in terms of geometric objects, as derived in \cref{sec::GeometricConstruction}.
With this ``Wilson line'', we can proceed to define the manifestly gauge-invariant collinear fields according to their tensor representation, as discussed in \cref{sec::CG::CollinearWilsonLine}.

The $R$ Wilson line was defined in the same fashion as a 
translation with the parameter $\theta_{\rm FLNC}$ \eqref{eq::SG::FLNCLLParameter}, given by 
\begin{equation}
    \theta^{\mu}_{\rm FLNC} = (\tensor{E}{^\mu_\rho} - \delta^\mu_\rho)(x-x_-)^\rho - \frac 12 (x-x_-)^\rho (x-x_-)^\sigma \tensor{E}{^\alpha_\rho}\tensor{E}{^\beta_\sigma}\chris{\mu}{\alpha\beta} + \dots\,,
\end{equation}
which depends on the original soft metric field $g_{s\mu\nu}$, as discussed in detail in \cref{sec::SG::FLNC}.
The $R$ Wilson line is then again defined as a translation \eqref{eq::SG::RFLNCDef}
\begin{equation}
    R^{-1} \equiv T_{\theta_{\rm FLNC}} = 1 + \theta_{\rm FLNC}^\alpha\partial_\alpha + \mathcal{O}(\lambda^2)\,.
\end{equation}
The collinear matter fields are redefined according to their representation as
\begin{equation}\label{eq::SG::CollinearRedefinition}
    \varphi_c = \lc RW_c^{-1}\hat\varphi_c\rc \,.
\end{equation}
Similarly for the collinear graviton $h_{\mu\nu}$, the new field 
$\hat{h}_{\mu\nu}$ is defined through
\begin{equation}\label{eq::SG::CollGravRedef}
h_{\mu\nu} = 
\brac{R\tensor{R}{_\mu^\alpha}\tensor{R}{_\nu^\beta}\left(
	\tensor{W}{^\rho_\alpha} \tensor{W}{^\sigma_\beta}\brac{W^{-1}_c(\hat{h}_{\rho\sigma} + \hat{g}_{s\rho\sigma})} - \hat{g}_{s\alpha\beta}\right)}\,,
\end{equation}
which generalizes \eqref{eq::ginv} of the purely collinear case. 
The original fields $\varphi_c$, $h_{\mu\nu}$ on the left-hand side are taken to be in collinear light-cone gauge, while the hatted fields $\hat{\varphi}_c$, $\hat{h}_{\mu\nu}$ are not collinear gauge-fixed.

\subsubsection{Constructing the soft-collinear Lagrangian}
\label{sec::SG::SCLagrangianDerivation}

We can now perform step (iv) and insert the redefinitions \eqref{eq::SG::CollinearRedefinition}, \eqref{eq::SG::CollGravRedef} into the Lagrangians \eqref{eq::SG::BFLagrangian0} -- \eqref{eq::SG::BFLagrangian2}, expressing the original fields $h_{\mu\nu}$, $\varphi_c$ in terms of $\hat{h}_{\mu\nu}$, $\hat{\varphi}_c$.
Note that the redefinitions \eqref{eq::SG::CollinearRedefinition}, \eqref{eq::SG::CollGravRedef} are only valid for the unhatted fields in light-cone gauge.
Thus, as a first step, we fix light-cone gauge $h_{\mu+}=0$ in the Lagrangian, and only then express it in terms of the hatted fields.

This process is quite involved, so we explain in detail how this is done for the leading $\mathcal{L}_{\varphi}$ from  \eqref{eq::SG::BFLagrangian0} for $\varphi_s=0$ and refer to \cref{sec::InsertingExpressions} for the complete derivation of the soft-collinear Einstein-Hilbert Lagrangian up to $\mathcal{O}(\lambda)$ and the  scalar matter Lagrangian up to $\mathcal{O}(\lambda^2)$, covering the terms \eqref{eq::SG::BFLagrangian1} and \eqref{eq::SG::BFLagrangian2}, which explicitly contain $h_{\mu\nu}$.
In $\mathcal{L}_{\varphi}$, the collinear graviton only appears through the Wilson line $W_c$.
A list of useful identities for dealing with the various Wilson lines can be found in \cref{sec::app::UsefulIdentities}. 
In this section, $g_{s\mu\nu}(x)$ and the R-Wilson line $R(x)$ are functions 
of $x$, but we will not write the argument explicitly. 
On the other hand, soft quantities that appear after 
the multipole expansion, such as $s_{\mu\nu}$ 
and the Riemann tensor, will be understood to be evaluated 
on the light-cone~$x_-^\mu$. 

We begin with inserting the redefinitions \eqref{eq::SG::CollinearRedefinition}, \eqref{eq::SG::CollGravRedef} into the first term of the Lagrangian \eqref{eq::SG::BFLagrangian0},
\begin{equation}
    \frac 12 \sqrt{-g_s}g_s^{\mu\nu}\partial_\mu\varphi_c\partial_\nu\varphi_c\,,
\end{equation}
which then reads
	\begin{equation}
	\begin{aligned}[b]
	    \frac 12 &\sqrt{-g_s}g_s^{\mu\nu}\brac{\partial_\mu RW_c^{-1}\hat\varphi_c}\brac{\partial_\nu RW_c^{-1}\hat\varphi_c}\\
	    &=\frac 12 \sqrt{-g_s}g_s^{\mu\nu} R \tensor{R}{_\mu^\alpha}\tensor{R}{_\nu^\beta}\brac{\partial_\alpha W_c^{-1}\hat{\varphi}_c}\brac{\partial_\beta W_c^{-1}\hat{\varphi}_c}\\
	    &= \frac 12 \det\!\lp \underline{R}\rp \brac{R^{-1}\sqrt{-g_s}}\brac{R^{-1}g_s^{\mu\nu}}\tensor{R}{_\mu^\alpha}\tensor{R}{_\nu^\beta}\brac{\partial_\alpha W_c^{-1}\hat{\varphi}_c}\brac{\partial_\beta W_c^{-1}\hat{\varphi}_c}\,.
	\end{aligned}\label{eq::SG::LagrangianWilsonlines1}
	\end{equation}
To go to the last line, we used the identities 
\eqref{appC:productrule}, \eqref{appC:detidentity}.
The square brackets indicate that the derivative acts on all terms inside the bracket. $\det\!\lp \underline{R}\rp$ denotes the determinant of the matrix $\tensor{[\underline{R}]}{^\mu_\alpha}\equiv \tensor{R}{^\mu_\alpha}$, and we integrated by parts in the last line, dropping the boundary term.
	Next, we add and subtract
	\begin{equation}
	    \frac 12 \sqrt{-\hat{g_s}}\hat{g}_s^{\mu\nu}\brac{\partial_\mu W_c^{-1}\hat{\varphi}_c}\brac{\partial_\nu W_c^{-1}\hat{\varphi}_c}\,,
	\end{equation}
	where the hatted background metric is given in \eqref{eq::SG::ResidualMetric1} -- \eqref{eq::SG::ResidualMetric3}, and obtain
	\begin{equation}
	\begin{aligned}
	    &\frac 12 \brac{\partial_\alpha W_c^{-1}\hat{\varphi}_c}
	    \brac{\partial_\beta W_c^{-1}\hat{\varphi}_c}\biggl(\sqrt{-\hat{g}_s}\hat{g}_s^{\alpha\beta}\\
	    &\quad
	    + \det\!\lp \underline{R}\rp  \brac{R^{-1}\sqrt{-g_s}} \tensor{R}{_\mu^\alpha}\tensor{R}{_\nu^\beta}\brac{R^{-1}g_s^{\mu\nu}(x)} - \sqrt{-\hat{g}_s}\hat{g}_s^{\alpha\beta}\biggr)\,.
	    \label{eq::SG::Lagrangiancalc1}
	\end{aligned}
	\end{equation}
The term containing the inverse metric $g_s^{\mu\nu}(x)$
	can be rewritten as
	\begin{align}
	    \tensor{R}{_\mu^\alpha}\tensor{R}{_\nu^\beta}\brac{R^{-1}g_s^{\mu\nu}(x)} &= \eta^{\alpha\beta} - \eta^{\alpha\rho}\eta^{\beta\sigma}\left(\tensor{R}{^\mu_\rho}\tensor{R}{^\nu_\sigma}\brac{R^{-1}s_{\mu\nu}(x) + \eta_{\mu\nu}} - \eta_{\rho\sigma}\right) + \dots\,,\label{eq::SG::Lagrangiancalc2}
	\end{align}
 	where the expression in the round brackets is just the soft background metric in fixed-line coordinates, $\check{g}_{s\mu\nu}$.
	We can rewrite this round bracket in \eqref{eq::SG::Lagrangiancalc2} as
	\begin{align}
	    \tensor{R}{^\mu_\rho}\tensor{R}{^\nu_\sigma}\brac{R^{-1}s_{\mu\nu}(x) + \eta_{\mu\nu}} = \hat{g}_{s\rho\sigma} + \mathfrak{g}_{s\rho\sigma}\,,
	\end{align}
 	introducing the manifestly soft-covariant metric tensor $\mathfrak{g}_{s\mu\nu}$ \eqref{eq::SG::BGMetricSplitRiemann} as the part that is expressed entirely in terms of Riemann tensor. Up to $\mathcal{O}(\lambda^2)$, it is given by 
    \begin{align}
        \mathfrak{g}_{s\mu\nu}(x) &= 
        -\frac{n_{+\mu}n_{+\nu}}{4} x_\perp^\alpha x_\perp^\beta R_{\alpha-\beta-}
        - \frac{n_{+\mu}}{2}\frac{2}{3}x_\perp^\alpha x_\perp^\beta R_{\alpha\nu_\perp\beta-} - \frac{n_{+\nu}}{2}\frac{2}{3}x_\perp^\alpha x_\perp^\beta R_{\alpha\mu_\perp\beta-}\nn \\
        &\quad
        - \lp \frac{n_{+\mu}n_{-\nu}}{4}
        + \frac{n_{+\nu}n_{-\mu}}{4}\rp \frac{2}{3}x_\perp^\alpha x_\perp^\beta R_{\alpha+\beta-}
        - \frac 13 x_\perp^\alpha x_\perp^\beta R_{\alpha\mu_\perp\beta\nu_\perp}
\label{eq::SG::Riemanntermsexpansion}\\
        &\quad- \frac{n_{-\mu}}{2} \frac{1}{3}x_\perp^\alpha x_\perp^\beta R_{\alpha\nu_\perp\beta+}
        - \frac{n_{-\nu}}{2}\frac{1}{3}x_\perp^\alpha x_\perp^\beta R_{\alpha\mu_\perp\beta+}
        - \frac{n_{-\mu}n_{-\nu}}{4} \frac{1}{3}x_\perp^\alpha x_\perp^\beta R_{\alpha+\beta+} + \mathcal{O}(\lambda^3)\,.\nn
    \end{align}
Due to the contraction with subleading collinear 
derivatives, only the $\mathfrak{g}_{s--}$ component 
will contribute to the effective Lagrangian at 
$\mathcal{O}(\lambda^2)$, while all others enter at higher orders in $\lambda$.
	Similarly, the determinant term in \eqref{eq::SG::Lagrangiancalc1},
	\begin{equation}\label{eq::DeterminantinFLNC}
	    \det\!\lp \underline{R}\rp  \brac{R^{-1}\sqrt{-g_s}}\,,
	\end{equation}
	is the soft determinant in fixed-line normal coordinates, i.e.~$\sqrt{-\check{g}_{s\mu\nu}}$, and we can write it as
	\begin{equation}
	     \det\!\lp \underline{R}\rp R^{-1}\sqrt{-g_s} = \sqrt{-\hat{g}_s}\,(1 - \frac16 x_\perp^\alpha x_\perp^\beta \tensor{R}{^\mu_{\alpha\mu\beta}} + \dots)\,,
	\end{equation}
	i.e.~as $\sqrt{-\hat{g_s}}$ times some Riemann-tensor traces. However, these Riemann terms do not contribute to the Lagrangian at $\mathcal{O}(\lambda^2)$.
	
	In summary, we find up to $\mathcal{O}(\lambda^2)$ for the leading term in the Lagrangian \eqref{eq::SG::LagrangianWilsonlines1}
	\begin{align}
	    \mathcal{L}_{\varphi} &= \frac 12 \brac{\partial_\alpha W_c^{-1}\hat{\varphi}_c}
	    \brac{\partial_\beta W_c^{-1}\hat{\varphi}_c}\left(\sqrt{-\hat{g}_s}\hat{g}_s^{\alpha\beta}
	     -\frac{1}{4} \sqrt{-\hat{g}_s} \np^\alpha \np^\beta x_\perp^\mu x_\perp^\nu R_{\mu-\nu-}\right)\nn\\
	     &\quad- \sqrt{-\hat{g}_s}\frac{\lambda_\varphi}{4!}\lc W_c^{-1}\hat{\varphi_c}\rc^4
	     + \mathcal{O}(\lambda^3)\,.\label{eq::SG::LagrangianResult1}
	\end{align}
	By introducing the gauge-invariant scalar field $\hat{\chi}_c = W^{-1}_c\hat{\varphi}_c$, we can write this as
	\begin{align}
	    \mathcal{L}_{\varphi} &= \frac 12 \brac{\partial_\alpha \hat{\chi}_c}
	    \brac{\partial_\beta \hat{\chi}_c}\left(\sqrt{-\hat{g}_s}\hat{g}_s^{\alpha\beta}
	     -\frac{1}{4} \sqrt{-\hat{g}_s} \np^\alpha \np^\beta x_\perp^\mu x_\perp^\nu R_{\mu-\nu-}\right)
	     -\sqrt{-\hat{g}_s}\frac{\lambda_\varphi}{4!}\hat{\chi}_c^4\,.
	\end{align}
	For the subleading Lagrangians \eqref{eq::SG::BFLagrangian1} and \eqref{eq::SG::BFLagrangian2}, we find \eqref{eq::app::L1LCgaugeresult} and \eqref{eq::app::L2LCgaugeresult}, and adding the self-interaction, we have
	\begin{align}
	\mathcal{L}_{\varphi h} &= \frac 12 \sqrt{-\hat{g}_s} \left(-\hat{g}_s^{\mu\alpha}\hat{g}_s^{\nu\beta}\hat{\mathfrak{h}}_{\alpha\beta} + \frac 12 \hat{g}_s^{\alpha\beta}\hat{\mathfrak{h}}_{\alpha\beta}\hat{g}_s^{\mu\nu}\right)\partial_\mu \hat{\chi}_c\partial_\nu \hat{\chi}_c
	-\sqrt{-\hat{g}_s} \frac 12 \hat{g}_s^{\alpha\beta}\hat{\mathfrak{h}}_{\alpha\beta}\frac{\lambda_\varphi}{4!}\hat{\chi}_c^4
	\,,\\[-0.4cm]
	\mathcal{L}_{\varphi hh}&= \frac 12 \sqrt{-\hat{g}_s}\left(\hat{g}_s^{\mu\alpha}\hat{g}_s^{\nu\beta}\hat{g}_s^{\rho\sigma}\hat{\mathfrak{h}}_{\alpha\rho}\hat{\mathfrak{h}}_{\beta\sigma} - \frac 12 \hat{g}_s^{\alpha\beta}\hat{\mathfrak{h}}_{\alpha\beta} \hat{g}_s^{\mu\rho}\hat{g}_s^{\nu\sigma}\hat{\mathfrak{h}}_{\rho\sigma} + \frac 18\hat{g}_s^{\mu\nu} (\hat{g}_s^{\alpha\beta}\hat{\mathfrak{h}}_{\alpha\beta})^2\right.\nn\\
	&\quad\phantom{\sqrt{-g_s}} \left.- \frac 14 \hat{g}_s^{\mu\nu}\hat{g}_s^{\rho\alpha}\hat{g}_s^{\sigma\beta} \hat{\mathfrak{h}}_{\rho\sigma} \hat{\mathfrak{h}}_{\alpha\beta}\right) \partial_\mu \hat{\chi}_c\partial_\nu \hat{\chi}_c\nn\\
	&\quad
	-\sqrt{-\hat{g}_s} \lp \frac{1}{8}(\hat{g}_s^{\alpha\beta}\hat{\mathfrak{h}}_{\alpha\beta})^2 - \frac 14 \hat{g}_s^{\mu\alpha}\hat{g}_s^{\nu\beta}\hat{\mathfrak{h}}_{\mu\nu}\hat{\mathfrak{h}}_{\alpha\beta}\rp \frac{\lambda_\varphi}{4!}\hat{\chi}_c^4\,.
	\end{align}
We introduced the manifestly gauge-invariant collinear graviton field $\mathfrak{\hat h}_{\mu\nu}$ via
	\begin{equation}
	    \hat{\mathfrak{h}}_{\mu\nu}(x) =  \tensor{W}{^{\rho}_\mu}\tensor{W}{^{\sigma}_\nu} \brac{W_c^{-1} (\hat{g}_{s\rho\sigma}(x) + \hat{h}_{\rho\sigma}(x))} - \hat{g}_{s\mu\nu}(x)\,.
	\end{equation}
Expressing the gauge-invariant building blocks $\hat{\chi}_c$ and $\mathfrak{\hat h}_{\mu\nu}$ in terms of $\hat{\varphi}_c$ and $\hat{h}_{\mu\nu}$ and the collinear Wilson line $W_c$, the Wilson lines cancel in the terms that do not contain the soft Riemann tensor, i.e. no $\mathfrak{g}_{s\mu\nu}$, as explained around \eqref{eq::app::Wilsoncancellation}. We can thus write the Lagrangian in terms of the covariant but non-invariant fields as
\begin{align}\label{eq::SG::FinalDressedLagrangian}
	\mathcal{L}_{\varphi} &= \frac 12 \sqrt{-\hat{g}_s} \hat{g}_s^{\mu\nu}\partial_\mu \hat{\varphi}_c \partial_\nu \hat{\varphi}_c
	-\frac{1}{8}\sqrt{-\hat{g_s}} x_\perp^\alpha x_\perp^\beta R_{\alpha-\beta-}\partial_+\hat{\varphi_c}\partial_+\hat{\varphi_c}
	-\sqrt{-\hat{g}_s}\frac{\lambda_\varphi}{4!}\hat{\varphi}_c^4
	\,,\\
	\mathcal{L}_{\varphi h} &= \frac 12 \sqrt{-\hat{g}_s} \left(-\hat{g}_s^{\mu\alpha}\hat{g}_s^{\nu\beta}\hat{h}_{\alpha\beta} + \frac 12 \hat{g}_s^{\alpha\beta}\hat{h}_{\alpha\beta}\hat{g}_s^{\mu\nu}\right)\partial_\mu \hat{\varphi}_c\partial_\nu \hat{\varphi}_c\nn\\
	&\quad
	-\sqrt{-\hat{g}_s} \lp \frac 12 \hat{g}_s^{\alpha\beta}\hat{h}_{\alpha\beta}\rp \frac{\lambda_\varphi}{4!}\hat{\varphi}_c^4
	\,,\\
	\mathcal{L}_{\varphi h h} &= \frac 12 \sqrt{-\hat{g}_s}\left(\hat{g}_s^{\mu\alpha}\hat{g}_s^{\nu\beta}\hat{g}_s^{\rho\sigma}\hat{h}_{\alpha\rho}\hat{h}_{\beta\sigma} - \frac 12 \hat{g}_s^{\alpha\beta}\hat{h}_{\alpha\beta} \hat{g}_s^{\mu\rho}\hat{g}_s^{\nu\sigma}\hat{h}_{\rho\sigma} + \frac 18\hat{g}_s^{\mu\nu} (\hat{g}_s^{\alpha\beta}\hat{h}_{\alpha\beta})^2\right.\nn\\
	&\quad\phantom{\sqrt{-g_s}} \left.- \frac 14 \hat{g}_s^{\mu\nu}\hat{g}_s^{\rho\alpha}\hat{g}_s^{\sigma\beta} \hat{h}_{\rho\sigma} \hat{h}_{\alpha\beta}\right) \partial_\mu \hat{\varphi}_c\partial_\nu \hat{\varphi}_c\nn\\
	&\quad-\sqrt{-\hat{g}_s}\lp \frac 18 (\hat{g}_s^{\alpha\beta}\hat{h}_{\alpha\beta})^2 - \frac 14 \hat{g}_s^{\mu\alpha}\hat{g}_s^{\nu\beta}\hat{h}_{\mu\nu}\hat{h}_{\alpha\beta}\rp\frac{\lambda_\varphi}{4!} \hat{\varphi}_c^4\label{eq::SG::FinalDressedLagrangian2}
	\,.
	\end{align}
Next, we introduce the soft-covariant derivative \eqref{eq::SG::softcovariantderivative} and specify that indices are raised with the background metric $\hat{g}_{s\mu\nu}$.
	Then, the Lagrangians \eqref{eq::SG::FinalDressedLagrangian} -- \eqref{eq::SG::FinalDressedLagrangian2} take the form
	\begin{equation}
	    \mathcal{L} = \sqrt{-\hat{g}_s} \lp\mathcal{L}^{(0)}_{D_s} + \mathcal{L}^{(1)}_{D_s} + \mathcal{L}^{(2)}_{D_s}\rp\,,
	\end{equation}
where the superscript indicates the leading $\lambda$-counting of each term. Up to $\mathcal{O}(\lambda^2)$, we find
		\begin{align}\label{eq::SG::LagrangianCovDevL0}
	    \mathcal{L}^{(0)}_{D_s} &= \frac 12 \partial_+\hat{\varphi}_c D_-\hat{\varphi}_c + \frac 12 \partial_{\alpha_\perp}\hat{\varphi}_c\partial^{\alpha_\perp}\hat{\varphi}_c
	    -\frac{\lambda_\varphi}{4!}\hat{\varphi}_c^4
	    \,,\\
	    \mathcal{L}^{(1)}_{D_s} &= -\frac 12 \hat{h}^{\mu\nu}\partial_{\mu}\hat{\varphi}_c\partial_{\nu}\hat{\varphi}_c
	    + \frac 14 \tensor{\hat{h}}{^{\beta_\perp}_{\beta_\perp}}\left(\partial_+\hat{\varphi}_c D_-\hat{\varphi}_c + \partial_{\alpha_\perp}\hat{\varphi}_c\partial^{\alpha_\perp}\hat{\varphi}_c\right)
	    -\frac 12 \tensor{\hat{h}}{^{\alpha_\perp}_{\alpha_\perp}}\frac{\lambda_\varphi}{4!}\hat{\varphi}_c^4
	    \,,\\
	    \mathcal{L}^{(2)}_{D_s} &=
	    - \frac 18 x_\perp^\alpha x_\perp^\beta R_{\alpha-\beta-}(\partial_+ W_c^{-1}\hat{\varphi}_c)^2
	    +\frac 12
	    \hat{h}^{\mu\alpha}\tensor{\hat{h}}{_{\alpha}^{\nu}}\partial_{\mu}\hat{\varphi}_c\partial_{\nu}\hat{\varphi}_c
	    -\frac 14 \tensor{\hat{h}}{^{\alpha_\perp}_{\alpha_\perp}} \hat{h}^{\mu\nu}\partial_\mu\hat{\varphi}_c\partial_\nu\hat{\varphi}_c
	    \nn\\
	    &\quad
	    + \frac{1}{16}\left((\tensor{\hat{h}}{^{\alpha_\perp}_{\alpha_\perp}})^2 - 2 \hat{h}^{\alpha\beta}\hat{h}_{\alpha\beta}\right)
	    \left(\partial_+\hat{\varphi}_c D_-\hat{\varphi}_c + \partial_{\mu_\perp}\hat{\varphi}_c\partial^{\mu_\perp}\hat{\varphi}_c\right)\nn\\
	    &\quad
	    -\lp \frac{1}{8}(\tensor{\hat{h}}{^{\alpha_\perp}_{\alpha_\perp}})^2 - \frac 14 \hat{h}^{\mu\nu}\hat{h}_{\mu\nu}\rp \frac{\lambda_\varphi}{4!}\hat{\varphi}_c^4
	    \,.\label{eq::SG::LagrangianCovDevL2}
	\end{align}
These expressions consistently contain all interactions of 
the scalar field with soft and collinear gravitons up to 
sub-subleading power $\mathcal{O}(\lambda^2)$, plus an 
infinite tower of higher-order terms in the covariant 
but inhomogeneous objects $W_c$ and $D_-=\nm D$. 
Note further that the collinear graviton field with upper 
indices is defined as
\begin{equation}
	    \hat{h}^{\mu\nu} \equiv \hat{g}^{\mu\alpha}_s \hat{g}^{\nu\beta}_s \hat{h}_{\alpha\beta}\,,
	\end{equation}
and that the soft background field also has a 
$\lambda$-expansion.

    For the conceptual understanding, the result \eqref{eq::SG::LagrangianCovDevL0} -- \eqref{eq::SG::LagrangianCovDevL2} is of the most intriguing form.
    In these expressions, one clearly recognises the strong formal similarity of soft-collinear gravity and soft-collinear gauge theory. One can identify a (non-linear) soft-covariant derivative and manifestly gauge-covariant terms that are expressed via the Riemann tensor, which describe the quadrupole and higher multipole interactions -- just as gauge theory, where the dipole and higher multipole interactions are expressed via the field-strength tensor.
    In addition, in this form one immediately sees why there are no collinear divergences in gravity, as there is no collinear covariant derivative and thus no leading-power interactions. The soft divergences are present, as is a soft-covariant derivative, and one can find a soft-decoupling transformation \cite{Beneke:2012xa} by a soft light-like Wilson 
line that converts $\nm D_s \to \nm\partial$ (at leading power in $\lambda$), just as in gauge theory.
    
However, for applications, like the computation of amplitudes, it is convenient to strictly expand the Lagrangian and all appearing contractions and non-linear objects in $\lambda$, to have interaction terms with definite $\lambda$ power-counting.
Reintroducing the gravitational coupling $\kappa$ by substituting $s_{\mu\nu}$ and $h_{\mu\nu}$ with $\kappa s_{\mu\nu}$ and $\kappa h_{\mu\nu}$, respectively, and fixing collinear light-cone gauge $\hat{h}_{\mu+}=0$, the result reads
	\begin{align}\label{eq::SG::SoftCollLagrangianExpanded}
	    \mathcal{L}^{(0)} &= \frac 12 \partial_+\hat{\varphi}_c \partial_-\hat{\varphi}_c + \frac 12 \partial_{\alpha_\perp}\hat{\varphi}_c\partial^{\alpha_\perp}\hat{\varphi}_c - \frac \kappa8 s_{--}(\partial_+\hat{\varphi}_c)^2
	    -\frac{\lambda_\varphi}{4!}\hat{\varphi}_c^4\,,\\
	    \mathcal{L}^{(1)} &= -\frac \kappa8 \lc\partial_\alpha s_{--} - \partial_- s_{\alpha-}\rc x_\perp^\alpha(\partial_+\hat{\varphi}_c)^2 - \frac \kappa4 s_{\mu_\perp-}\partial^{\mu_\perp}\hat{\varphi}_c\partial_+\hat{\varphi}_c\
\nn\\
	    &\quad-\frac \kappa2 \left(\hat{h}^{\mu_\perp\nu_\perp}\partial_{\mu_\perp}\hat{\varphi}_c\partial_{\nu_\perp}\hat{\varphi}_c
	    + \hat{h}^{\mu_\perp-}\partial_{\mu_\perp}\hat{\varphi}_c\partial_+\hat{\varphi}_c
	    + \frac 14 \hat{h}^{--}(\partial_+\hat{\varphi}_c)^2\right)\nonumber\\
	    &\quad
	    + \frac \kappa4 \tensor{\hat{h}}{^{\alpha_\perp}_{\alpha_\perp}}\left(\partial_+\hat{\varphi}_c \partial_-\hat{\varphi}_c - \frac \kappa4 s_{--}(\partial_+\hat{\varphi}_c)^2 + \partial_{\alpha_\perp}\hat{\varphi}_c\partial^{\alpha_\perp}\hat{\varphi}_c\right)
	    -\frac \kappa 2 \hat{h}^{\alpha_\perp}_{\alpha_\perp} \frac{\lambda_\varphi}{4!}\hat{\varphi}_c^4\,,
\\
	    \mathcal{L}^{(2)} &= -\frac{\kappa}{16} \lc\partial_{[+} s_{-]-}\rc n_-x (\partial_+\hat{\varphi}_c)^2
	    - \frac \kappa4 \lc \partial_{[\alpha_\perp} s_{\mu_\perp]-}\rc x_\perp^\alpha \partial^{\mu_\perp}\hat{\varphi}_c\partial_+\hat{\varphi}_c
	    \nonumber\\
	    &\quad+ \frac{\kappa^2}{32}s_{--}s_{+-}(\partial_+\hat{\varphi}_c)^2 
	    + \frac{\kappa^2}{32}s_{-\alpha_\perp}s^{\alpha_\perp}_- (\partial_+\hat{\varphi}_c)^2
	    - \frac 18 x_\perp^\alpha x_\perp^\beta R_{\alpha-\beta-}(\partial_+\hat{\varphi}_c)^2\nonumber\\
	    &\quad
	    + \frac{\kappa}{8} s_{+-}\partial_{\alpha_\perp}\hat{\varphi}_c\partial^{\alpha_\perp}\hat{\varphi}_c -\frac \kappa4 s_{+-}\frac{\lambda_\varphi}{4!}\hat{\varphi}_c^4
	    \,,\nn\\
	    &\quad+ \frac{\kappa^2}{2} \left(
	    \hat{h}^{\mu_\perp\alpha_\perp}\hat{h}_{\alpha_\perp}^{\nu_\perp}\partial_{\mu_\perp}\hat{\varphi}_c\partial_{\nu_\perp}\hat{\varphi}_c + \hat{h}^{\mu_\perp\alpha_\perp}\hat{h}_{\alpha_\perp-}\partial_{\mu_\perp}\hat{\varphi}_c\partial_+\hat{\varphi}_c + \frac 14 \hat{h}^{-\alpha_\perp}\hat{h}_{\alpha_\perp-}(\partial_+\hat{\varphi}_c)^2\right)\nonumber\\
	    &\quad-\frac{\kappa^2}{4} \tensor{\hat{h}}{^{\alpha_\perp}_{\alpha_\perp}} \lp
	    \hat{h}^{\mu_\perp\nu_\perp}\partial_{\mu_\perp}\hat{\varphi}_c\partial_{\nu_\perp}\hat{\varphi}_c 
	    + \hat{h}^{\mu_\perp-}\partial_{\mu_\perp}\hat{\varphi}_c\partial_+\hat{\varphi}_c
	    + \frac 14 \hat{h}_{--}\partial_+\hat{\varphi}_c\partial_+\hat{\varphi}_c\rp 
	    \nn\\
	    &\quad+ \frac{\kappa^2}{16}\left((\hat{h}^{\alpha_\perp}_{\alpha_\perp})^2 - 2 \hat{h}^{\alpha_\perp\beta_\perp}\hat{h}_{\alpha_\perp\beta_\perp}\right)
	    \left(\partial_+\hat{\varphi}_c\partial_-\hat{\varphi}_c - \frac \kappa4 s_{--}(\partial_+\hat{\varphi}_c)^2 + \partial_{\mu_\perp}\hat{\varphi}_c\partial^{\mu_\perp}\hat{\varphi}_c\right)\nonumber\\
	    &\quad+ \frac{\kappa^2}{4} \hat{h}^{\mu_\perp\alpha_\perp}s_{\alpha_\perp-}\partial_+\hat{\varphi}_c\partial_{\mu_\perp}\hat{\varphi}_c + \frac{\kappa^2}{8} \hat{h}^{-\alpha_\perp}s_{\alpha_\perp-}(\partial_+\hat{\varphi}_c)^2\nonumber\\
	    &\quad - \frac {\kappa^2}{8} \hat{h}^{\alpha_\perp}_{\alpha_\perp} s_{\mu_\perp-} \partial_+\hat{\varphi}_c \partial^{\mu_\perp}\hat{\varphi}_c - \frac{\kappa^2}{16} \hat{h}^{\alpha_\perp}_{\alpha_\perp}\lc \partial_{[\mu_\perp} s_{-]-}\rc x_\perp^\mu (\partial_+\hat{\varphi}_c)^2\nn\\
	    &\quad
	    -\kappa^2\frac{\lambda_\varphi}{4!}\hat{\varphi_c}^4 \lp 
	    (\hat{h}^{\alpha_\perp}_{\alpha_\perp})^2 - \frac 14 \hat{h}^{\alpha_\perp\beta_\perp}\hat{h}_{\alpha_\perp\beta_\perp}\rp
	    \,,\label{eq::SG::SoftCollLagrangianExpanded2}\\
	    \mathcal{L}^{(1)}_{\varphi_s} &= -\frac{\lambda_\varphi}{3!} \hat{\varphi}_c^3\varphi_s\,,\\
	    \mathcal{L}^{(2)}_{\varphi_s} &= 
	    \frac \kappa 4 \hat{h}^{\alpha_\perp}_{\alpha_\perp} \partial_+\hat{\varphi}_c \partial_-\varphi_s
	    -\frac{\lambda_\varphi}{4}\hat{\varphi_c}^2\varphi_s^2
	    -\frac{\lambda_\varphi}{3!}\frac{\kappa}{2}\hat{h}^{\alpha_\perp}_{\alpha_\perp}\hat{\varphi}_c^3\varphi_s\,.\label{eq::SG::SoftCollLag3}
	\end{align}
We observe from \eqref{eq::SG::SoftCollLagrangianExpanded2}, \eqref{eq::SG::SoftCollLag3} that unlike for the purely collinear Lagrangian the $\lambda$ and the $\kappa$ expansions are not identical for the full soft-collinear theory. 

    Performing the same steps for the Einstein-Hilbert action, one similarly derives a Lagrangian that is manifestly covariant with respect to $\hat{g}_{s\mu\nu}$. Since here we are only interested in the result up to $\mathcal{O}(\lambda)$, this derivation is a lot simpler, as the Riemann terms start contributing only at $\mathcal{O}(\lambda^2)$.
    For the bilinear collinear Lagrangian $\mathcal{L}_{hh}$ in \eqref{eq::SG::BilinearEH}, we can simply write it with respect to the residual background metric $\hat{g}_{s\mu\nu}$. The result then reads
    \begin{align}
    \mathcal{L}_{hh} &= \sqrt{-\hat{g}_s}\,\biggl(
    \frac 12 \nabla_\mu \hat{\mathfrak{h}}_{\alpha\beta} \nabla^\mu \hat{\mathfrak{h}}^{\alpha\beta} 
    - \frac 12 \nabla_\mu \hat{\mathfrak{h}} \nabla^\mu \hat{\mathfrak{h}}
    + \nabla_\alpha \hat{\mathfrak{h}}^{\alpha\beta}\nabla_\beta \hat{\mathfrak{h}}
    - \nabla_\alpha\hat{\mathfrak{h}}^{\alpha\beta}\nabla_\mu \hat{\mathfrak{h}}^\mu_\beta \nn\\
    &\quad - 4 R_{\alpha\beta} \hat{\mathfrak{h}}^{\alpha\mu} \hat{\mathfrak{h}}^\beta_\mu
    + 2 R_{\alpha\beta\mu\nu} \hat{\mathfrak{h}}^{\alpha\mu}\hat{\mathfrak{h}}^{\beta\nu} 
    + R_{\alpha\beta}\hat{\mathfrak{h}} \hat{\mathfrak{h}}^{\alpha\beta}
    - \frac{1}{4}R\,(\hat{\mathfrak{h}}^2 - 2 \hat{\mathfrak{h}}^{\alpha\beta}\hat{\mathfrak{h}}_{\alpha\beta})
    \biggr)\,.
\end{align}
Here, the soft background field $\hat{g}_{s\mu\nu}$ is used to raise and lower indices, and the covariant derivative $\nabla_\mu$ is defined with respect to $\hat{g}_{s\mu\nu}$, as are the Ricci and Riemann tensors. 
Note that the last three terms in fact contribute only at $\mathcal{O}(\lambda^4)$, but appear already at $\mathcal{O}(h^2)$ in this covariant construction.
To simplify further, we drop these last three terms, and note that the leading term in the soft Christoffel symbol $\chris{\mu}{\alpha\beta}\sim\lambda^4$, so we can ignore its contribution in the soft-covariant derivative. We also explicitly introduce the soft metric $\hat{g}_{s\mu\nu}$ when contracting derivatives.
The Lagrangian then reads
\begin{equation}
       \mathcal{L}_{hh} = \sqrt{-\hat{g}_s}\biggl(
    \frac 12 \hat{g}_{s}^{\mu\nu} \partial_\mu \hat{\mathfrak{h}}_{\alpha\beta} \partial_\nu \hat{\mathfrak{h}}^{\alpha\beta} 
    - \frac 12 \hat{g}_s^{\mu\nu}\partial_\mu \hat{\mathfrak{h}} \partial_\nu \hat{\mathfrak{h}}
    + \partial_\alpha \hat{\mathfrak{h}}^{\alpha\beta}\partial_\beta \hat{\mathfrak{h}}
    - \partial_\alpha\hat{\mathfrak{h}}^{\alpha\beta}\partial_\mu \hat{\mathfrak{h}}^\mu_\beta\biggr) + \mathcal{O}(\lambda^3)\,.
\end{equation}
Finally, we can introduce the (scalar) soft-covariant derivative $\nm D_s$ \eqref{eq::SG::softcovariantderivative} and obtain
\begin{align}
       \mathcal{L}_{hh} &= \sqrt{-\hat{g}_s}\biggl(
    \frac 12 \partial_+ \hat{\mathfrak{h}}_{\alpha\beta} D_- \hat{\mathfrak{h}}^{\alpha\beta} 
    +\frac 12 \partial_{\mu_\perp} \hat{\mathfrak{h}}_{\alpha\beta} \partial^{\mu_\perp} \hat{\mathfrak{h}}^{\alpha\beta} 
    - \frac 12 \partial_+ \hat{\mathfrak{h}} D_- \hat{\mathfrak{h}}
    - \frac 12 \partial_{\mu_\perp} \hat{\mathfrak{h}} \partial^{\mu_\perp} \hat{\mathfrak{h}}\nn\\
    &\quad
    + \partial_\alpha \hat{\mathfrak{h}}^{\alpha\beta}\partial_\beta \hat{\mathfrak{h}}
    - \partial_\alpha\hat{\mathfrak{h}}^{\alpha\beta}\partial_\mu \hat{\mathfrak{h}}^\mu_\beta\biggr) + \mathcal{O}(\lambda^3)\,.
\end{align}
For the trilinear Lagrangian $\mathcal{L}_{hhh}$, we take the full-theory result \eqref{eq::GR::TrilinearLagrangian} and replace all $\partial_-$ by the covariant $D_-$.
It is not necessary to compute the fully-covariant expansion with respect to an arbitrary background, as the purely collinear theory is already of $\mathcal{O}(\lambda)$. Hence we only need to incorporate the soft-covariant derivative.
The expanded Lagrangians homogeneous in $\lambda$ and in collinear light-cone gauge are then given by
    \begin{align}
        \mathcal{L}^{(0)}_{\rm EH} &= 
        \frac 12 \partial_\mu \hat{h}_{\alpha_\perp\beta_\perp} \partial^\mu \hat{h}^{\alpha_\perp\beta_\perp}
        - \frac 12 \partial_\mu \hat{h} \partial^\mu \hat{h} \nn\\
        &\quad
        + \lp 
        \partial_{\alpha_\perp} \hat{h}^{\alpha_\perp\beta_\perp}\partial_{\beta_\perp} \hat{h}
        +  \frac 12 \partial_{\alpha_\perp} \hat{h}^{\alpha_\perp-}\partial_{+} \hat{h}
        + \frac 12 \partial_{+} \hat{h}^{-\beta_\perp}\partial_{\beta_\perp} \hat{h}
        + \frac 14 \partial_{+} \hat{h}_{--}\partial_{+} \hat{h}\rp\nn\\
        &\quad
        - \lp 
        \partial_{\alpha_\perp} \hat{h}^{\alpha_\perp\mu_\perp} \partial^{\beta_\perp} \hat{h}_{\beta_\perp\mu_\perp}
        + \partial_+ \hat{h}^{-\mu_\perp} \partial^{\beta_\perp} \hat{h}_{\beta_\perp\mu_\perp}
        + \frac 14  \partial_+ \hat{h}^{-\mu_\perp} \partial_+ \hat{h}_{-\mu_\perp}
        \rp
        \nn\\
        &\quad - \frac{\kappa}{8}s_{--}\partial_+ \hat{h}_{\alpha\beta} \partial_+ \hat{h}^{\alpha\beta} + \frac{\kappa}{8}s_{--}\partial_+ \hat{h}\partial_+ \hat{h}\,,
\\[0.2cm]
        \mathcal{L}^{(1)}_{\rm EH} &= 
    		    \frac{\kappa}{4}\hat{h}^{\alpha\beta}(-2\partial_\mu \hat{h}^{\mu}_\alpha\partial_\beta \hat{h}
	            + 4 \partial_\mu \hat{h}_{\alpha\nu}\partial_\beta \hat{h}^{\mu\nu}
            	+ 2 \partial_\nu \hat{h}^{\mu\nu}\partial_\mu \hat{h}_{\alpha\beta}
            	- 2 \partial_\alpha \hat{h}^{\mu}_\beta\partial_\mu \hat{h}\nn\\
            	&\qquad
            	+ \partial^\mu \hat{h}_{\alpha\nu}\partial^\nu \hat{h}_{\beta\mu}
            	- 2 \partial^\mu \hat{h}_{\beta\nu}\partial_\mu \hat{h}_{\alpha}^\nu)\nn\\
            	&\quad + \frac{\kappa}{8}\partial_\alpha (\hat{h}^2) \partial^\alpha \hat{h}
            	+ \frac{\kappa}{4}\partial_\alpha(\hat{h}^{\mu\nu}\hat{h}_{\mu\nu})\partial^\alpha \hat{h}
            	- \frac{\kappa}{8}\partial_\alpha(\hat{h}^2)\partial_\beta \hat{h}^{\alpha\beta}
            	- \frac{3\kappa}{4}\partial_\alpha(\hat{h}^{\mu\nu}\hat{h}_{\mu\nu})\partial_\beta \hat{h}^{\alpha\beta}\nn\\
            	&\quad + \frac{\kappa}{2}\partial_\alpha(\hat{h}^{\mu\rho}\hat{h}_\rho^\nu)\partial^\alpha \hat{h}_{\mu\nu}
            	- \partial_\alpha(\hat{h}^{\mu\rho}\hat{h}_\rho^\nu)\partial_\mu \hat{h}^\alpha_\nu
            	+ \frac{\kappa}{2}\partial_\mu(\hat{h}^{\mu\rho}\hat{h}_{\rho}^\nu)\partial_\nu \hat{h} 
            	- \frac{\kappa}{8}\hat{h}\partial_\alpha \hat{h} \partial^\alpha \hat{h}\nn\\
            	&\quad - \frac{\kappa}{8}\hat{h}\partial_\alpha \hat{h}_{\mu\nu}\partial^\alpha \hat{h}^{\mu\nu}
            	+ \frac{\kappa}{4} \hat{h} \partial_\mu \hat{h}_{\alpha\beta}\partial^\alpha \hat{h}^{\mu\beta}
            	-\frac{\kappa}{4} s_{-\mu_\perp}\partial^{\mu_\perp} \hat{h}^{\alpha\beta} \partial_+ \hat{h}_{\alpha\beta}
            	+ \frac{\kappa}{4}s_{-\mu_\perp} \partial^{\mu_\perp}\hat{h}\partial_+ \hat{h}\nn\\
            	&\quad
            	-\frac{\kappa}{8}\lc\partial_\alpha s_{--} - \partial_- s_{\alpha-}\rc x_\perp^\alpha \partial_+ \hat{h}_{\mu\nu} \partial_+ \hat{h}^{\mu\nu} 
            	+ \frac{\kappa}{8}\lc\partial_\alpha s_{--} - \partial_- s_{\alpha-}\rc x_\perp^\alpha \partial_+ \hat{h}\partial_+\hat{h}\nn\\
            	&\quad
            	-\frac{\kappa^2}{64}s_{--} \partial_+(\hat{h}^2)\partial_+ \hat{h}
            	- \frac{\kappa^2}{32} s_{--}\partial_+ (\hat{h}^{\mu\nu} \hat{h}_{\mu\nu})\partial_+\hat{h}
            	- \frac{\kappa^2}{16} s_{--} \partial_+ (\hat{h}^{\mu\rho}\hat{h}^\nu_\rho)\partial_+ \hat{h}_{\mu\nu}\nn\\
            	&\quad
            	+ \frac{\kappa^2}{64}s_{--}\hat{h}\partial_+ \hat{h}\partial_+ \hat{h}
            	+ \frac{\kappa^2}{64}s_{--} \hat{h}\partial_+ \hat{h}_{\mu\nu} 
            	\partial_+ \hat{h}^{\mu\nu} 
            	\,.
    \end{align}
In $\mathcal{L}^{(1)}_{\rm EH}$ we kept the four-vector contractions and did not introduce the decomposition into $+,-,\perp$ components to keep the result more compact.

\subsection{Operator basis}

In this section, we discuss the minimal operator basis for the hard process that produces collinear and soft scalar particles or gravitons, following closely the outline presented in \cref{sec::SQCD::OperatorBasis}.
Again, the structure of this basis is heavily constrained by soft and collinear gauge-covariance.

The generic $N$-jet operator takes the form \eqref{eq::SQCD::Njet} as in QCD, 
\begin{equation}
    \mathcal{J}(0) = \int [dt]_N\: C(t_{i_1},t_{i_2},\dots) J_s(0) \prod_{i=1}^N J_i(t_{i_1},t_{i_2},\dots)\,,
\end{equation}
where 
$[dt]_N = \prod_{i_k} dt_{i_k}$.
Here, $J_i$ are collinear gauge-invariant operators, built from the matter and metric fields, which transform covariantly under soft gauge transformations, and $J_s$ are gauge-covariant soft building blocks.
Notably, since this current is evaluated at $x=0$, it transforms under the soft gauge symmetry \eqref{eq::SG::infinitesimalresidualtransformation} only with $\varepsilon_\mu(0)$ and $\omega_{\mu\nu}(0)$.

Sitting at $x=0$, the operator $\mathcal{J}(0)$ is not translationally invariant, and hence, not soft gauge-invariant.
However, we can translate this operator to the coordinate $x$ and then integrate over space-time. In this way, we obtain the manifestly translationally (hence also gauge) invariant operator
\begin{equation}\label{eq::SG::TranslatedCurrent}
    \mathcal{J} = \int d^4x \,T_x \mathcal{J}(0) T_x^{-1}\,.
\end{equation}
Once a scattering matrix element of $J$ is evaluated, the translation $T_x = e^{ix\hat{p}}$ together with the space-time integral turns into the momentum-conserving $\delta$-function.
Hence, in any practical computation, we can work with $\mathcal{J}(0)$ and impose momentum conservation by hand, just as one imposes colour conservation in QCD.

Note that the current \eqref{eq::SG::TranslatedCurrent} does not need a $\sqrt{-g}$ to render the measure invariant.
The reason for this is that ultimately we perform an expansion around Minkowski space, i.e.~we describe a spin-2 fluctuation on top of Minkowski space-time with metric $\eta_{\mu\nu}$.
The $d^4x$ integration is related to this underlying space-time, and a non-trivial metric determinant would only appear if one quantised around a curved space-time, e.g.~a de Sitter background.
We can also see this from effective gauge invariance.
The collinear parts are already manifestly gauge-invariant, and soft transformations are evaluated at $x=0$, i.e.~for the operator $\mathcal{J}(0)$, they correspond to global transformations.
The integration over Minkowski space-time then corresponds to hard momentum (and angular momentum) conservation, and this is enough to render the current \eqref{eq::SG::TranslatedCurrent} manifestly soft gauge-invariant.
If the current were put at $x_0$ instead of $x=0$, the multipole expansion would have to be performed about $x_0$, and 
$x_-$ replaced by $x_0 + x_-$. The soft expansion is 
hence the same irrespective of the position of the hard scattering vertex.

Let us now proceed to constrain the possible operators that can appear in $J_i$ and $J_s$.
As the operators must be collinear gauge-invariant, we make use of the Wilson line $W_c$ and construct the collinear gauge-invariant fields $\hat{\chi}_c$ and $\hat{\mathfrak{h}}_{\mu\nu}$ as
\begin{equation}
    \hat{\chi}_c = \lc W^{-1}_c\hat{\varphi}_c\rc \,,\quad \hat{\mathfrak{h}}_{\mu\nu} = \tensor{W}{^\alpha_\mu}\tensor{W}{^\beta_\nu} \lc W^{-1}_c \hat{h}_{\alpha\beta}\rc + \lp \tensor{W}{^\alpha_\mu}\tensor{W}{^\beta_\nu}\lc W^{-1}_c\hat{g}_{s\alpha\beta}\rc - \hat{g}_{s\mu\nu}\rp\,,
\end{equation}
which correspond to the fields in collinear light-cone gauge.
In particular, they satisfy $\hat{\mathfrak{h}}_{\mu+}=0$, hence eliminate the dangerous power-enhanced $h_{++}\sim \mathcal{O}(\lambda^{-1})$ and the unsuppressed $h_{\perp+}\sim\mathcal{O}(\lambda^0)$ components of the collinear graviton field, which are gauge-artefacts.
These components can only appear inside the Wilson line $W_c$.
Note also that we can use the graviton equations of motion to eliminate $\hat{\mathfrak{h}}_{--}$ and $\hat{\mathfrak{h}}_{\perp-}$ in favour of the physical transverse component $\hat{\mathfrak{h}}_{\perp\perp} \sim \mathcal{O}(\lambda)$.
Hence, the elementary collinear building blocks are restricted to
\begin{equation}
    J_i^{A0}(t_i) \in \left\{ \hat{\chi}_{c_i}(t_i\nip)\,,\hat{\mathfrak{h}}_{i\perp\perp}(t_i\nip)\right\}\,.
\end{equation}
Note the similarity to the QCD building blocks \eqref{eq::SQCD::BuildingBlocks}. We only need to trade the collinear gluon $\hat{\mathcal{A}}_{c\perp}$ for the collinear graviton components $\hat{\mathfrak{h}}_{\perp\perp}$.
Just as in QCD, we can now construct subleading operators in the same two ways, by (i) adding $\partial_\perp$ derivatives, and (ii) using multiple building blocks of the same direction. Also as in QCD, we can use the collinear matter and graviton equations of motion to eliminate the covariant $\nm D_s$ derivative in favour of the collinear building blocks.
Hence, the soft-covariant derivative is not a necessary building block.
This immediately implies that the first purely soft building block that can contribute to $J_s$ is the Riemann tensor, which is quadratic in derivatives and thus counts as $\mathcal{O}(\lambda^6)$, i.e.~it is suppressed by three orders in the soft expansion. These soft building blocks in the hard sources cause process-dependence of soft emission. Hence, up to 
order $\mathcal{O}(\lambda^6)$, soft emission is entirely 
determined by the soft-collinear interaction Lagrangian 
constructed in this paper, and hence universal.
An immediate consequence  \cite{Beneke:2021umj} 
of this statement is the existence 
of a next-to-next-to-soft term in the gravitational 
soft theorem, which was found explicitly in 
\cite{Cachazo:2014fwa,Bern:2014vva,Schwab:2014xua,Broedel:2014fsa,Zlotnikov:2014sva}.


\section{Conclusion}

In this paper, we completed the construction of soft-collinear gravity. The main motivation for this analysis has been understanding the structure of the collinear and soft limits, and its comparison to the already well-understood case of gauge theory.  Our main result is that despite the apparently very different leading-power collinear physics, SCET for gauge theory and gravity is much more similar than might have been expected:
\begin{itemize}
\item After multipole expansion, the soft-collinear Lagrangian is invariant under two separate, collinear and soft gauge symmetries, called emergent, since they differ from the original symmetry of the full theory. The soft gauge symmmetry lives on the light-cone of the collinear fields. The collinear gauge symmmetry is defined in the background of two soft gauge 
fields on the light cone, $e_{-\alpha}(x_-)$ and $\lc\Omega_-\rc_{\alpha\beta}(x_-)$.
\item The soft-collinear Lagrangian is expressed to all orders in the $\lambda$-expansion in terms of collinear gauge-invariant collinear matter and metric fields, which reduce to the elementary fields in light-cone gauge. The soft interactions are expressed directly in terms of the Riemann tensor, and the covariant derivative with respect to the above gauge symmetries, which appears only in the leading-power Lagrangian.
\item The source operators representing hard scattering are expressed in terms of a few gauge-invariant building blocks, which scale with positive powers of $\lambda$, allowing for a well-defined power expansion.
\end{itemize}
An important difference between gravity and gauge theory 
results from the fact that in gauge theory, the charges, which source the interactions, are unrelated to the kinematic expansion in the soft and collinear limits. As a consequence, the gauge-invariant objects are all-order quantities in the gauge coupling $g$, but can be constructed to be strictly homogeneous in the SCET expansion parameter $\lambda$. On the contrary, the charge in gravity is momentum, which itself has a non-trivial scaling in $\lambda$, hence the gauge-invariant objects are unavoidably all-order quantities in $\lambda$, and inhomogeneous. Nevertheless, the construction achieves the goal of a controlled and systematic expansion in $\lambda$, since all the gauge-invariant building blocks scale with 
positive powers of $\lambda$ and control the unsuppressed fields to all orders. In other words, the unavoidable inhomogeneity is constrained completely to all orders through the inhomogeneous gauge symmetries. Calculations to a desired order can therefore be performed by expanding all objects in $\lambda$ to the required order.

With this in mind, the rules for constructing and working with soft-collinear gravity are remarkably similar to QCD. We illustrated this with a new derivation of the gravitational soft theorem up to sub-subleading order in \cite{Beneke:2021umj}, which also shows that the existence of a universal next-to-next-to-soft term in gravity 
\cite{Cachazo:2014fwa,Bern:2014vva,Schwab:2014xua,Broedel:2014fsa,Zlotnikov:2014sva} is a consequence 
of the fact that the soft Riemann tensor is second order 
in derivatives (rather than first order as is the field 
strength in gauge theory), such that soft source operators, 
which violate universality are suppressed by three 
(rather than two) powers 
in the soft expansion. Besides this, SCET gravity now 
allows for the systematic study of collinear graviton 
physics, which arises first at subleading power.
 
\vspace*{0.5em}
\noindent
\subsubsection*{Acknowledgement}
This work was supported by the Excellence Cluster ORIGINS
funded by the Deutsche Forschungsgemeinschaft (DFG, German Research Foundation)
under Ger\-many's Excellence Strategy - EXC-2094 - 390783311.
RS is supported by the United States Department of Energy under Grant Contract DE-SC0012704.


\appendix 

\section{Fixed-point gauge}
\label{sec::App::FixedPointGauge}

When performing a multipole expansion of the gauge field, we can express all subleading terms in a manifestly gauge-invariant fashion via the field-strength tensor by employing fixed-line gauge.
Fixed-point or Fock-Schwinger gauge is obtained by requiring 
\begin{equation}\label{eq::app::FSGauge}
    x^\mu A_\mu(x) = 0\,.
\end{equation}
This gauge can always be attained. To see this, we have to find the gauge transformation
\begin{equation}
    A_\mu \to A_\mu^\prime = U A_\mu U^\dagger - \frac{i}{g} \partial_\mu U U^\dagger
\end{equation}
such that $x^\mu A_\mu^\prime(x) = 0$. 
This gives the relation
\begin{equation}
    x^\mu U(x) A_\mu U^\dagger(x) = \frac{i}{g}x^\mu \partial_\mu U(x) U^\dagger(x)\,,
\end{equation}
i.e. we have
\begin{equation}
    x^\mu \partial_\mu U = -ig x^\mu U A_\mu\,,
\end{equation}
which can be solved by integrating along radial lines 
in $x^\mu$.
We solve for $V \equiv U^\dagger$ with initial condition 
$V(0)=1$ and find
\begin{align}\label{eq::App::FPWilsonLine}
    V(x) = P\exp\lp ig\int_0^1 ds \,x^\mu A_\mu(sx)\rp\,,
\end{align}
where $P$ denotes path-ordering. This solution is the 
fixed-point gauge analogue of the $R$-Wilson line. 
Under a gauge transformation, 
\begin{equation}
    V(x) \to U(x) V(x) U^\dagger(0)\,.
\end{equation}
Hence, a generic field configuration $A_\mu(x)$ can be moved to fixed-point gauge according to
\begin{equation}
    \tilde{A}_\mu(x) = V^\dagger A_\mu(x) V + \frac{i}{g} V^\dagger \partial_\mu V\,.
\end{equation}
The field $\tilde{A}_\mu(x)$ transforms only with the 
global transformation $U(0)$.
This gauge allows us to express the gauge field in terms of the field-strength tensor as
\begin{equation}\label{eq::APP::FixedPointExpansion}
    \tilde A_\nu(x) = \int_0^1 ds\: s x^\mu F_{\mu\nu}(sx)\,.
\end{equation}
To see this, write (now dropping the tilde)
\begin{align}
    A_\mu(x) &= \int_0^1 ds\: \frac{d}{ds} \lp s A_\mu(sx)\rp= \int_0^1 ds\: \lp A_\mu(sx) + sx^\nu \frac{\partial}{\partial (sx)^\nu} A_\mu(sx)\rp\nn\\
    &= \int_0^1 ds\: \lp (A_\mu(sx) - sx^\nu F_{\mu\nu}(sx) + sx^\nu \frac{\partial}{\partial (sx)^\mu}A_\nu(sx) + ig sx^\nu \lc A_\mu\,,A_\nu\rc(sx)\rp\,.
\end{align}
The gauge condition
$x^\mu A_\mu(x) = sx^\mu A_\mu(sx) = 0$ 
eliminates the commutator term.
Next, observe that after integration by parts,
\begin{equation}
    sx^\nu \frac{\partial}{\partial (sx)^\mu} A_\nu (sx) = \frac{\partial}{\partial(s x)^\mu} (sx^\nu A_\nu(sx)) - A_\mu(sx) = -A_\mu(sx)\,,
\end{equation}
which cancels with the first term.
Hence we are left with \eqref{eq::APP::FixedPointExpansion}.
By continuity of $A_\mu(x)$, the gauge condition \eqref{eq::app::FSGauge} implies $A_\mu(0)=0$, so there is no $x$-independent term in \eqref{eq::APP::FixedPointExpansion}.

This gauge condition arises quite naturally in the context of multipole expansion. Consider some (hard) matter field $\psi(x)$, which transforms under the gauge symmetry as
\begin{equation}
    \psi(x) \to U(x)\psi(x)\,,
\end{equation} 
coupled to a (soft) vector potential $A_\mu(x)$ with the standard transformation
\begin{equation}
    A_\mu(x) \to U(x)A_\mu(x)U^\dagger(x) + \frac{i}{g}U(x) \lc\partial_\mu U^\dagger\rc(x) \,.
\end{equation}
For example, let $\psi$ be a charged fermion with Lagrangian
\begin{equation}\label{eq::app:FPExampleLagrangian}
    \mathcal{L} = \bar{\psi}i\gamma^\mu \partial_\mu \psi + g \bar\psi \gamma^\mu A_\mu(x)\psi\,.
\end{equation}
In the situation that the vector potential 
describes a slowly varying background field 
compared to the variation of $\psi(x)$ with $x$, one can 
perform the multipole expansion
\begin{equation}\label{eq::app::GaugeFieldExpansion}
    A_\mu(x) = A_\mu(0) + x^\alpha\lc\partial_\alpha A_\mu\rc\!(0) + \mathcal{O}(x^2)\,.
\end{equation}
The Lagrangian up to first order in $x$ is
\begin{align}
    \mathcal{L}^{(0)} &= \bar{\psi}i\gamma^\mu \partial_\mu \psi + g \bar\psi \gamma^\mu A_\mu\psi\,,\\
    \mathcal{L}^{(1)} &= g \bar\psi \gamma^\mu x^\alpha\lc\partial_\alpha A_\mu\rc\psi\,,
\end{align}
where $A_\mu = A_\mu(0)$ and derivatives of $A_\mu$ are taken before setting $x=0$.
The gauge transformation of $\psi$ now mixes different orders of $x$, since the parameter of the transformation still depends on $x$, hence
\begin{equation}
    \psi \to U(0)\psi + x^\alpha\lc\partial_\alpha U\rc\!(0)\psi + \mathcal{O}(x^2)\,.
\end{equation}
We must homogenise the transformation of $\psi$ so that it respects the multipole expansion.
In this case, the homogenised transformation is the global transformation $\psi \to U(0)\psi$.
The fermion field that transforms with $U(0)$ is given by
\begin{equation}
    \hat{\psi}(x) = V^\dagger(x) \psi(x)\,.
\end{equation}
Expressing the Lagrangian in terms of $\hat{\psi}$ and $V(x)$, we find
\begin{align}
    \mathcal{L} &= \bar{\hat\psi} i\gamma^\mu \partial_\mu \hat\psi + g\bar{\hat\psi}\lp V \gamma^\mu A_\mu(x) V^\dagger + \frac{i}{g} V\lc\partial_\mu V^\dagger\rc\rp\hat\psi\,,
\end{align}
and note that $\hat{\psi}$ couples to the ``dressed'' vector field $\tilde{A}_\mu(x)$. Inserting \eqref{eq::APP::FixedPointExpansion} and writing the Lagrangian (dropping the hats of $\psi$) as
\begin{equation}\label{eq::app::LagrangianFPgauge}
    \mathcal{L} = \bar{\psi} i\gamma^\mu \partial_\mu \psi + g \bar{\psi}\gamma^\nu \lp\int_0^1 ds\: s x^\mu F_{\mu\nu}(sx) \rp \psi\,,
\end{equation}
we see that the interactions to all orders in $x$ can be written in a closed form in terms of only the manifestly gauge-covariant field-strength tensor $F_{\mu\nu}$.
Multipole-expanding this integral in $x$, we find
\begin{equation}
    \int_0^1 ds\: s x^\mu F_{\mu\nu}(sx) = \frac 12 x^\mu F_{\mu\nu} + \frac 13 x^\mu x^\alpha \lc \partial_\alpha F_{\mu\nu}\rc + \mathcal{O}(x^3)\,,
\end{equation}
where the field-strength tensor on the right-hand side is evaluated at $x=0$.

In conclusion, fixed-point gauge arises quite naturally when homogenising the gauge symmetry with respect to the multipole expansion of the gauge field. The  interactions are the expressed in terms of the field-strength tensor and its derivatives, and the vector potential no longer appears.

\section{Fixed-line gauge}
\label{sec::App::FixedLineGauge}

We review the fixed-line gauge introduced in \cite{Beneke:2002ni}.
In SCET, the power-counting in $\lambda$ enforces a multipole expansion around the light ray $x_-^\mu = \np x \,\frac{\nm^\mu}{2}$ rather than the point $x^\mu=0$ as for the standard multipole expansion, i.e.~there is a left-over dependence on $x_-$ in all soft fields. We wish to express the gauge field in terms of the manifestly gauge-invariant field-strength tensor, but fixed-point gauge is now too constraining.
The more general and less constraining fixed-line gauge is defined by the condition
\begin{equation}\label{eq::app::FLgauge}
    (x-x_-)^\mu A_{s\mu}(x) = 0\,.
\end{equation}
Following the discussion in \cref{sec::App::FixedPointGauge}, the Wilson line corresponding to \eqref{eq::App::FPWilsonLine} is now given by
\begin{equation}\label{eq::app::FLWilsonLine}
    R(x) = P\exp\lp ig \int_0^1 ds\: (x-x_-)^\mu A_{s\mu}\lp x + s(x-x_-)\rp\rp\,.
\end{equation}
The first major difference to fixed-point gauge is that the condition \eqref{eq::app::FLgauge} does not constrain $\nm A_s(x_-)$, hence there exists a residual gauge symmetry, consistent with the multipole expansion about $x_-$. 
For $\np A_s$ and $A_{s\perp}$, the gauge condition is equivalent to fixed-point gauge, and we find the same expressions, following \eqref{eq::APP::FixedPointExpansion},
\begin{align}\label{eq::SQCD::FixedLineA+Identity}
    \np A_s(x) &= \int_0^1 ds\: s(x-x_-)^\mu n_+^\nu F_{s\mu\nu}(y(s))\,,\\
    A_{s\nu_\perp}(x) &= \int_0^1 ds\: s(x-x_-)^\mu F_{s\mu\nu_\perp}(y(s))\,,
\end{align}
where $y(s) = x_- + s(x-x_-)$. Similar to fixed-point gauge, $\np A_{s}(x_-) = A_{s\perp}(x_-) = 0$.
The analogous formula for $\nm A_s$ is, however, given by
\begin{equation}\label{eq::SQCD::FixedLineA-Identity}
    \nm A_s(x) - \nm A_s(x_-) = \int_0^1 ds\: (x-x_-)^\mu \nm^\nu F_{s\mu\nu}(y(s))\,.
\end{equation}
This allows us to rewrite $\nm A_s(x)$ in terms of the homogeneous soft background field $\nm A_s(x_-)$, as well as subleading terms proportional to the field-strength tensor.
The crucial difference to fixed-point gauge is that now there is a residual background field $\nm A_s(x_-)$ associated with the residual, unfixed gauge symmetry described by $U_s(x_-)$.
This is consistent with the multipole expansion, which now allows for dependence on $x^\mu_-$.
We use these fixed-line gauge identities to expand the gauge-covariant light-front multipole expansion of the soft fields in the soft-collinear interaction Lagrangian.

Let us go back to the example of a charged fermion with Lagrangian \eqref{eq::app:FPExampleLagrangian} and generalise the discussion from \eqref{eq::app::GaugeFieldExpansion} to \eqref{eq::app::LagrangianFPgauge} to the expansion about $x_-$.
Performing the multipole expansion about $x^\mu_-$, we obtain the generalisation of \eqref{eq::app::GaugeFieldExpansion},
\begin{equation}
    A_{s\mu}(x) = A_{s\mu}(x_-) + x_\perp^\alpha\lc\partial_\alpha A_{s\mu}\rc(x_-) + \mathcal{O}(\lambda^2 A_{s\mu})\,, 
\end{equation}
where now the counting in $\lambda$ is relevant, and not the power in $x$.
In this case, the soft gauge transformation mixes different orders in $\lambda$, and the Wilson line $R$ defined in \eqref{eq::app::FLWilsonLine} appears in the redefinition of collinear fields. In light-cone gauge for the collinear fields, 
the new fields are defined as
\begin{equation}\label{eq::app::FLRedef}
    \hat{\phi}_c(x) = R^\dagger(x) \phi_c(x)\,,
\end{equation}
and transform homogeneously with $U_s(x_-)$,
\begin{equation}
    \hat{\phi}_c(x) \to U_s(x_-)\hat{\phi}_c(x)\,.
\end{equation}
As an example, we insert the redefinition \eqref{eq::app::FLRedef} in the leading term
\begin{align}
\bar{\phi}_c\frac{\slashed n_+}{2}\nm D_s\phi_c &= \bar{\hat{\phi_c}}\frac{\slashed n_+}{2} R^\dagger \nm D_s R\hat{\phi_c} \nn\\
     &= \bar{\hat{\phi_c}}\frac{\slashed n_+}{2}\lp \nm\partial + g\lp R^\dagger \nm A_s R + \frac{i}{g} R \lc\nm \partial R^\dagger\rc \rp\rp \hat{\phi_c}
     \,,
\end{align}
where all fields are evaluated at $x$.
To make use of the fixed-line gauge identities \eqref{eq::SQCD::FixedLineA+Identity}-\eqref{eq::SQCD::FixedLineA-Identity}, we add and subtract the homogeneous background field $\nm A_s(x_-)$, to find 
\begin{equation}
    \bar{\phi}_c\frac{\slashed n_+}{2}\nm D_s\phi_c = \bar{\hat{\phi_c}}\frac{\slashed n_+}{2}\lp \nm D_s + \big[R^\dagger i\nm D_s(x) R - i\nm D_s\big]\rp \hat{\phi_c}\,.
\end{equation}
The identities \eqref{eq::SQCD::FixedLineA+Identity}-\eqref{eq::SQCD::FixedLineA-Identity} in fixed-line gauge can be 
promoted to the general case by un-doing the transformation to fixed-line gauge with the gauge transformation $U=R^\dagger$, resulting in \cite{Beneke:2002ni}
\begin{align}\label{eq::app::FLA-R}
    R^\dagger i\nm D_s(x) R - i\nm D_s &= \int_0^1 ds\: (x-x_-)^\mu \nm^\nu R^\dagger(y(s)) g F_{s\mu\nu}(y(s))R(y(s))\,,\\
    R^\dagger iD_{s\nu_\perp}(x)R - i\partial_{\nu_\perp} &= \int_0^1 ds\: s(x-x_-)^\mu R^\dagger(y(s)) gF_{s\mu\nu_\perp}(y(s))R(y(s))\,,\\
    R^\dagger \np D_{s}(x) R - i\np \partial &= \int_0^1 ds\: s(x-x_-)^\mu \np^\nu R^\dagger(y(s))gF_{s\mu\nu}(y(s))R(y(s))\,,\label{eq::app::FLA+R}
\end{align}
They give closed all-order expressions for the soft fields. Once expanded in $\lambda$, they generate an infinite tower of subleading terms. The Lagrangian expressed in terms of the hatted collinear fields contains the homogeneous soft background field $\nm A_s(x_-)$ in the covariant derivative $\nm D_s(x_-)$, as well as subleading interactions, expressed entirely in terms of $F_{s\mu\nu}$.
Once the integrals are expanded in $\lambda$, these interactions will depend on $F_{s\mu\nu}$ and its (covariant) derivatives at $x_-$.

\section{Useful gauge transformation identities}
\label{sec::app::UsefulIdentities}

The gauge transformation $U(x)$ introduced in \eqref{eq:scalargaugetrafofull} and the Jacobians $\tensor{U}{_\mu^\alpha}$ \eqref{eq::GR::JacobianDef} derived from it satisfy a list of useful identities.
Naming and definition of the operations stems from the translation $x \to x + \varepsilon(x)$, where in the following $\varepsilon(x) \equiv \varepsilon$ and $U(x) \equiv U$, suppressing the argument. We have
\begin{itemize}
    \item the gauge transformation $U$ \eqref{eq:scalargaugetrafofull},
    \begin{equation}
        U = 1 - \varepsilon^\alpha\partial_\alpha + \frac 12 \varepsilon^\alpha\varepsilon^\beta\partial_\alpha\partial_\beta + \varepsilon^\alpha\partial_\alpha\varepsilon^\beta\partial_\beta + \mathcal{O}(\varepsilon^3)\,,
    \end{equation}
    \item its inverse $U^{-1}$,
    \begin{equation}
        U^{-1} = 1 + \varepsilon^\alpha\partial_\alpha + \frac 12 \varepsilon^\alpha\varepsilon^\beta\partial_\alpha\partial_\beta + \mathcal{O}(\varepsilon^3)\,,
    \end{equation}
    \item the Jacobian matrix $\tensor{U}{^\mu_\alpha}$,
    \begin{equation}
        \tensor{U}{^\mu_\alpha} = \delta^\mu_\alpha + \partial_\alpha\varepsilon^\mu\,,
    \end{equation}
    \item the inverse Jacobian matrix $\tensor{U}{_\alpha^\mu}$,
    \begin{equation}
        \tensor{U}{_\alpha^\mu} = \delta_\alpha^\mu - \partial_\alpha\varepsilon^\mu + \partial_\alpha\varepsilon^\beta\partial_\beta\varepsilon^\mu + \mathcal{O}(\varepsilon^3)\,,
    \end{equation}
    \item the Jacobian determinant $\det\!\lp \underline{U}\rp$ of the Jacobian $\tensor{\lc\underline{U}\rc}{^\mu_\alpha} = \tensor{U}{^\mu_\alpha}$,
    \begin{equation}
        \det\!\lp \underline{U}\rp = 1 + \partial_\alpha\varepsilon^\alpha + \frac 12 \partial_\alpha\varepsilon^\alpha\partial_\beta\varepsilon^\beta - \frac 12 \partial_\alpha\varepsilon^\beta\partial_\beta\varepsilon^\alpha + \mathcal{O}(\varepsilon^3)\,,
    \end{equation}
    \item the inverse Jacobian determinant $\det\!\lp \underline{U}^{-1}\rp$,
    \begin{equation}
        \det\!\lp \underline{U}^{-1}\rp = 1 - \partial_\alpha\varepsilon^\alpha + \frac 12 \partial_\alpha\varepsilon^\alpha\partial_\beta\varepsilon^\beta + \frac 12 \partial_\alpha\varepsilon^\beta\partial_\beta\varepsilon^\alpha + \mathcal{O}(\varepsilon^3)\,.
    \end{equation}
\end{itemize}

These objects satisfy a number of useful identities, that we employ in the following.
\begin{itemize}
    \item They are inverse with respect to each other, i.e.
    \begin{align}
        UU^{-1} &= 1\,, \quad \tensor{U}{^\mu_\alpha}\tensor{U}{_\mu^\nu} = \delta_\alpha^\nu\,,\quad 
        \det\!\lp \underline{U}\rp\det\!\lp\underline{U}^{-1}\rp = 1\,.
    \end{align}
    \item We can move translation and inverse translation past derivatives,
    \begin{align}
        \brac{\partial_\mu U\phi} &= U \tensor{U}{_\mu^\alpha}\partial_\alpha\phi\,,\\
        \brac{\partial_\mu U^{-1}\phi} &= \tensor{U}{^\alpha_\mu}U^{-1} \partial_\alpha\phi\,,
    \end{align}
    which is consistent with the gauge transformation of a covariant vector.
    \item There is a ``product rule'' for the translation operator
    \begin{align}
\label{appC:productrule}
        \brac{U\phi\psi} &= U\phi\psi U^{-1}
        = U\phi U^{-1}U\psi U^{-1}
        = \brac{U\phi}\brac{U\psi}\,,
    \end{align}
    and the same holds for $U^{-1}$.
    \item Scalar densities $\sqrt{-g}\phi$ are gauge-invariant up to total derivatives. These transform as
    \begin{align}
        \int d^4x\:\sqrt{-g}\phi &\to \int d^4x\:\sqrt{-g}\phi\,.\label{eq::app::invarianceIdentity}
    \end{align}
    With the gauge transformation of the metric determinant
    \begin{equation}
        \sqrt{-g} \to U\det\!\lp\underline{U}^{-1}\rp\sqrt{-g}\,,
    \end{equation}
    one can infer
    \begin{equation}
        \int d^4x\: U\det\!\lp \underline{U}^{-1}\rp\sqrt{-g}\phi= \int d^4x\: \sqrt{-g}\phi\,,\label{eq::app::UsefulFullTranslation}
    \end{equation}
    which is the active point of view of the invariance \eqref{eq::app::invarianceIdentity}.
    \item There is an integration by parts identity based on \eqref{eq::app::invarianceIdentity}. Moving the inverse translation $U(x)$ from one term to another via integration by parts, we generate $U^{-1}$ and a determinant, namely
    \begin{align}
\label{appC:detidentity}
        \phi\, U\psi &= \det\!\lp \underline{U}\rp \brac{U^{-1}\phi}\psi\,.
    \end{align}
    \end{itemize}

		\section{Geometric construction of light-cone gauge}\label{sec::app::GeometricConMain}
	
	        The second approach to constructing the gauge-invariant collinear fields is geometrically motivated, and follows \cite{Donnelly:2015hta}. This construction has also been used in \cite{Chakraborty:2019tem}.
            This construction is equivalent to fixing light-cone gauge, as shown in \ref{sec::app::equivalenceofParameters}.
    		One begins by parametrising a geodesic in a general coordinate system as
    		\begin{equation}
    		\bar{x}^\mu(s) = x^\mu + s n_+^\mu + v^\mu(s) \equiv y^\mu(s) + v^\mu(s)\,,
    		\end{equation}
    		where $v^\mu(s)$ can be viewed as the displacement of the general coordinate system from the light-cone gauge one, and we introduced $y^\mu(s) = x^\mu + s n_+^\mu$.
    		Then, one uses the geodesic equation to compute $v^\mu(s)$ iteratively in powers of $h_{\mu\nu}$, and performs an inverse translation by $v^\mu(0)$ of the point $x$. 
    		This, of course, only works if higher orders of $v^\mu$ are suppressed in $\lambda$.
    		As it turns out, $v^\mu$ starts at $\mathcal{O}(\lambda)$, and higher orders are indeed suppressed.
    		In this new coordinate system, the gauge transformation of the collinear fields at $x$ is then evaluated at $x-\infty n_+$, where it vanishes.
    		
    		\subsection{Derivation of the Wilson line}
    		\label{sec::GeometricConstruction}
    		
    		The geodesic equation is given by
    		\begin{equation}\label{eq::app::GeometricGeodesicEq1}
    		0 = \frac{\de^2 \bar{x}^\mu}{\de s^2} + \chris{\mu}{\alpha\beta}(\bar{x}(s))\frac{\de \bar{x}^\alpha}{\de s}\frac{\de \bar{x}^\beta}{\de s}\,,
    		\end{equation}
    		or, in terms of $v^\mu(s)$, as
    		\begin{align}\label{eq::app::GeometricGeodesicEq2}
    		\frac{\de^2}{\de s^2}v^{\mu}(s)
    		&= -\chris{\mu}{\alpha\beta}\left(y(s) + v(s)\right) \Bigl( n_+^\alpha + v^{\alpha}(s)\Bigr) \Bigl( n_+^\beta + v^{\beta}(s)\Bigr) + \dots\,.
    		\end{align}
    		We expand $v^\mu(s)$ as
    		\begin{equation}
    			v^\mu(s) = v^{(1)\mu}(s) + v^{(2)\mu}(s) + \dots\,,
    		\end{equation}
insert this in the geodesic equation, and obtain
    		\begin{align}\label{eq::app::GeometricGeodesicEq3}
    			\frac{\de^2}{\de s^2}\left( v^{(1)\mu}(s) + v^{(2)\mu}(s) + \dots\right)
    			&= -\chris{\mu}{\alpha\beta}(y(s) + v(s)) \Bigl( n_+^\alpha + v^{(1)\alpha}(s) + v^{(2)\alpha}(s)\Bigr)\nn\\
    			&\quad\times\Bigl( n_+^\beta + v^{(1)\beta}(s) + v^{(2)\beta}(s)\Bigr) + \dots\,.
    		\end{align}
    		At leading order, this yields
    		\begin{align}
    		\frac{\de^2}{\de s^2}v^{(1)\mu}(s) &= -\chris{\mu}{\alpha\beta}(y(s)) n_+^\alpha n_+^\beta\,.
    		\end{align}
    		Integrating this twice with boundaries at $-\infty$ and $s$ and the initial conditions $v^\mu(-\infty) = v^{\prime\,\mu}(-\infty)=0$, gives 
    		\begin{equation}\label{eq::app::Geometricvs}
    		v^{(1)\mu}(s) = -\int_{-\infty}^{s}\de s^\prime\: \int_{-\infty}^{s^\prime}\de s^{\prime\prime}\: \chris{\mu}{++}(y(s^{\prime\prime}))
    		= -\frac{1}{(n_+\partial)^2}\chris{\mu}{++}(y(s))\,.
    		\end{equation}
    		Note that this expression inhomogeneous in $\lambda$ and $h$, as we have not yet expanded $\chris{\mu}{++}$.
    		
    		To compute $v^{(2)\mu}(s)$, one reinserts $v^{(1)\mu}(s)$ to compute the next order of the geodesic equation \eqref{eq::app::GeometricGeodesicEq3}, which reads
    		\begin{align}
    			\frac{\de^2}{\de s^2}v^{(2)\mu}(s) &= 
    			-2\chris{\mu}{\alpha\beta}(y(s))\,n_+^\alpha \frac{d v^{(1)\beta}}{ds} - 
    			v^{(1)\rho}\brac{\partial_\rho \chris{\mu}{\alpha\beta}}\!(y(s))\,n_+^\alpha n_+^\beta\,,
    		\end{align}
    		where the second term on the right-hand side is due to the Taylor expansion of $\Gamma^\mu_{\alpha\beta}(y(s) + v(s))$.
    		Inserting $v^{(1)\mu}(s)$, we find
    		\begin{align}
    			\frac{\de^2}{\de s^2}v^{(2)\mu}(s) &=
    			-2\chris{\mu}{\alpha\beta}(y(s))n_+^\alpha \left( -\frac{1}{\delp}\chris{\beta}{++}(y(s))\right)
    			-\left(-\frac{1}{(\delp)^2}\chris{\rho}{++}\right)\brac{\partial_\rho \chris{\mu}{++}}(y(s))\nn\\
    			&= 2\chris{\mu}{\beta+}(y(s)) \left(\frac{1}{\delp}\chris{\beta}{++}(y(s))\right)
    			+ \left(\frac{1}{(\delp)^2}\chris{\rho}{++}\right)\brac{\partial_\rho \chris{\mu}{++}}(y(s))\,.
    		\end{align}
Integrating twice with the same initial conditions, we find
    		\begin{equation}
    			v^{(2)\mu}(0) = \frac{1}{(\delp)^2}\left(
    			2\chris{\mu}{\beta+} \frac{1}{\delp}\chris{\beta}{++}
    			+\brac{\partial_\rho \chris{\mu}{++}}\frac{1}{(\delp)^2}\chris{\rho}{++}\right)\,,
    		\end{equation}
    		where all Christoffel symbols are evaluated at $x$, and are then acted on by the inverse derivative.
    		
    		The gauge-invariant scalar and graviton field in covariant light-cone gauge can then be obtained by performing a translation by $v^\mu(0)$.
    		As an example, the gauge-invariant scalar field is given by
    		\begin{equation}
    			\chi(x) = T_{v(0)}\phi(x) = \phi(x) + v^\alpha(0)\partial_\alpha\phi(x) + \frac 12 v^{\alpha}(0)v^{\beta}(0)\partial_\alpha\partial_\beta\phi(x) + \dots\,,
    		\end{equation}
    		where we can insert the expansion for $v^\mu(0)$ and obtain an invariant field to the desired order in $\lambda$.
    		
    		While this construction seems somewhat orthogonal to the previous explicit gauge-fixing,
    		the quantities $\theta_c^\mu$ and $v^\mu(0)$ obtained from the two approaches, in fact, coincide, as verified in \ref{sec::app::equivalenceofParameters} below. 
The two approaches offer slightly different points of view on the construction.
    		The first one, explicitly fixing the gauge, makes transparent that one is working in light-cone gauge $h_{\mu+}=0$, and immediately shows the irrelevance of the power-enhanced and 
$\mathcal{O}(1)$ metric components $h_{++}$ and $h_{+\mu_\perp}$, 
respectively. The second approach clarifies why these constructions are invariant under diffeomorphisms, as the gauge transformation gets transported from $x$ to $-\infty$, where it vanishes.
    		This is very similar to what the Wilson line does in gauge theory.
    		
    \subsection{Equivalence of gauge-fixing and geometric approach}\label{sec::app::equivalenceofParameters}
    
    We show that $v^{\mu} \equiv v^{\mu}(0)$, as computed in \ref{sec::GeometricConstruction} and also in \cite{Chakraborty:2019tem} coincides with $\theta_c^\mu$ used to fix light-cone gauge. The geometrically derived parameter $v^\mu$ is given by
	\begin{align}\label{eq::app::v1}
	    v^{(1)\mu} &= -\frac{1}{\partial_+^2}\Gamma^\mu_{++}\,,\\
	    v^{(2)\mu} &= -\frac{1}{\partial_+^2}\left( 2 \Gamma^\mu_{\nu+} \frac{1}{\partial_+}\Gamma^\nu_{++} + 
	    \partial_\nu \Gamma^\mu_{++} \frac{1}{\partial_+^2}\Gamma^\nu_{++}\right)\,.\label{eq::app::v2}
	\end{align}
We compare this to the light-cone gauge parameter $\theta_c^\mu$ given by \eqref{eq::CG::colltheta}~--~\eqref{eq::CG::colltheta2}, 
		\begin{align}
		\theta^{(0)}_{c\mu} &= -\frac{1}{\delp}\left(h_{\mu+} - 
		\frac 12 \frac{\partial_\mu}{\delp}h_{++}\right)\,,\\
		\theta_{c+}^{(1)} &= -\frac 12 \frac{1}{\delp}\left( 
		2 \delp \theta_c^{(0)\alpha}h_{\alpha+} + 
		\theta_c^{(0)\alpha}\partial_\alpha h_{++} + 
		\delp \theta_c^{(0)\alpha}\delp \theta^{(0)}_{c\alpha}\right)\label{eq::app::Theta1+}\,,\\
		\theta^{(1)}_{c\mu} &= -\frac{1}{\delp}\biggl(
		\partial_\mu \theta_{c+}^{(1)} + 
		\partial_\mu \theta_c^{(0)\alpha}h_{\alpha+} + 
		\delp \theta_c^{(0)\alpha}h_{\alpha\mu} + 
		\theta_c^{(0)\alpha}\partial_\alpha h_{\mu+}\nonumber\\
		&\quad + 
		\partial_\mu \theta_c^{(0)\alpha}\delp \theta^{(0)}_{c\alpha}\biggr)\,.\label{eq::app::Theta1}
	\end{align}
First, one expands the Christoffel symbol to second order  in 
$h_{\mu\nu}$ as
	\begin{align}
	    \Gamma^{(1)\mu}_{\alpha\beta} &= -\frac 12 \left( \partial^\mu h_{\alpha\beta} - \partial_\alpha h_{\beta}^{\mu} - \partial_\beta h^{\mu}_\alpha\right)\,,\\
	    \Gamma^{(2)\mu}_{\alpha\beta} &= \frac 12 \left( h^{\mu\rho} \partial_\rho h_{\alpha\beta} - h^{\mu\rho}\partial_\alpha h_{\beta\rho} - h^{\mu\rho}\partial_\beta h_{\rho\alpha}\right)\,.
	\end{align}
With this, one finds  straightforwardly for $v^{(1)\mu}$ at $\mathcal{O}(1)$\footnote{$\mathcal{O}(1)$ here means the leading power of this object, e.g. $v_+\sim\lambda^{-1}$, $v_\perp\sim 1$ and $v_- \sim \lambda$ in this case. The next order is suppressed relative to the leading order by $\lambda$.}
	\begin{align}
	    \left.v^{(1)\mu}\right\rvert_{\mathcal{O}(1)} &= 
	    -\frac{1}{\partial_+^2}\lp -\frac 12 \partial^\mu h_{++} + \partial_+ h^\mu_+\rp
	    = \theta_c^{(0)\mu}\,.
	\end{align}
At NLO, we have to consider the non-linear terms $\Gamma^{(2)\mu}_{\alpha\beta}$ in $v^{(1)\mu}$ as well as the leading order of $v^{(2)\mu}$. For $v^{(1)\mu}$, we find
	\begin{align}
	    \left. v^{(1)\mu} \right\rvert_{\mathcal{O}(\lambda)} &= -\frac{1}{\partial_+^2}\lp 
	    -h^{\mu\nu}\partial_+ h_{\nu+} + \frac 12 h^{\mu\nu}\partial_\nu h_{++}\rp \nn\\
	    &= \frac{1}{\partial_+^2} \lp h^{\mu\nu} \partial_+^2 \lp \frac{1}{\partial_+}h_{\nu+} - \frac 12 \frac{\partial_\nu }{\partial_+^2}h_{++}\rp \rp 
	    = -\frac{1}{\partial_+^2} \lp h^{\mu\nu} \partial_+^2 \theta^{(0)}_{c\nu}\rp\,, 
	\end{align}
and for $v^{(2)\mu}$, 
	\begin{align}
	    \left.v^{(2)\mu}\right\rvert_{\mathcal{O}(\lambda)} &= \frac{1}{\partial_+^2}\left( 
	    \lp -\partial^\mu h_{\nu+} + \partial_\nu h^\mu_+ + \partial_+ h^\mu_\nu \rp 
	    \lp -\frac 12 \frac{\partial^\nu}{\partial_+} h_{++} + h^\nu_+\rp\right.\nn\\
	    &\quad\phantom{\frac{1}{\partial_+^2}\Bigl(}
	    \left.+ \frac 12 \partial_\nu \lp -\partial^\mu h_{++} + 2 \partial_+ h^\mu_+ \rp 
	    \frac 12 \lp -\frac{\partial^\nu}{\partial_+^2}h_{++} + 2 \frac{1}{\partial_+}h^\nu_+\rp \right)\nn\\
	    &= \frac{1}{\partial_+^2}\lp
	    \lp \partial^\mu h_{\nu+} - \partial_\nu h^\mu_+ - \partial_+ h^\mu_\nu \rp \partial_+ \theta_c^{(0)\nu}
	    + \frac 12 \partial_\nu\partial^\mu h_{++}  \theta_c^{(0)\nu} - \partial_\nu\partial_+ h^\mu_+\theta_c^{(0)\nu} \rp\,.
	\end{align}
Adding both contributions gives
	\begin{align}
	   \left. v^{(1)\mu} \right\rvert_{\mathcal{O}(\lambda)} + \left. v^{(2)\mu} \right\rvert_{\mathcal{O}(\lambda)}
	   &= \frac{1}{\partial_+^2}\Bigl( 
	   -h^{\mu\nu} \partial_+^2 \theta^{(0)}_{c\nu}
	   -\partial_+ h^\mu_\nu \partial_+\theta_c^{(0)\nu}
	   + \partial^\mu h_{\nu+} \partial_+\theta_c^{(0)\nu}\nn\\
	   &\phantom{\frac{1}{\partial_+^2}\Bigl[}
	   - \partial_\nu h^\mu_+ \partial_+\theta_c^{(0)\nu}
	   - \partial_\nu\partial_+h^\mu_+ \theta_c^{(0)\nu}
	   + \frac 12 \partial_\nu \partial^\mu h_{++}\theta_c^{(0)\nu}
	   \Bigr)\,.\label{eq::app::SumX}
	\end{align}
The first two terms of the first line, and the first two terms of the second line of \eqref{eq::app::SumX} combine into total derivatives as
	\begin{align}
	    -h^{\mu\nu} \partial_+^2 \theta^{(0)}_{c\nu} - \partial_+ h^\mu_\nu \partial_+\theta_c^{(0)\nu} &= -\partial_+ \lp 
	    h^{\mu\nu}\partial_+\theta^{(0)}_{c\nu}\rp\\
	    - \partial_\nu h^\mu_+ \partial_+\theta_c^{(0)\nu}
	    - \partial_\nu\partial_+h^\mu_+ \theta_c^{(0)\nu} &=
	    - \partial_+ \lp \partial_\nu h^{\mu_+}\theta_c^{(0)\nu}\rp\,,
	\end{align}
	and these already correspond to the third and fourth term of $\theta_c^{(1)\mu}$ \eqref{eq::app::Theta1}, as we find
	\begin{align}
	    \left. v^{(1)\mu} \right\rvert_{\mathcal{O}(\lambda)} + \left. v^{(2)\mu} \right\rvert_{\mathcal{O}(\lambda)} &\supset 
	    -\frac{1}{\partial_+}\lp h^{\mu\nu}\partial_+\theta^{(0)}_{c\nu} + \partial_\nu h^{\mu_+}\theta_c^{(0)\nu}\rp\,.
	\end{align}
The remaining two terms from \eqref{eq::app::SumX} can be rewritten as
\begin{align}
	    \partial^\mu h_{\nu+} \partial_+\theta_c^{(0)\nu}
	    &  + \frac 12 \partial_\nu \partial^\mu h_{++}\theta_c^{(0)\nu}
	    = \partial^\mu \lp h_{\nu+}\partial_+\theta_c^{(0)\nu}
	    + \frac 12 \partial_\nu h_{++}\theta_c^{(0)\nu} + \frac 12  \partial_+\theta_c^{(0)\nu} \partial_+ \theta^{(0)}_{c\nu}\rp\nn\\
	    &\hspace*{-1cm}- h_{\nu+}\partial^\mu \partial_+\theta_c^{(0)\nu} 
	    - \frac12 \partial_\nu h_{++} \partial^\mu \theta_c^{(0)\nu} - \lc\partial^\mu\partial_+ \theta^{(0)}_{c\nu}\rc \partial_+\theta_c^{(0)\nu}\,.\label{eq::app::EquivalenceCalc1}
	\end{align}
	where we added and subtracted the term $[\partial^\mu\partial_+ \theta^{(0)}_{c\nu}]\, \partial_+\theta_c^{(0)\nu}$. The first bracket, together with $\frac{1}{\partial_+}$, is just $\partial^\mu \theta^{(1)}_{c+}$ \eqref{eq::app::Theta1+}.
	For the remaining terms in \eqref{eq::app::EquivalenceCalc1}, note that the first term can be rewritten as
	\begin{align}
	    - h_{\nu+}\partial^\mu \partial_+ \theta_c^{(0)\nu} = -\partial_+\lp h_{\nu+}\partial^\mu \theta_c^{(0)\nu}\rp + \partial_+ h_{\nu+}\partial^\mu \theta_c^{(0)\nu}\,.
	\end{align}
In this expression, we keep the first term, and combine the second with the second-to-last term in \eqref{eq::app::EquivalenceCalc1} to
\begin{align}
	    \partial_+ \lp h_{\nu+} - \frac 12 \frac{\partial_\nu}{\partial_+}h_{++}\rp \partial^\mu\theta_c^{(0)\nu}
	    = -\partial_+^2 \theta^{(0)}_{c\nu} \partial^\mu\theta_c^{(0)\nu}\,.\label{eq::app::EquivalenceCalc2}
	\end{align}
	Finally, this term and the last one of \eqref{eq::app::EquivalenceCalc1} combine to yet another total derivative
	\begin{align}
	   -\partial_+^2 \theta^{(0)}_{c\nu} \partial^\mu\theta_c^{(0)\nu} - \lc\partial^\mu\partial_+ \theta^{(0)}_{c\nu}\rc \partial_+\theta_c^{(0)\nu} 
	   &= - \partial_+ \lp \partial_+\theta_c^{(0)\nu} \partial^\mu \theta^{(0)}_{c\nu}\rp\,.
	\end{align}
Finally, we add all terms to find for $v^{(1)} + v^{(2)}$ at $\mathcal{O}(\lambda)$ 
\begin{eqnarray}
\left. v^{(1)\mu} \right\rvert_{\mathcal{O}(\lambda)} + \left. v^{(2)\mu} \right\rvert_{\mathcal{O}(\lambda)} && =
\frac{1}{\partial_+^2}\biggl( 
	    -\partial_+ \lp 
	    h^{\mu\nu}\partial_+\theta^{(0)}_{c\nu}\rp
	    - \partial_+ \lp \partial_\nu h^{\mu_+}\theta_c^{(0)\nu}\rp \nn\\
	    &&\hspace*{-2cm}+\,\partial^\mu \lp h_{\nu+}\partial_+\theta_c^{(0)\nu} + \frac 12 \partial_\nu h_{++}\theta_c^{(0)\nu}  + \frac 12  \partial_+\theta_c^{(0)\nu} \partial_+ \theta^{(0)}_{c\nu}\rp\nn\\
	    &&\hspace*{-2cm}-\,\partial_+\lp h_{\nu+}\partial^\mu \theta_c^{(0)\nu}\rp
	    - \partial_+ \lp \partial_+\theta_c^{(0)\nu} \partial^\mu \theta^{(0)}_{c\nu}\rp\biggr)\nn\\
	    &&\hspace*{-2.5cm}= \,-\frac{1}{\partial_+}\biggl( 
	    -\frac{\partial^\mu}{\partial_+} \lp h_{\nu+}\partial_+\theta_c^{(0)\nu} + \frac 12 \partial_\nu h_{++}\theta_c^{(0)\nu}  + \frac 12  \partial_+\theta_c^{(0)\nu} \partial_+ \theta^{(0)}_{c\nu}\rp\nn\\
	    &&\hspace*{-2cm}
	    +\,h^{\mu\nu}\partial_+\theta^{(0)}_{c\nu} + \partial_\nu h^{\mu_+}\theta_c^{(0)\nu}
	    + \partial^\mu \theta_c^{(0)\nu} h_{\nu+}
	    + \partial_+\theta_c^{(0)\nu} \partial^\mu \theta^{(0)}_{c\nu}\biggr)\,.\label{eq::app::EquivalenceCalc3}
\end{eqnarray}
The first expression in the bracket is $\partial^\mu \theta_{c+}^{(1)}$, and we recover $\theta_c^{(1)\mu}$ given in \eqref{eq::app::Theta1}.
    
    \section{Computing the effective Lagrangian}\label{sec::InsertingExpressions}

We provide the complete derivation of the soft-collinear 
Lagrangian to $\mathcal{O}(\lambda^2)$.
In \cref{sec::SG::SCLagrangianDerivation}, the derivation is presented for the leading term \eqref{eq::SG::BFLagrangian0}. Here, we treat the terms containing $h_{\mu\nu}$ explicitly.

The starting points are the Lagrangians \eqref{eq::SG::BFLagrangian1}, \eqref{eq::SG::BFLagrangian2}, which we copy here again,
\begin{align}
    \mathcal{L}_{\varphi h} &= \frac 12 \sqrt{-g_s} \left(-g_s^{\mu\alpha}g_s^{\nu\beta}h_{\alpha\beta} + \frac 12 g_s^{\alpha\beta}h_{\alpha\beta}g_s^{\mu\nu}\right)\partial_\mu \varphi_c\partial_\nu \varphi_c\,,\label{eq::app::BFLagrangian1}
    \\
	\mathcal{L}_{\varphi hh} &= \frac 12 \sqrt{-g_s}\left(g_s^{\mu\alpha}g_s^{\nu\beta}g_s^{\rho\sigma}h_{\alpha\rho}h_{\beta\sigma} - \frac 12 g_s^{\alpha\beta}h_{\alpha\beta} g_s^{\mu\rho}g_s^{\nu\sigma}h_{\rho\sigma} + \frac 18 g_s^{\mu\nu} (g_s^{\alpha\beta}h_{\alpha\beta})^2\right.\nn\\
	&\quad\phantom{\sqrt{-g_s}} \left.- \frac 14 g_s^{\mu\nu}g_s^{\rho\alpha}g_s^{\sigma\beta} h_{\rho\sigma} h_{\alpha\beta}\right) \partial_\mu \varphi_c\partial_\nu \varphi_c\,,\label{eq::app::BFLagrangian2}
\end{align}
leaving out the scalar self-interactions, which are simple. 
These Lagrangians were obtained by performing the split
$g_{\mu\nu}(x) = g_{s\mu\nu}(x) + h_{\mu\nu}(x)$ in the full theory.
Next, we introduce the hatted fields $\hat{\varphi}_c$ and $\hat{h}_{\mu\nu}$ as in \eqref{eq::SG::CollinearRedefinition}, \eqref{eq::SG::CollGravRedef}, which are given by
\begin{align}\label{eq::app::hattedphi}
    \varphi_c &= \lc RW_c^{-1}\hat\varphi_c \rc\,,\\
    h_{\mu\nu} &= \brac{R\tensor{R}{_\mu^\alpha}\tensor{R}{_\nu^\beta} \tensor{W}{^{\rho}_\alpha}\tensor{W}{^{\sigma}_\beta} \brac{W_c^{-1} (\hat{g}_{s\rho\sigma}(x) + \hat{h}_{\rho\sigma})} - \hat{g}_{s\mu\nu}(x)}\,.
    \label{eq::app::hattedh}
\end{align}
Here, $\hat{g}_{s\mu\nu}$ is the residual soft metric \eqref{eq::SG::ResidualMetric1} -- \eqref{eq::SG::ResidualMetric3}.
The Wilson lines $W_c$ and $R$ are defined in \eqref{eq::SG::CollWilsonLine} and \eqref{eq::SG::RFLNCDef}, respectively. Throughout this section, we suppress the arguments of $\hat{g}_{s\mu\nu}(x)$ and $R(x)$. Other soft fields without argument are understood to be evaluated at $x_-$.

We first write the Lagrangians in collinear light-cone gauge, and then insert the relations  \eqref{eq::app::hattedphi}, \eqref{eq::app::hattedh} between the original and hatted fields. We do this order by order.
	The derivation for $\mathcal{L}_{\varphi}$ is presented in detail in \cref{sec::SG::SCLagrangianDerivation}.
	Thus, we proceed with $\mathcal{L}_{\varphi h}$ given in \eqref{eq::app::BFLagrangian1}, which contains two terms. The first one is given by
	\begin{equation}
	    \mathcal{L}_{\varphi h,\, 1} = -\frac 12 \sqrt{-g_s}g_s^{\mu\alpha}g_s^{\nu\beta}h_{\alpha\beta}\partial_\mu\varphi_c\partial_\nu\varphi_c\,.
	\end{equation}
	Inserting the redefinitions \eqref{eq::app::hattedphi}, \eqref{eq::app::hattedh}, it turns into 
	\begin{align}
	    \mathcal{L}_{\varphi h,\,1} &= -\frac 12 \sqrt{-g_s}g_s^{\mu\alpha}g_s^{\nu\beta}
	    \brac{R\tensor{R}{_\alpha^\kappa}\tensor{R}{_\beta^\lambda}\left(
	    \tensor{W}{^\rho_\kappa} \tensor{W}{^\sigma_\lambda}\brac{W^{-1}_c(\hat{h}_{\rho\sigma} + \hat{g}_{s\rho\sigma})} - \hat{g}_{s\kappa\lambda}\right)}\nn\\
	    &\quad\times\brac{\partial_\mu (RW^{-1}_c\hat{\varphi}_c)}\brac{\partial_\nu(RW^{-1}_c\hat{\varphi}_c)}\,.
	\end{align}
Integrating by parts to remove the $R$ Wilson lines from 
the scalar field gives
	\begin{equation}
	    \mathcal{L}_{\varphi h,\,1} = -\frac 12 \det\!\lp \underline{R}\rp  \brac{R^{-1}\sqrt{-g_s}}\, \tensor{R}{_\alpha^\rho}\tensor{R}{_\mu^\kappa} \brac{R^{-1}g^{\mu\alpha}_s}\,
	    \tensor{R}{_\beta^\sigma}\tensor{R}{_\nu^\lambda}\brac{R^{-1}g^{\nu\beta}_s}\,
	    \mathcal{M}_{\rho \sigma \kappa \lambda}\,,
	\end{equation}
	where we defined 
	\begin{equation}\label{eq::app::MRestterm}
	    \mathcal{M}_{\rho\sigma\kappa\lambda} \equiv \left(
	    \tensor{W}{^\alpha_\rho} \tensor{W}{^\beta_\sigma}\brac{W^{-1}_c(\hat{h}_{\alpha\beta} + \hat{g}_{s\alpha\beta})} - \hat{g}_{s\rho \sigma}\right)\brac{\partial_\kappa W^{-1}_c\hat{\varphi}_c}\brac{\partial_\lambda W^{-1}_c\hat{\varphi}_c}\,.
	\end{equation}
We add and subtract
	\begin{equation}
	    -\frac 12 \sqrt{-\hat{g}_s}\hat{g}_s^{\mu\alpha}\hat{g}_s^{\nu\beta}\mathcal{M}_{\alpha\beta\mu\nu}\,,
\label{eq:addsubterm}
	\end{equation}
	and identify the combination of terms that is covariant with respect to the background field $\hat{g}_{s\mu\nu}$ and the combination proportional to the manifestly covariant piece $\mathfrak{g}_{s\mu\nu}$.
	This yields
	\begin{align}
	    \mathcal{L}_{\varphi h,\,1} &= -\frac 12 \sqrt{-\hat{g}_s}\hat{g}_s^{\mu\alpha}\hat{g}_s^{\nu\beta}\mathcal{M}_{\alpha\beta\mu\nu}\nn\\
	    &\quad -\frac 12 \biggl( \det\!\lp \underline{R}\rp  \brac{R^{-1}\sqrt{-g_s}}\, \tensor{R}{_\alpha^\rho}\tensor{R}{_\mu^\kappa} \brac{R^{-1}g^{\mu\alpha}_s}\,
	    \tensor{R}{_\beta^\sigma}\tensor{R}{_\nu^\lambda}\brac{R^{-1}g^{\nu\beta}_s}\nn\\
	    &\qquad - \sqrt{-\hat{g}_s}\hat{g}_s^{\rho \kappa}\hat{g}_s^{\sigma \lambda}\biggr)\mathcal{M}_{\rho\sigma \kappa\lambda}\,.
	\end{align}
Note that the second and third line generate terms of higher order than $\mathcal{O}(\lambda^2)$, so we can already ignore them for the following discussion. Proceeding the same way for the second term in $\mathcal{L}_{\varphi h}$, we find
	\begin{align}
	    \mathcal{L}_{\varphi h,\,2} &= \frac 14 \sqrt{g}_s g_s^{\alpha\beta}g_s^{\mu\nu}
	    \brac{ R \tensor{R}{_\alpha^\kappa}\tensor{R}{_\beta^\lambda}\left(
	    \tensor{W}{^\rho_\kappa}\tensor{W}{^\sigma_\lambda}\brac{W^{-1}_c(\hat{h}_{\rho\sigma} + \hat{g}_{s\rho\sigma})} - \hat{g}_{s\kappa\lambda}\right)}\nn\\
	    &\quad\times\brac{\partial_\mu RW^{-1}_c\hat{\varphi}_c}\brac{\partial_\nu RW^{-1}_c\hat{\varphi}_c}\\
	    &= \frac 14 \det\!\lp \underline{R}\rp  \brac{R^{-1}\sqrt{-g_s}}\,
	    \tensor{R}{_\alpha^\rho}\tensor{R}{_\beta^\sigma}\brac{R^{-1}g_s^{\alpha\beta}}\,
	    \tensor{R}{_\mu^\kappa}\tensor{R}{_\nu^\lambda}\brac{R^{-1}g_s^{\mu\nu}}\,
	    \mathcal{M}_{\rho\sigma\kappa\lambda}\,,\nn
	\end{align}
with the same $\mathcal{M}$ as before, \eqref{eq::app::MRestterm}. We again add and subtract a term of the form 
\eqref{eq:addsubterm},
	and find
	\begin{align}
	    \mathcal{L}_{\varphi h,\,2} &= \frac 14 \sqrt{-\hat{g}}_s\hat{g}_s^{\alpha\beta}\hat{g}_s^{\mu\nu}\mathcal{M}_{\alpha\beta\mu\nu}\nn\\
	    &\quad + \frac 14 \biggl(\det\!\lp \underline{R}\rp  \brac{R^{-1}\sqrt{-g_s}}\,
	    \tensor{R}{_\alpha^\rho}\tensor{R}{_\beta^\sigma}\brac{R^{-1}g_s^{\alpha\beta}}\,
	    \tensor{R}{_\mu^\kappa}\tensor{R}{_\nu^\lambda}\brac{R^{-1}g_s^{\mu\nu}}\nn\\
	    &\qquad - \sqrt{-\hat{g}_s}\,\hat{g}_s^{\rho\sigma}\hat{g}^{\kappa\lambda}_s\biggr)\mathcal{M}_{\rho\sigma\kappa\lambda}\,.\label{eq::app::WLagrangian1}
	\end{align}
As for the first part of the Lagrangian, only the first term 
is relevant at $\mathcal{O}(\lambda^2)$. Dropping the others, writing out $\mathcal{M}$, and absorbing all Wilson lines inside the gauge-invariant building blocks, one obtains for the sum of both parts the expression 
	\begin{equation}\label{eq::app::L1LCgaugeresult}
	    \mathcal{L}_{\varphi h} = \frac 12 \sqrt{-\hat{g}_s} \left(-\hat{g}_s^{\mu\alpha}\hat{g}_s^{\nu\beta}\hat{\mathfrak{h}}_{\alpha\beta} + \frac 12 \hat{g}_s^{\alpha\beta}\hat{\mathfrak{h}}_{\alpha\beta}\hat{g}_s^{\mu\nu}\right)\partial_\mu \hat{\chi}_c\partial_\nu \hat{\chi}_c\,. 
	\end{equation}

Performing exactly the same manipulations, we find for $\mathcal{L}^{(2)}$ given in \eqref{eq::app::BFLagrangian2},
	\begin{align}
	    \mathcal{L}_{\varphi hh} &= \frac 12 \sqrt{-\hat{g}_s}\left(\hat{g}_s^{\mu\alpha}\hat{g}_s^{\nu\beta}\hat{g}_s^{\rho\sigma}\hat{\mathfrak{h}}_{\alpha\rho}\hat{\mathfrak{h}}_{\beta\sigma} - \frac 12 \hat{g}_s^{\alpha\beta}\hat{\mathfrak{h}}_{\alpha\beta} \hat{g}_s^{\mu\rho}\hat{g}_s^{\nu\sigma}h_{\rho\sigma} + \frac 18 \hat{g}_s^{\mu\nu}(\hat{g}_s^{\alpha\beta}\hat{\mathfrak{h}}_{\alpha\beta})^2\right.\nn\\
	&\quad\phantom{\sqrt{-g_s}} \left.- \frac 14 \hat{g}_s^{\mu\nu}\hat{g}_s^{\rho\alpha}g_s^{\sigma\beta} \hat{\mathfrak{h}}_{\rho\sigma} \hat{\mathfrak{h}}_{\alpha\beta}\right) \partial_\mu \hat{\chi}_c\partial_\nu \hat{\chi}_c\,,
	\label{eq::app::L2LCgaugeresult}
	\end{align}
	as stated in \eqref{eq::SG::FinalDressedLagrangian} -- \eqref{eq::SG::FinalDressedLagrangian2}, where the scalar 
self-interactions are included.
	
Finally, we show how the $W_c$ Wilson line cancels out in all terms that do not contain the manifestly covariant $\mathfrak{g}_{s\mu\nu}$ field.
	These terms are the leading terms in the $\lambda$-expansion at each order in the collinear $\hat{h}_{\mu\nu}$ expansion. The first three of these terms are given by
	\begin{align}
	    \left.\mathcal{L}_{\varphi}\right\rvert_{\mathfrak{g}_{s\mu\nu}=0} &= \frac{1}{2}\sqrt{-\hat{g}_s} \hat{g}_s^{\mu\nu} \lc \partial_\mu W_c^{-1}\hat{\varphi}_c\rc \lc\partial_\nu W^{-1}_c\hat{\varphi}_c\rc\,,\\ 
	     \left.\mathcal{L}_{\varphi h}\right\rvert_{\mathfrak{g}_{s\mu\nu}=0} &= \left(-\hat{g}_s^{\mu\alpha}\hat{g}_s^{\nu\beta}\hat{\mathfrak{h}}_{\alpha\beta} + \frac 12 \hat{g}_s^{\alpha\beta}\hat{\mathfrak{h}}_{\alpha\beta}\hat{g}_s^{\mu\nu}\right)
	     \lc \partial_\mu W_c^{-1}\hat{\varphi}_c\rc \lc\partial_\nu W^{-1}_c\hat{\varphi}_c\rc\,,\\
	    \left.\mathcal{L}_{\varphi hh}\right\rvert_{\mathfrak{g}_{s\mu\nu}=0} &=
	    \frac 12 \sqrt{-\hat{g}_s}\left(\hat{g}_s^{\mu\alpha}\hat{g}_s^{\nu\beta}\hat{g}_s^{\rho\sigma}\hat{\mathfrak{h}}_{\alpha\rho}\hat{\mathfrak{h}}_{\beta\sigma} - \frac 12 \hat{g}_s^{\alpha\beta}\hat{\mathfrak{h}}_{\alpha\beta} \hat{g}_s^{\mu\rho}\hat{g}_s^{\nu\sigma}h_{\rho\sigma} + \frac 18 \hat{g}_s^{\mu\nu}(\hat{g}_s^{\alpha\beta}\hat{\mathfrak{h}}_{\alpha\beta})^2\right.\nn\\
	&\quad\phantom{\sqrt{-g_s}} \left.- \frac 14 \hat{g}_s^{\mu\nu}\hat{g}_s^{\rho\alpha}\hat{g}_s^{\sigma\beta} \hat{\mathfrak{h}}_{\rho\sigma} \hat{\mathfrak{h}}_{\alpha\beta}\right) \lc \partial_\mu W_c^{-1}\hat{\varphi}_c\rc \lc\partial_\nu W^{-1}_c\hat{\varphi}_c\rc\,.
	\end{align}
The terms $\left.\mathcal{L}_{\varphi h^i}\right\rvert_{\mathfrak{g}_{s\mu\nu}=0}$, when summed to all orders in $\hat{h}_{\mu\nu}$, can be expressed in the closed form
	\begin{equation}\label{eq::app::WcancelLagrangian}
	    \sum_{i=0}^{\infty}\left.\mathcal{L}_{\varphi h^i}\right\rvert_{\mathfrak{g}_{s\mu\nu}=0} =
	    \sqrt{-\bar{g}}\bar{g}^{\mu\nu} \lc\partial_\mu \hat{\chi}_c\rc \lc\partial_\nu \hat{\chi}_c\rc\,,
	\end{equation}
in terms of the metric tensor
	\begin{equation}
	    \bar{g}_{\mu\nu} = \hat{g}_{s\mu\nu} + \hat{\mathfrak{h}}_{\mu\nu}\,.
	\end{equation}
    This metric describes a fluctuation $\hat{\mathfrak{h}}_{\mu\nu}$ around the background $\hat{g}_{s\mu\nu}$.
    The Lagrangian density \eqref{eq::app::WcancelLagrangian} thus describes a scalar field $\hat{\chi}_c$ coupled to this background, and is gauge-invariant under arbitrary diffeomorphisms, hence in particular under the inherited transformations of the collinear fluctuations $\hat{\chi}_c$ and $\hat{\mathfrak{h}}_{\mu\nu}$.
    Next, note that the definitions
    \begin{align}
        \hat{\chi}_c &= \lc W_c^{-1} \hat{\varphi}_c\rc \,,\\
        \hat{\mathfrak{h}}_{\mu\nu} &= \tensor{W}{^\alpha_\mu}\tensor{W}{^\beta_\nu}\lc W^{-1}_c \lp\hat{h}_{\alpha\beta} + \hat{g}_{s\alpha\beta}\rp\rc  - \hat{g}_{s\mu\nu}\,,
    \end{align}
    already take the form of an (inverse) gauge transformation of the fields $\hat{\varphi}_c$ and $\hat{h}_{\mu\nu}$.
    These collinear Wilson lines thus cancels out in the terms $\sum_{i=0}^{\infty}\left.\mathcal{L}_{\varphi h^i}\right\rvert_{\mathfrak{g}_{s\mu\nu}=0}$.
    To see this explicitly, first observe that
    \begin{equation}
        \bar{g}_{\mu\nu} = \hat{g}_{s\mu\nu} + \hat{\mathfrak{h}}_{\mu\nu} = \tensor{W}{^\alpha_\mu}\tensor{W}{^\beta_\nu}\lc W^{-1}_c\tilde{g}_{\alpha\beta}\rc\,,
    \end{equation}
    where $\tilde{g}_{\alpha\beta} = \hat{g}_{s\alpha\beta} + \hat{h}_{\alpha\beta}$ is the unfixed metric.
    Then, its inverse is given by
    \begin{equation}
        \bar{g}^{\mu\nu} = \tensor{W}{_\alpha^\mu}\tensor{W}{_\beta^\nu}\lc W^{-1}_c\tilde{g}^{\alpha\beta}\rc\,.
    \end{equation}
    Inserting this into the Lagrangian \eqref{eq::app::WcancelLagrangian}, one finds
    \begin{align}
        \mathcal{L}_{\bar{g}}
        &= \det\!\lp \underline{W}\rp\lc W^{-1}_c\sqrt{-\tilde{g}}\rc \tensor{W}{_\alpha^\mu}\tensor{W}{_\beta^\nu}\lc W^{-1}_c\tilde{g}^{\alpha\beta}\rc
        \lc \partial_\mu W_c^{-1}\hat{\varphi}_c\rc 
        \lc \partial_\nu W_c^{-1}\hat{\varphi}_c\rc\nn\\
        &= \det\!\lp \underline{W}\rp\lc W^{-1}_c\sqrt{-\tilde{g}}\rc \tensor{W}{_\alpha^\mu}\tensor{W}{_\beta^\nu}\lc W^{-1}_c\tilde{g}^{\alpha\beta}\rc
        \tensor{W}{^\rho_\mu} \lc W_c^{-1} \partial_\rho \hat{\varphi}_c\rc 
        \tensor{W}{^\sigma_\nu}\lc W_c^{-1}\partial_\sigma \hat{\varphi}_c\rc\nn\\
        &= \det\!\lp \underline{W}\rp\lc W^{-1}_c\sqrt{-\tilde{g}}
        \tilde{g}^{\mu\nu}
        \lc\partial_\mu \hat{\varphi}_c\rc 
        \lc\partial_\nu \hat{\varphi}_c\rc\rc\nn\\
        &= \sqrt{-\tilde{g}} \tilde{g}^{\mu\nu}\lc\partial_\mu\hat{\varphi}_c\rc\lc\partial_\nu\hat{\varphi}_c\rc + \mathrm{t.d.}\label{eq::app::Wilsoncancellation}
    \end{align}
    In the last line, we used that the combination $\det\!\lp \underline{W}\rp W^{-1}_c$ changes the Lagrangian only by a total derivative, similar to \eqref{eq::app::UsefulFullTranslation}.
    Alternatively, we can use that the Lagrangian is integrated over $d^4x$ to drop the (inverse) translation, by performing the translation on $x$. This argument shows that for the terms that do not contain $\mathfrak{g}_{s\mu\nu}$, we can drop the collinear Wilson line $W_c$.
    Only if a soft-covariant building block, i.e. $\mathfrak{g}_{s\mu\nu}$, resp. the Riemann tensor after multipole expansion, is present, we cannot remove $W_c$.
	
\bibliography{gravity}

\end{document}